\documentclass[twoside,12pt]{article}  
\usepackage[numbers,square,sort&compress]{natbib}
\usepackage[dvips]{graphicx}
\usepackage{hyperref}
\hypersetup{
 colorlinks=true,     
 linkcolor=blue,      
 citecolor=cyan,      
 filecolor=magenta,   
 urlcolor=cyan        
}
\usepackage{lineno}
\usepackage{soul,upgreek}
\usepackage{booktabs}
\usepackage{multirow,longtable}
\usepackage{amssymb}                 
\usepackage{amsmath}                 
\usepackage{bm}                      
\usepackage{xcolor}

\def\text#1{\mathrm{#1}}
\def\off#1{}

\def\be{\begin{equation}} \def\ee{\end{equation}}
\def\bal#1\eal{\begin{align}#1\end{align}}
\def\pv{\bm{p}}
\def\kv{\bm{k}}
\def\qv{\bm{q}}
\def\rv{\bm{r}}
\def\sv{\bm{\sigma}}
\def\bra{\langle}
\def\ket{\rangle}
\def\ra{\rightarrow}
\def\eps{\varepsilon}
\def\de{\Delta}
\def\la{\Lambda}
\def\om{\omega}
\def\al{\alpha}
\def\ms{\,M_\odot}
\def\fm3{\,\text{fm}^{-3}}
\def\gc3{\,\text{g/cm}^3}
\def\lgi{L_\gamma^\infty}

\begin{document}


\title{Neutron Stars and the Nuclear Equation of State}

\author{
G.F.\ Burgio, H.-J.\ Schulze, I.\ Vida\~na, and J.-B.\ Wei
\\[2mm]
INFN Sezione di Catania, Via S. Sofia 64, I-95123 Catania, Italy
}


\maketitle

\begin{abstract}
We review the current status and recent progress of microscopic
many-body approaches and phenomenological models,
which are employed to construct the equation of state of neutron stars.
The equation of state is relevant for the description of their
structure and dynamical properties,
and it rules also the dynamics of core-collapse supernovae
and binary neutron star mergers.
We describe neutron star matter assuming that the main degrees of freedom
are nucleons and hyperons,
disregarding the appearance of quark matter.
We compare the theoretical predictions of the different equation-of-state models
with the currently available data coming from both
terrestrial laboratory experiments and recent astrophysical observations.
We also analyse the importance of the nuclear strong interaction
and equation of state for the cooling properties of neutron stars.
We discuss the main open challenges in the description of the equation of state,
mainly focusing on the limits of the different many-body techniques,
the so-called ``hyperon puzzle,"
and the dependence of the direct URCA processes on the equation of state.
\end{abstract}
\vfill\eject

\tableofcontents
\vfill\eject





\section{Introduction}
\label{s:intro}

Neutron stars (NS) harbor unique conditions that challenge the physical theories
of dense matter \cite{Haensel07}.
Beneath a thin stellar atmosphere,
a NS interior consists of three main layers,
i.e.~an outer crust, an inner crust, and a core,
each one characterized by different physical conditions.
At densities larger than $1.5\times10^{14}\gc3$,
nuclear matter forms a homogeneous liquid composed of neutrons
plus a certain fraction of protons, electrons, and muons
that maintain the system in beta-equilibrium,
whereas in the deep interior,
at still higher densities,
strange baryons and even deconfined quarks may appear.
On the other hand, moving from the core to the exterior,
density and pressure decrease.
When the density becomes lower than approximately $1.5\times10^{14}\gc3$,
clusterized nuclear matter sets in.
The positive charges concentrate in individual clusters of charge $Z$
and form a solid lattice to minimize the Coulomb repulsion among them.
The lattice is embedded in a gas of free neutrons and a background of electrons
such that charge neutrality is maintained \cite{Lattimer96,Haensel07}.
This layer is the so-called inner crust,
and the nuclear structures may adopt non-spherical shapes
(usually termed ``nuclear pasta'') in order to minimize their energy
\cite{Baym71,Ravenhall83,Oyamatsu93}.
At lower densities,
neutrons are finally confined within the nuclear clusters
and matter is made of a lattice of neutron-rich nuclei
permeated by a degenerate electron gas.
This layer is known as the outer crust \cite{Baym71}
and extends from the neutron drip density
$\rho_\text{drip}\approx 4\times10^{11}\gc3$ to about $10^{4}\gc3$
at the surface.

It is usually believed that asymmetric nuclear matter
forms the interior bulk part of NSs.
Despite infinite nuclear matter being obviously an idealized physical system,
the theoretical determination of the corresponding equation of state (EoS)
is an essential step towards the understanding of the NS physical properties,
and the comparison of the theoretical predictions with the
experimental observations can provide serious constraints on the nuclear EoS.
Unfortunately, only indirect observations of NS structure,
including their masses and radii, are possible.
However, the astrophysics of NSs is a rapidly developing research field,
in view of the observations coming from the currently operating satellites
(NICER, Neutron star Interior Composition Explorer, \cite{Gendreau17})
and the gravitational-wave (GW) laser interferometers
(AdvLIGO \cite{Aasi15,Abbott17}, and Virgo \cite{Acernese14}),
and it is now possible to confront the theoretical predictions
with more and more stringent phenomenological data.

Heavy ion reactions is another field of research where the nuclear EoS
is a relevant issue.
In this case, the difficulty of extracting the EoS is due to the
complexity of the processes,
since the interpretation of the data is necessarily
linked to the analysis of the reaction mechanism.
An enormous amount of work has been done in the last two decades in this field,
and some indications about the main characteristics of the EoS emerged,
but limited to a density range slightly above the nuclear saturation density,
and therefore far away from the high density range typical of the NS interior.
However, a direct link between the two fields of research
can also be impeded by the fact that the typical time scale of
heavy ion reactions is enormously different from the typical NS time scale.
Indeed, nuclear matter inside NSs is completely catalized,
i.e.~it is quite close to the ground state,
reachable also by weak processes.
In heavy ion reactions the rapidity of the dynamical evolution can hinder the
weak processes which relax the system towards such a catalized state,
and therefore the tested EoS can differ from the NS one,
especially at high density.

Further astrophysical scenarios characterized by high nuclear density
and large temperature,
and where the EoS plays a relevant role are
i) core-collapse supernova explosions (CCSN),
which lead to the formation of a very hot proto-neutron star (PNS)
and subsequently to a cold NS or to a black hole (BH)
\cite{Mezzacappa05,Janka12,Prakash97,Pons99},
ii) merging of NS in close binary systems,
NS-NS and NS-BH
\cite{Shibata11,Rosswog15,Baiotti17}.
The dynamical evolution and the structure of the forming final compact object
are strongly determined by the nuclear-matter EoS,
as well as the nucleosynthesis conditions and the emitted neutrino spectra.
Therefore there is a deep connection between the gross features
of these astrophysical phenomena and the underlying microphysics
ingredients which arise from the interactions among the particles.

Within this review we will consider only nucleonic and hyperonic
degrees of freedom,
and disregard the appearance of quark matter,
since this is a very complex topic which deserves a review by itself;
we refer the interested reader to Ref.~\cite{Blaschke18,Baym18}.

The main goal of this work is to review current theoretical approaches for
the description of NS matter that can yield EoSs relevant for compact stars.
After an extensive reexamination of the baryon-baryon interaction
in Sec.~\ref{s:bbinteraction},
where all the different theoretical approaches are discussed,
we concentrate in Sec.~\ref{s:eos} on the different theoretical methods
currently used for the construction of the EoS.
We illustrate both ab-initio and phenomenological models,
at zero and finite temperature, and confront them with experiments.
After a short review of the crust EoS in Sec.~\ref{s:crust},
we will illustrate in Sec.~\ref{s:hyp} the role of hyperonic
degrees of freedom in nuclear matter,
and its consequences on the NS structure.
In particular, the so-called ``hyperon puzzle'' is reviewed.
Sec.~\ref{s:constr} is devoted to  illustrate the current nuclear physics
and astrophysical constraints available so far,
and show how the different NS models behave
with respect to the laboratory and observational data.
Finally, in Sec.~\ref{s:cool} we discuss how the EoS determines the
cooling behaviour of NSs,
emphasizing the roles of superfluidity and nuclear pairing gaps.
Conclusions are drawn in Sec.~\ref{s:conc}.

\section{Baryon-baryon interaction}
\label{s:bbinteraction}

The features of the baryon-baryon interaction
[nucleon-nucleon (NN), hyperon-nucleon (YN), hyperon-hyperon (YY)],
in particular the presence of a hard repulsive core,
strongly determine the behaviour of the nuclear medium,
and are the basic ingredient of calculations performed within
microscopic approaches.
The dominant component is the two-body interaction,
but higher-order forces can be important,
e.g.~the three-body force (TBF) which is
crucial in reproducing the saturation properties of nuclear matter.
The first model of the strong force between hadrons was proposed by Yukawa in
1935 \cite{Yukawa35},
and this model was used to describe the nuclear
force between nucleons mediated by the $\pi$-meson,
whose mass was related to the range of interaction,
thus restricting the potential to a finite range,
which is the essential point.
Later on, with the discovery that the
fundamental theory of strong interactions is quantum chromodynamics (QCD)
and not meson theory,
all attempts to derive a proper theory of the nuclear
force had to be formulated again.
Soon it became clear that the nuclear
Hamiltonian should in principle be derived from QCD,
but this is still a very difficult task to solve,
and therefore we have to resort to alternative approaches.
Around 1990, a major breakthrough occurred when the
concept of an Effective Field Theory (EFT) was applied to low-energy QCD
\cite{Weinberg90,Weinberg91}.
This scheme is also known as Chiral Perturbation Theory (ChPT)
and allows to calculate the various terms that
make up the baryon force systematically order by order,
thus generating not only the force between two baryons,
but also many-baryon forces, all on the same footing.
Nowadays, the bare baryon interactions can be basically
divided into four classes:
i) phenomenological interactions;
ii) chiral effective field theories ($\chi$EFT);
iii) resonating group method (RGM) models that include explicitly
the quark-gluon degrees of freedom, and
iv) renormalization group methods.
A major progress has been also achieved recently
on derivation of the baryon-baryon interaction from lattice QCD calculations.
We will briefly discuss all of them in the following.

\subsection{Phenomenological models}

Theoretical approaches to construct phenomenological forces include
meson-exchange models and potential models.
In both cases, quark degrees of freedom are not treated explicitly
but are replaced by hadrons -- baryons mesons and their resonances --
in which quarks are confined.

\subsubsection{Meson-exchange models}

Meson-exchange models are based on the Yukawa idea according to which the
strong interaction among the different baryons is assumed to be mediated by
the exchange of mesons between the baryons.
The long-range part of the interaction is mediated by the exchange of
pseudoscalar mesons ($\pi$, $K$, $\eta$, $\eta'$),
whereas scalar mesons ($\sigma$, $\kappa$, $\delta$)
describe the intermediate-range attractive part,
and the vector mesons ($\rho$, $K^*$, $\om$, $\phi$)
mediate the short-range repulsive contribution.
The various models differ mainly in the mesonic content
and the treatment of two-meson exchange contributions,
but they are very successful in describing NN scattering data and phase shifts
at laboratory energies $\lesssim 350\,$MeV,
and also the deuteron properties,
even if discrepancies between the results of different groups do still exist
\cite{Machleidt01}.
Very refined and complete phenomenological models
have been constructed for the NN interactions,
e.g.~the Paris potential \cite{Lacombe80},
the Bonn potential \cite{Machleidt87,Machleidt01},
and the Nijmegen potentials \cite{Nagels77,Nagels78}.
Meson-exchange theory has been employed by the Nijmegen
\cite{Maessen89,Rijken99,Stoks99,Rijken08,Rijken10}
and J\"ulich \cite{Holzenkamp89,Juelich2}
groups to construct also YN and YY interactions.

Following symmetry principles,
simplicity and physical intuition,
the most general interaction Lagrangians that couple the meson fields
to the baryon ones can be written as
\bal
 \mathcal{L}_s &= g_s\bar\Psi\Psi \Phi^{(s)} \:,
\\
 \mathcal{L}_{ps} &= g_{ps}\bar\Psi i\gamma^5\Psi \Phi^{(ps)} \:,
\\
 \mathcal{L}_v &=
 g_v\bar\Psi \gamma^\mu\Psi \Phi^{(v)}_\mu
 + g_t\bar\Psi \sigma^{\mu\nu}\Psi
 \left( \partial_\mu\Phi^{(v)}_\nu-\partial_\nu\Phi^{(v)}_\mu\right) \:,
\eal
for scalar, pseudoscalar and vector coupling, respectively.
Alternatively, for the pseudoscalar field one can write also the so-called
pseudovector (pv) or gradient coupling,
which is suggested as an effective coupling by chiral symmetry
\begin{equation}
 \mathcal{L}_{pv} =
 g_{pv}\bar\Psi \gamma^5\gamma^\mu\Psi\partial_\mu\Phi^{(ps)} \:.
\end{equation}
In the above expressions $\Psi$ denotes the baryon fields for spin-1/2 baryons,
$\Phi^{(s)}, \Phi^{(ps)}$ and $\Phi^{(v)}$ are the corresponding
scalar, pseudoscalar and vector fields,
and the $g$'s are the corresponding coupling constants that must be constrained,
when possible, by scattering data.
Note that the above Lagrangians are for isoscalar mesons.
To obtain the Lagrangians describing the couplings with the isovector mesons,
the fields $\Phi$ should be simply replaced by $\bm\tau\cdot\bm\Phi$
in the previous expressions,
with $\bm\tau$ being the usual isospin Pauli matrices.

A typical contribution to the baryon-baryon scattering amplitude
arising from the exchange of a certain meson $\Phi$ is given by
\begin{equation}
 \bra p_1'p_2'|V_\Phi|p_1p_2\ket =
 \frac{\bar u (p_1')g_1\Gamma_1u(p_1)P_\Phi \bar u (p_2')g_2\Gamma_2u(p_2)}
 {(p_1-p_1')^2-m_\Phi^2} \:,
\end{equation}
where $P_\Phi/[(p_1-p_1')^2-m_\Phi^2]$
represents the meson propagator,
$m_\Phi$ is the mass of the exchanged meson,
$u$ and $\bar u$ are the usual Dirac spinor and its adjoint
($\bar uu=1, \bar u=u^\dagger\gamma^0$),
$g_1$ and $g_2$ are the coupling constants at the two vertices,
being the $\Gamma$'s their corresponding Dirac structures
\be
\Gamma_{s} =1 \ , \,\,
\Gamma_{ps}=i\gamma^5 \ , \,\,
\Gamma_{v}=\gamma^\mu \ , \,\,
\Gamma_{t}=\sigma^{\mu\nu} \ , \,\,
\Gamma_{pv}=\gamma^5\gamma^\mu\partial_\mu \:.
\ee

In the case of scalar and pseudoscalar meson exchanges,
the numerator $P_\Phi$ of the propagator is just~1.
For vector-meson exchange, however, it is the rank-2 tensor
\begin{equation}
 P_\Phi\equiv P_{\mu\nu} = -g_{\mu\nu}+\frac{q_\mu q_\nu}{m_\Phi^2} \:,
\label{e:numerator}
\end{equation}
where $g_{\mu\nu}=\mbox{diag}(1,-1,-1,-1)$ is the usual Minkowski metric tensor
and $q_\mu=(p_1-p_1')_\mu$ is the four-momentum transfer.

In general, when all types of baryons are included,
the interaction potential will be simply the sum of all the partial contributions
\begin{equation}
 \bra p_1'p_2'|V|p_1p_2\ket =
 \sum_{\Phi} \bra p_1'p_2'|V_\Phi|p_1p_2\ket \:.
\end{equation}

Expanding the free Dirac spinor in terms of $1/M$
(where $M$ is the mass of the relevant baryon)
to lowest order leads to the familiar non-relativistic expressions
for the baryon-baryon potentials,
which through Fourier transformation give the configuration-space version
of the interaction.
The general expression for the local approximation
of the baryon-baryon interaction in configuration space is
\bal
 V(\rv) &= \sum_{\Phi}\left\{
 C_{C_\Phi} + C_{\sigma_\Phi}\sv_1\cdot\sv_2
 + C_{LS_\Phi}\left(\frac{1}{m_\Phi r}+\frac{1}{(m_\Phi r)^2} \right)
 \bm L \cdot \bm S \right.
\nonumber \\ &\hskip15mm
 \left. +\, C_{T_\Phi}\left(1+\frac{3}{m_\Phi r}
 + \frac{3}{(m_\Phi r)^2} \right)S_{12}(\hat\rv) \right\}
 \frac{e^{-m_\Phi r}}{r} \:,
\label{e:rspace}
\eal
where $C_{C_\Phi}, C_{\sigma_\Phi}, C_{LS_\Phi}$ and $C_{T_\Phi}$
are numerical factors containing the baryon-baryon-meson coupling constants
and the baryon masses,
$\bm L$ is the total orbital angular momentum,
$\bm S$ is the total spin,
and $S_{12}(\hat\rv)$ is the usual tensor operator in configuration space,
\begin{equation}
 S_{12}(\hat\rv) = 3(\sv_1\cdot \hat\rv)(\sv_2 \cdot \hat\rv)
 - (\sv_1\cdot \sv_2) \ , \,\,\,
 \hat\rv = \frac{\rv}{|\rv|} \:.
\label{e:tensor}
\end{equation}

Finally, one has to remember that in the meson exchange theory all
meson-baryon vertices must be necessarily modified by the introduction
of the so-called form factors.
Each vertex is multiplied by a form factor of the type
\begin{equation}
 F_\al(|\kv|^2) =
 \left(\frac{\la_\al^2-m_\al^2}{\la_\al^2+|\kv|^2} \right)^{n_\al}
\label{e:ff}
\end{equation}
or by
\begin{equation}
 F_\al(|\kv|^2) = \exp\left(-\frac{|\kv|^2}{2\la_\al^2}\right) \:.
\label{e:ff2}
\end{equation}
In Eq.~(\ref{e:ff}) the quantity $n_\al$ is usually taken equal
to 1 (monopole form factor) or 2 (dipole form factor).
The vector $\kv$ denotes the three-momentum transfer,
whereas $\la_\al$ is the so-called cut-off mass,
typically of the order $1.2 - 2\,$GeV.
Originally the form factors were introduced for purely mathematical reasons,
namely to avoid divergences in the scattering equation.
Nowadays our present knowledge of the (quark) substructure of baryons and mesons
provides a physical reason for their presence.
Obviously, it does not make sense to take the meson-exchange picture seriously
in a region in which modifications due to the extended structure of hadrons
come into play.

\subsubsection{Potential models}
\label{s:tbf}

As far as the potential models are concerned,
they have a quite complex structure expressed via operator invariants
consistent with the symmetries of the strong interactions,
i.e.~translational invariance, Galilean invariance, rotational invariance,
space-reflection invariance, time-reversal invariance,
invariance under the interchange of particles 1 and 2, isospin symmetry,
and hermiticity.
The most widely known potential models are the Urbana \cite{Lagaris81}
and Argonne potential \cite{Wiringa95},
in which a sufficiently generic form of the NN potential
acting between a nucleon pair $ij$,
able to reproduce the abundance of NN scattering data,
is expressed in operatorial form as
\begin{equation}
 \hat V_{ij} = \sum_{u=1}^{18} V_u(r_{ij}) \hat O_{ij}^u \:,
\label{e:av18}
\end{equation}
with
\bal
 \hat O_{ij}^{u=1,...14} =&
 1,\, \bm{\tau}_i \cdot \bm{\tau}_j,\, \sv_i \cdot \sv_j,\,
 (\sv_i \cdot \sv_j) (\bm{\tau}_i \cdot \bm{\tau}_j),\,
 \hat S_{ij},\, \hat S_{ij} (\bm{\tau}_i \cdot \bm{\tau}_j),
\nonumber \\ &
 \hat{\bm{L}} \cdot \hat{\bm{S}},\,
 \hat{\bm{L}} \cdot \hat{\bm{S}} (\bm{\tau}_i \cdot \bm{\tau}_j),\, \hat L^2,\,
 \hat L^2 (\bm{\tau}_i \cdot \bm{\tau}_j),\,
 \hat L^2 (\sv_i \cdot \sv_j),
\nonumber \\ &
 \hat L^2 (\sv_i \cdot \sv_j) (\bm{\tau}_i \cdot \bm{\tau}_j),\,
 (\hat {\bm{L}} \cdot \hat {\bm{S}})^2,\,
 (\hat {\bm{L}} \cdot \hat {\bm{S}})^2 (\bm{\tau}_i \cdot \bm{\tau}_j) \:.
\label{e:av18_op}
\eal
The notation $\rv_{ij} = \rv_i - \rv_j$
indicates the relative position vector.
The operators $\sv_i$ and $\sv_j$ are spins, whereas
$\bm{\tau}_i$ and $\bm{\tau}_j$ are isospins
(both in units of $\hbar/2$),
and the tensor operator is given in Eq.~(\ref{e:tensor}).
The relative momentum is denoted by
$\hat{\pv}_{ij} = \hat{\pv_i}-\hat{\pv_j}$;
$\hat{\bm{L}} = \rv_{ij} \times \hat{\pv}_{ij}$
is the total orbital angular momentum,
and $\hat{L}^2 $ its square in the centre-of-mass system.
The spin-orbit coupling enters via
$\hat{\bm{L}} \cdot \hat{\bm{S}}$,
being $\hat{\bm{S}} = (\sv_i + \sv_j)/2$ the total spin
(in units of $\hbar$).
The first fourteen operators are charge-independent,
whereas the terms with $\hat{O}_{ij}^{u=15,...18}$
are small and break charge independence,
and they correspond to $V_{np}(T=1) = V_{nn} = V_{pp}$,
while the charge symmetry implies only that $V_{nn} = V_{pp}$.
Modern fits to very precise nucleon scattering data indicate the existence
of charge-independence breaking,
but the effect of such forces on the energy of nucleonic matter
is much smaller than the uncertainties of many-body calculations
and therefore can be neglected while constructing the EoS.

Calculations show that the two-body interactions which satisfactorily reproduce
NN scattering and deuteron properties,
give the binding energies of $^3$H and $^4$He systematically lower
than experimental ones.
This indicates the necessity of introducing TBFs into the nuclear Hamiltonian.
Moreover, realistic two-body forces saturate the nuclear matter
at a density significantly higher than $0.16\fm3$.
The TBF can correct this,
because it supplies a repulsion which increases rapidly
with the growth of nucleon density.
The TBF $V_{ijk}$ depends on spatial, spin and isospin coordinates
of three nucleons and cannot be reduced to a sum of two-body interactions
involving these coordinates.
Widely used phenomenological models of TBF are dubbed as Urbana UVII and UIX
\cite{Schiavilla86,Pudliner95},
which are expressed by the following decomposition
\begin{equation}
 \hat V_{ijk} = \hat V_{ijk}^{2\pi}  + \hat V_{ijk}^{IS} \:,
\end{equation}
where the two-pion exchange part is dominant at low densities
and provides the additional binding for the $^3$H and $^4$He nuclei,
whereas the IS part gives a repulsion at higher densities that is needed
to saturate correctly the symmetric nuclear matter (SNM);
this part does not contribute much in low-density systems such as light nuclei.

Besides phenomenological models,
microscopic approaches to the calculations of TBF are available,
in which the same meson-exchange parameters as the two-body potentials
are employed \cite{Grange89,Zuo02a,Zuo02b,Li08a,Li08}.
Although this approach has been frequently adopted in the EoS calculations of
nuclear matter,
the theory still needs to be refined,
and further studies are required.

\subsection{Chiral perturbation expansion}

A different approach to the study of the baryon-baryon interactions is the
one based on quark and gluon degrees of freedom,
thus connecting the low-energy hadron physics phenomena
with the underlying QCD structure of the baryons.
However, deriving nuclear forces directly from QCD is problematic,
because the whole hadron sector is in the non-perturbative regime,
due to confinement.
In fact each nucleon is, by itself,
a complicated many-body system consisting of quarks, quark-antiquark pairs
and gluons,
thus rendering the two-nucleon problem an even more complex many-body problem.
Second, the interaction among quarks,
which is due to the exchange of gluons,
is very strong at the low energies involved in nuclear physics processes.
For this reason,
it is difficult to find converging perturbative solutions.

A new era for the theory of nuclear forces started when Steven Weinberg
\cite{Weinberg90,Weinberg91} worked out an effective field theory (EFT)
for low-energy QCD.
He argued that all one needs to do is to write the most general Lagrangian
consistent with all the properties of low-energy QCD,
as this would render the theory equivalent to low-energy QCD.
A crucially important property for this discussion is the chiral symmetry,
which is spontaneously broken,
since the bare $u$ and $d$ quark masses in non-strange matter
are just a few MeV.
According to a theorem first proven by Goldstone,
the spontaneous symmetry breaking implies the existence of a pseudoscalar meson,
the pion.
Thus, the pion plays an outstanding role in generating the nuclear force.
Along this line, Weinberg \cite{Weinberg90,Weinberg91} proposed a scheme
for including in the interaction a series of operators
which reflect the partially broken chiral symmetry of QCD.
The strength parameters associated to each operator are
then determined by fitting the properties of few-body nuclear systems,
i.e.~deuteron and NN phase shifts.
The method is then implemented in the framework of the Effective Field Theory
(EFT) by ordering the terms according to their dependence
on the physical parameter $q/m$,
where $m$ is the nucleon mass and $q$ a generic momentum that appears
in the Feynman diagram for the considered process.
This parameter is assumed to be small and each term is dependent
on a given power of this parameter, thus fixing its relevance.

In this way the chiral perturbation expansion (ChPE) allows to calculate
the various contributions to the potential systematically order by order,
where each order refers to a particular power of the momentum.
In particular, the pion-exchange term is treated explicitly
and is considered the lowest-order (LO) term of the expansion.
Furthermore, the ChPE can generate not only the force between two nucleons,
but also many-nucleon forces in a consistent manner;
in fact the TBFs so introduced are of higher order than the simplest
two-body forces and they are treated on an equal footing.
They arise first at next-to-next-to-leading order (N2LO) and,
as a consequence,
because of the hierarchy intrinsic in the chiral expansion,
TBFs are expected to be smaller than two-body forces,
at least within the range of validity of the expansion,
whereas four-body forces appear only at
next-to-next-to-next-to-leading order (N3LO) level, and so on.
This ChPE can be used to construct NN interactions that are of reasonably good
quality in reproducing the two-body data \cite{Entem02,Holt09}.
The assumption of a small $q/m$ parameter in principle restricts the
applications of these forces to not too large momenta,
and therefore to a not too large density of nuclear matter.
It turns out that the safe maximum density is around
the saturation value, $n_0$.
During the last years the NN interaction has been described to high precision
using ChPE \cite{Entem03,Epelbaum05,Entem17}.
At present the nucleonic interaction has been calculated up to N4LO \cite{Hu17}.
An exhaustive list of higher-order diagrams up to N3LO can be found
in review papers \cite{Meissner05, Epelbaum09}.
This method has been refined along the
years and many applications can be found in the literature.

There are very few studies of the YN interaction using ChPE in
comparison to the NN case.
A recent extension of the scheme used in Ref.~\cite{Epelbaum05}
to the YN and the YY interactions has been performed
by the J\"ulich-Bonn-Munich group
\cite{Polinder06,Haidenbauer13,Haidenbauer20,Petschauer20}.
In the following, we briefly describe the ChPE approach to the baryon-baryon
interaction at the two lowest orders (LO and NLO) of the chiral expansion
and refer the interested reader to the original works of this approach for
details.

\begin{figure}[t]
\begin{center}
\includegraphics[width=10cm]{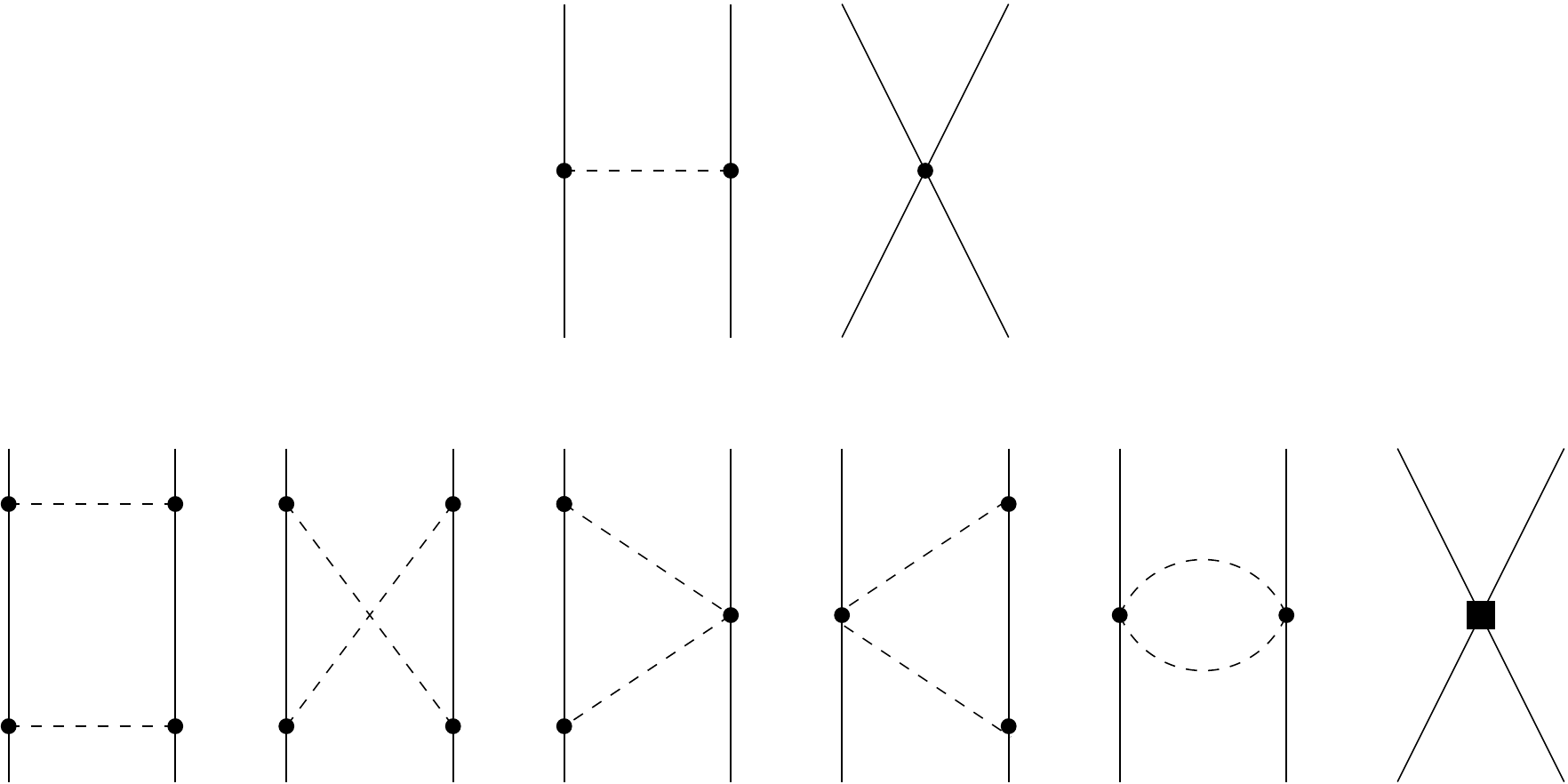}
\end{center}
\caption{
Leading-order (upper diagrams)
and next-to-leading-order (lower diagrams)
contributions to the baryon-baryon interaction.
Figure adapted from Ref.~\cite{Haidenbauer13}.}
\label{fig:xeft}
\end{figure}

At leading order in the power counting the baryon-baryon potential
consist of one pseudoscalar-meson exchange and of four-baryon contact terms,
where each of these two contributions is constrained via SU(3)-flavor symmetry
(see the upper diagrams of Fig.~\ref{fig:xeft}).
The contribution from the one pseudoscalar-meson exchange term
is obtained from the Lagrangian density
\begin{equation}
 {\mathcal{L}} = \bra i \bar B \gamma^\mu D_\mu B-M_0\bar B B
 + \frac{D}{2}\bar B \gamma^\mu\gamma_5\{u_\mu,B\}
 +\frac{F}{2}\bar B\gamma^\mu\gamma_5[u_\mu,B] \ket \:,
\label{e:loopsx}
\end{equation}
where the brackets denote the trace in flavor space,
$B$ is the irreducible baryon octet representation of SU(3)-flavor given by
\be
 B = \left(
\begin{array}{ccc}
\frac{\Sigma^0}{\sqrt{2}}+\frac{\la}{\sqrt{6}} & \Sigma^+ &  p \\
\Sigma^- &  -\frac{\Sigma^0}{\sqrt{2}}+\frac{\la}{\sqrt{6}} & n \\
 -\Xi^-& \Xi^0 & -\frac{2\la}{\sqrt{6}}
\end{array}
\right) \:,
\label{e:b}
\ee
where $D_\mu$ is the covariant derivative,
$M_0$ is the octet baryon mass in the chiral limit,
$F$ and $D$ are coupling constants satisfying the relation
$F+D=g_A\simeq 1.26$ with $g_A$ the axial-vector strength and
$u_\mu=i(u^\dag\partial_\mu u - u\partial_\mu u^\dag)$ with
\begin{equation}
 u = \exp\left(\frac{iP}{\sqrt{2}F_\pi}\right) \:,
\label{e:u}
\end{equation}
being $F_\pi= 92.4$ MeV the weak pion decay constant, and
\be
 P = \left(
\begin{array}{ccc}
\frac{\pi^0}{\sqrt{2}}+\frac{\eta}{\sqrt{6}} & \pi^+ &  K^+ \\
\pi^- &  -\frac{\pi^0}{\sqrt{2}}+\frac{\eta}{\sqrt{6}} & K^0 \\
 -K^-& \bar K^0 & -\frac{2\eta}{\sqrt{6}}
\end{array}
\right)
\label{e:p}
\ee
the SU(3)-flavor irreducible octet representation of the pseudoscalar mesons.
The form of the baryon-baryon potentials obtained from this contribution
is similar to the ones derived from the meson-exchange approach
and in momentum space reads
\begin{equation}
 V^{BB}_\text{OBE}=-f_{B_1B_2P}f_{B_2B_4P}
 \frac{(\sv_1\cdot\qv)(\sv_2\cdot\qv)}{\qv^2+m_{ps}^2}
 \mathcal{I}_{B_1B_2 \ra B_3B_4}
\label{e:obe}
\end{equation}
with $f_{B_1B2P}$ and $f_{B_2B_4P}$ the coupling constants of the two vertices,
$m_{ps}$ the mass of the exchanged pseudoscalar meson,
$\qv$ the transferred momentum,
and $\mathcal{I}_{B_1B_2 \ra B_3B_4}$ the corresponding isospin factor.

The contribution from the four-baryon contact interactions can be derived
from the following minimal set of Lagrangian densities
\bal
 \mathcal{L}^1 &=
 C_i^1\bra \bar B_a\bar B_b (\Gamma_i B)_b(\Gamma_i B)_a\ket \:,
\\
 \mathcal{L}^2 &=
 C_i^2\bra \bar B_a (\Gamma_i B)_a \bar B_b (\Gamma_i B)_b \ket \:,
\\
 \mathcal{L}^3 &=
 C_i^3\bra \bar B_a (\Gamma_i B)_a\ket \bra \bar B_b (\Gamma_i B)_b \ket \:.
\eal
Here, the labels $a$ and $b$ are the Dirac indices of the particles
and $\Gamma_i$ denotes the five elements of the Clifford algebra,
$\Gamma_1=1, \Gamma_2=\gamma^\mu, \Gamma_3=\sigma^{\mu\nu},
\Gamma_4=\gamma^\mu\gamma^5, \Gamma_5=\gamma^5$,
which are actually diagonal $3\times3$ matrices in the flavor space.
In LO these Lagrangian densities give rise to six independent
low-energy coefficients (LECs):
$C_C^1,C_S^1,C_C^2,C_S^2,C_C^3$ and $C_S^3$,
where the labels $C$ and $S$ refer to the central and spin-spin parts
of the potential, respectively.
The LO contact potentials for the different baryon-baryon interactions
resulting from these Lagrangians have the general form
\begin{equation}
 V^{BB}_{L0} = C_C^{BB} + C_S^{BB} \sv_1\cdot\sv_2 \:,
\label{e:cpot}
\end{equation}
where the coefficients $C_C^{BB}$ and $C_S^{BB}$ are linear combinations
of $C_C^1,C_S^1,C_C^2,C_S^2,C_C^3$ and $C_S^3$.

At next-to-leading order
(see the lower diagrams of Fig.~\ref{fig:xeft}),
the contact terms read
\bal
 V^{BB}_{NLO} &=
 C_1\qv^2 + C_2\kv^2 +(C_3\qv^2+C_4\kv^2) \sv_1\cdot\sv_2
 + \frac{i}{2} C_5(\sv_1+\sv_2) \cdot (\qv\times\kv)
\nonumber \\ &
 + C_6(\qv\cdot\sv_1)(\qv\cdot\sv_2)
 + C_7(\kv\cdot\sv_1)(\kv\cdot\sv_2)
 + C_8(\sv_1-\sv_2) \cdot (\qv\times\kv) \:,
\label{e:cpot2}
\eal
where $C_i$ $(i=1,\cdots,8)$ are additional LECs.
The momenta $\qv$ and $\kv$ are defined
in terms of the initial $\pv$ and final $\pv'$ baryon momenta
in the center-of-mass frame as
$\qv =\pv'-\pv$ and $\kv =(\pv+\pv')/2$, respectively.
The expressions for the two-pseudoscalar meson-exchange contributions
are rather cumbersome and we refer the interested reader
to the original works for details
(see e.g.~Ref.~\cite{Haidenbauer13}).

The baryon-baryon potentials constructed in this way are then inserted
in the Lippmann--Schwinger equation which is regularized
with a cut-off regulator function of the type
\begin{equation}
 F(p,p') = \exp\left(-\frac{p^4+p'^4}{\la^4}\right)
\end{equation}
in order to remove high-energy components of the
baryon and pseudoscalar meson fields.
The cut-off $\la$ is usually taken in the range $450-700\,$MeV.

\subsection{Resonating-group method}

The resonating-group method (RGM) has been used to develop a further
approach to the study of the NN interaction
\cite{Fujiwara06,Oka86,Shimizu89,Valcarce05}.
In this approach,
inspired by the QCD theory of strong interaction,
the quark degree of freedom is explicitly introduced and the NN interaction
is constructed from gluon and meson exchange between quarks,
the latters being confined inside the nucleons.
Due to the RGM formalism, the resulting interaction is highly
non-local and contains a natural cut-off in momentum.
The most recent model, named fss2 \cite{Fujiwara01,Fujiwara06},
reproduces closely the experimental phase shifts,
and fairly well the data on the few-body systems,
e.g.~the triton binding energy is reproduced within 300 keV.
This model has been used in nuclear matter calculations performed with the
non-relativistic BHF approach,
and in particular it has been found that it is able to reproduce correctly
the nuclear matter saturation point without
the TBF contribution \cite{Baldo14}.
This finding is quite relevant, and surely deserves further investigation.

\subsection{Renormalization-group methods}

A further class of NN interactions is based on renormalization group
(RG) methods
(see e.g.~\cite{Entem02,Bogner10} for a complete review).
The main effect of the hard core in the NN interaction is to produce
scattering to high momenta of the interacting particles.
A possible way to soften the hard core from the beginning
is by integrating out all the momenta larger than a certain cut-off $\la$
and ``renormalize'' the interaction to an effective interaction
$V_\text{low\,k}$ in such a way
that it is equivalent to the original interaction for momenta $q<\la$.
This results in a modified Lippmann--Schwinger equation
with a cutoff-dependent effective potential $V_\text{low\,k}$
\begin{equation}
 T(k',k;k^2) =
 V_\text{low\,k}(k',k) + \frac{2}{\pi} P\int_0^\la dqq^2
 \frac{V_\text{low\,k}(k',q)T(q,k;k^2)}{k^2-q^2} \:.
\end{equation}
By demanding $dT(k',k;k^2)/d\la=0$,
an exact Renormalization Group flow equation
for $V_\text{low\,k}$ can be obtained
\begin{equation}
 \frac{dV_\text{low\,k}(k',k)}{d\la} =
 \frac{2}{\pi}\frac{V_\text{low\,k}(k',\la)T(\la,k;\la^2)}{1-k^2/\la^2} \:.
\end{equation}
Integrating this flow equation one can obtain a phase-shift,
energy independent, soft (i.e.~without hard core)
and hermitian low-momentum potential $V_\text{low\,k}$.
The $V_\text{low\,k}$ interaction turns out to be much softer,
since no high momentum components are present and, as a consequence,
three- and many-body forces emerge automatically from a pure two-body force.
The short range repulsion is replaced by the
non-local structure of the interaction.
The cut-off $\la$ is taken above $300$ MeV in the laboratory,
corresponding to relative momentum $q\approx2.1\,$fm$^{-1}$,
that is the largest energy where the experimental data are established.
The fact that $V_\text{low\,k}$ is soft has the advantage to be
much more manageable than a hard-core interaction,
in particular it can be used in perturbation expansion and in nuclear structure
calculations in a more efficient way \cite{Bogner10,Furnstahl13}.

Following the same idea that in the NN case made possible to calculate a
``universal'' effective low-momentum potential $V_\text{low\,k}$
by using renormalization-group techniques,
in \cite{Schaefer06} this method was generalized to the YN sector.
Unfortunately, contrary to the NN case there exist only few YN scattering data
and hence the YN interaction is not well constrained.
It was found (see Figs.~1--6 of Ref.~\cite{Schaefer06})
that the YN phase shifts have approximately the same shape
but different heights,
and the diagonal matrix elements differ for lower momenta,
although they collapse for momenta near the cut-off.
In conclusion, however,
one can still say that in general the results seem to indicate
a similar convergence to an ``universal'' softer low-momentum YN interaction
as for the NN case.

\subsection{Baryon-baryon interactions from lattice QCD}

Recently, a further possibility of constructing the baryon-baryon
interaction based on lattice QCD has been explored,
see \cite{Aoki11,Aoki12,Beane11} for a review.
Lattice QCD, from which one should be able in principle to calculate
the hadron properties directly from the QCD Lagrangian,
is however extremely expensive from the numerical point of view
and current simulations can be performed only with large quark masses
\cite{Inoue13}.
In fact, an accurate simulation has to be made on a fine grid
spacing and large volumes,
thus requiring high-performance computers.
Nonetheless, a big progress to derive baryon-baryon interactions from
lattice QCD has been made by the NPLQCD \cite{Beane12} and HALQCD
\cite{Nemura18} collaborations.
It is worthwhile to point out, however,
that there exist some discrepancies between these two collaborations
regarding the methods employed in their studies.
The NPLQCD collaboration combines calculations of correlation functions at
several light-quark-mass values with the low-energy effective field theory.
This approach is particularly interesting since it allows to match
lattice QCD results with low-energy effective field theories providing the
means for first predictions in the physical quark mass limit.
The HALQCD collaboration, on the other hand,
follows a method to extract the different baryon-baryon potentials
from the Nambu--Bethe--Salpeter wave function measured on the lattice.
Recently this collaboration managed to approach the region of physical masses,
obtaining results for various NN, NY and YY interaction channels
at a single value of the lattice volume and of the lattice spacing
\cite{Iritani19,Iritani19b,Sasaki20}.

Using the NN forces thus obtained,
the properties of nuclei such as
$^4$He, $^{16}$O and $^{40}$Ca
and the EoS of nuclear matter were investigated \cite{Inoue15,Inoue2016nuclear},
finding that these nuclei and the symmetric nuclear matter are bound at a
quark mass corresponding to a pseudo-scalar meson (pion) mass of 469 MeV.
The obtained binding energy per nucleon has a uniform mass-number $A$ dependence,
which is qualitatively consistent with the Bethe-Weizs\"acker mass formula,
thus demonstrating that the HALQCD strategy works well
to investigate various properties of atomic nuclei and nuclear matter
starting from QCD.
We would also like to note that in the strangeness sector
the NPLQCD collaboration has been able to determine the binding energies
of light hypernuclei including
$^3_\la$He, $^4_\la$He and $^4_{\la\la}$He \cite{Beane13};
to compute the magnetic moment of the baryon octet \cite{Parreno17}
and to constrain the interactions of two-baryon octets
at the SU(3)-flavor symmetric-point \cite{Wagman17}.

\section{Theoretical approaches to the EoS of the core}
\label{s:eos}

Once the interaction between nucleons is established,
one can try to solve the many-body problem for nuclear matter.
Since there are currently no {\em ab initio} QCD calculations of dense matter
available at the thermodynamic conditions typical of compact stars,
one has to rely on interaction models,
which can be constrained by laboratory measurements and NS observations.
In the following we discuss mostly theoretical approaches
based on nucleonic degrees of freedom,
whereas later on we will treat approaches where hyperons or even quarks
are included, since they are essential at large densities.
The currently used many-body methods are divided into two classes,
i.e.~i) {\em Ab-initio microscopic methods}
which start from bare two- and three-nucleon interactions
able to reproduce nucleon scattering data
and properties of bound few-nucleon systems.
The main drawback of the ab-initio methods is that they can be employed
only for the description of homogeneous matter in the NS core,
and not for the clustered matter typical of the NS crust;
ii) {\em Phenomenological approaches}
which use effective interactions
with a simple structure dependent on a limited number of parameters,
which are fitted to different properties of finite nuclei and nuclear matter.
The main problem of these approaches is their extrapolation to
high-density conditions,
because those models are fitted to the ground-state properties of finite nuclei.

In the following we will briefly discuss all of them,
referring the interested reader to Refs.~\cite{Oertel17,Burgio18a}
and the quoted references in each subsection.

\subsection{Ab-initio microscopic methods}
\label{s:models-abinitio}

\subsubsection{(Dirac) Brueckner-Hartree-Fock}
\label{BBG}

The Bethe-Brueckner-Goldstone (BBG) many-body theory is based
on the re-summation of the perturbation expansion of the ground-state energy
of nuclear matter
\cite{Baldo99,Baldo07}.
The original bare NN interaction is systematically replaced by the so-called
$G$-matrix,
an effective interaction that describes the in-medium scattering processes,
and that takes into account the effect of the Pauli principle
on the scattered particles.
A further essential ingredient is the in-medium single-particle
potential $U(k)$ felt by each nucleon, where $k$ is the momentum.
The key point is the solution of the Bethe-Goldstone integral equation
for the $G$-matrix,
which can be written as
\be
 \bra k_1 k_2 \vert G(\om) \vert k_3 k_4 \ket =
 \bra k_1 k_2 \vert V \vert k_3 k_4 \ket +
 \sum_{k'_3 k'_4} \bra k_1 k_2 \vert V \vert k'_3 k'_4 \ket
 \frac{\left[1 - \Theta_F(k'_3)\right] \left[1 - \Theta_F(k'_4)\right]}
 {\om - e_{k'_3} + e_{k'_4} + i \eta}
  \, \bra k'_3 k'_4 \vert G(\om) \vert k_3 k_4 \ket \:,
\label{e:bruin}
\ee
where $V$ is the bare NN interaction,
$\om$ is the starting energy,
the two factors $[1 - \Theta_F(k)]$
force the intermediate momenta to be above the Fermi momentum
(``particle states''),
the single-particle energy being $e_k=k^2/2m+U(k)$,
with $m$ the particle mass,
and the summation includes spin-isospin variables.
The main feature of the $G$-matrix is that it does not have the hard core
of the original bare NN interaction
and it is defined even for bare interactions with an infinite hard core.
The introduction and choice of the in-medium single-particle potential
are essential to make the re-summed expansion convergent,
and is calculated self-consistently
with the $G$-matrix itself in order to incorporate as much
higher-order correlations as possible.
The resulting nuclear EoS can be calculated with good accuracy
at the so-called Brueckner-Hartree-Fock (BHF) level
(``two hole-line'' approximation)
with the continuous choice for the single-particle potential,
the results in this scheme being quite close to the calculations
which include also the three hole-line contribution
\cite{Song98,Lu17,Lu18}.

It should be noted that all non-relativistic many-body approaches
fail to reproduce the correct saturation point of nuclear matter.
For this purpose, TBFs are introduced in the calculations,
and are reduced to a density dependent two-body force by
averaging over the generalized coordinates
(position, spin, and isospin) of the third particle.
In the BHF calculations for nuclear matter,
the so-called Urbana model is often introduced as TBF,
and this produces a shift in the binding energy
of about $+1$~MeV and of $-0.01\fm3$ in density.
The problem of such a procedure is that the TBF
is dependent on the two-body force.
The connection between two-body and TBFs within the meson-nucleon theory
of nuclear interaction is extensively discussed and developed
in \cite{Zuo02a,Zuo02b,Li08}.
At present the theoretical status of microscopically derived TBFs
is still quite rudimentary; however, a tentative approach has been proposed
using the same meson-exchange parameters as the underlying NN potential.
Results have been obtained with the Argonne $V_{18}$ \cite{Wiringa95},
the Bonn~B \cite{Brockmann90},
and the Nijmegen~93 potentials \cite{Li08a,Li08}.
However, the role of TBF also depends on the model adopted for TBF.
In fact, if the NN potential is based on a realistic constituent quark model
\cite{Baldo14} then the role of TBF is strongly reduced.
On the other hand, if we consider a new class of chiral-inspired TBF,
then the BHF calculations are not able to reproduce simultaneously
the correct saturation point and the properties
of three- and four-nucleon systems \cite{Logoteta15}.

The relativistic BHF formalism,
i.e.~the Dirac-Brueckner approach \cite{Machleidt89},
has been developed in analogy with the non-relativistic case,
where the two-body correlations are described
by introducing the in-medium relativistic $G$-matrix.
The novel and most striking feature of the DBHF theory is its ability
to describe the saturation properties of nuclear matter,
a fundamental aspect which reflects the saturating nature of the nuclear force.
The DBHF method contains important relativistic features
through the description of the nuclear mean field
in terms of a scalar and a vector component,
strong and of opposite sign.
In their combination,
they provide an explanation for the binding of nucleons
and the spin-orbit splitting in nuclear states.
The main relativistic effect is due to the use of the spinor formalism,
which has been shown \cite{Brown87} to be equivalent
to introducing a particular TBF,
the so-called $Z$-diagram,
which is repulsive and consequently gives a better saturation point
than the BHF method,
and actually including in BHF only these particular TBFs,
one gets results close to DBHF calculations \cite{Li06}.
In conclusion, the EoS calculated within the DBHF method turns out to be stiffer
above saturation than the ones calculated from the BHF+TBF method.
A further point of difference is that the relativistic DBHF produces
a superluminal EoS at higher densities than the BHF approach.
The reader is referred to Ref.~\cite{Muther17}
for a relatively recent review of the DBHF method
and a variety of applications to both nuclear matter and nuclei.

\subsubsection{Approaches based on the variational method}
\label{s:Var}

The variational approach to the many-body problem
is based on the Ritz-Raleigh variational principle,
according to which the trial ground state energy
\be
 E_\text{trial} = \frac{\bra\Psi \vert H \vert \Psi\ket}
 {\bra\Psi\vert \Psi\ket} \:,
\ee
calculated from the system's Hamiltonian $H$
with a trial many-body wave function $\Psi$,
gives an upper bound for the true ground state energy of the system.
In the variational method one assumes that the ground state wave function $\Psi$
can be written as
\be
 \Psi(r_1,r_2,......) = \prod_{i<j} f(r_{ij}) \Phi(r_1,r_2,.....) \:,
\ee
where $\Phi$ is the unperturbed ground-state wave function,
properly antisymmetrised,
and the product runs over all possible distinct pairs of particles.
The correlation factors $f$,
which are intended to transform the uncorrelated wave function $\Phi$
to the correlated one,
are determined by the Ritz-Raleigh variational principle,
i.e.~by assuming that the mean value of the Hamiltonian reaches a minimum:
\be
 \frac{\delta}{\delta f} \frac{\bra\Psi\vert H \vert \Psi\ket}
 {\bra\Psi \vert \Psi\ket} = 0 \:.
\ee
Therefore the main task in the variational method
is to find a suitable ansatz for the correlation factors~$f$.
Several different methods exist for the calculation of $f$,
e.g.~in the nuclear context the Fermi-Hyper-Netted-Chain (FHNC)
\cite{Fantoni75,Pandharipande79}
calculations have been proved to be efficient.

For nuclear matter it is necessary to introduce channel-dependent
correlation factors,
which is equivalent to assume that
$f$ is actually a two-body operator $\hat{F}_{ij}$.
One then assumes that $\hat{F}$ can be expanded in the same spin-isospin,
spin-orbit, and tensor operators appearing in the NN interaction
\cite{Fantoni98,Carlson15}.
Due to the formal structure of the Argonne NN forces,
most variational calculations have been performed
with this class of NN interactions,
often supplemented by the Urbana TBFs.
The best known and most used variational nuclear matter EoS
is the one of Akmal-Pandharipande-Ravenhall (APR) \cite{Akmal98}.
Many excellent review papers exist in the literature
on the variational method and its extensive use for
the determination of the nuclear matter EoS, e.g.~\cite{Pandharipande79}.

Among other methods based on the variational principle
and widely used in nuclear physics,
we mention the coupled-cluster theory,
proposed in \cite{Coester58,Coester60}, in which
the correlation operator is represented in terms of the cluster operators.
In recent nuclear matter calculations with chiral NN interactions
\cite{Hagen07,Hagen14}
and also in nuclear structure calculations
\cite{Hagen12,Hagen10},
the method has been proven to be successful.
The variational Monte Carlo (VMC) approach is also commonly used,
especially for light nuclei, including two and three-body correlations
\cite{Wiringa14},
but the EoS of homogeneous nuclear matter is hard to obtain,
due to the large number of nucleons which increases the computational effort
\cite{Navarro02,Carlson15}.

\subsubsection{Self-consistent Green's function}
\label{s:scgf}

Another way to approach the many-body problem is through the
many-body Green's functions formalism \cite{Dickhoff08}.
In this approach one performs a diagrammatic analysis of the
many-body propagators in terms of free one-body Green's functions
and two-body interactions.
The perturbative expansion results in an infinite series of diagrams,
among which one has to choose those which are relevant
for the considered physical problem.
Depending on the approximation,
one can either choose a given number of diagrams
or sum an infinite series of them,
in analogy with the BBG approach.
In the description of nuclear matter,
the method is conventionally applied at the ladder-approximation level,
which encompasses at once particle-particle and hole-hole propagation,
and this represents the main difference with respect to the $G$-matrix,
where only particle-particle propagators are included.
At a formal level,
the comparison between the BBG and the SCGF approaches is not straightforward.
Even though both approaches arise from a diagrammatic expansion,
the infinite subsets of diagrams considered in the two approaches
are not the same,
and the summation procedures are also somewhat different.
Whereas the BHF formalism in the continuous choice can be derived
from the ladder SCGF formalism after a series of approximations,
this is not the case for the full BBG expansion.
In principle, if both BBG and SCGF were carried out to all orders,
they should yield identical results.
BBG theory, however, is an expansion in powers of density (or hole-lines),
and the three-hole line results seem to indicate that it converges quickly.
The error in the SCGF expansion is more difficult to quantify,
as one cannot directly compute (or even estimate)
which diagrams have to be included in the expansion.

The energy per particle of nuclear matter is obtained within the SCGF approach
through the Galitskii--Migdal--Koltun sum-rule \cite{Galitskii58,Koltun74}:
\begin{equation}
 \frac{E}{A} = \frac{\nu}{\rho} \int\frac{d^3k}{(2\pi)^3}
 \int\frac{d\om}{2\pi}\frac12\left(\frac{k^2}{2m}+\om\right)
 {\cal A}(k,\om)f(\om) \:,
\label{e:GMK}
\end{equation}
where $\nu=4(2)$ is the spin-isospin degeneracy of nuclear (neutron) matter,
$\rho$ is the total density and
$f(\om)=[1+e^{\om-\mu}/T]^{-1}$ is the Fermi--Dirac distribution.
The key quantity of the approach is the one-body spectral function
${\cal A}(k,\om)$ which, in few words,
represents the probability density of removing from or adding to the system
a nucleon with momentum $k$ and energy $\om$.
The single-particle spectral function gives access to the calculation
of all the one-body properties of the system
and it can be obtained from the imaginary part of the one-body propagator
${\cal G}(k,\om)$,
\begin{equation}
 {\cal A}(k,\om) = -\frac{1}{\pi}\mbox{Im}\,{\cal G}(k,\om) \:,
\label{e:SF}
\end{equation}
or, alternatively,
from the proper or irreducible self-energy $\Sigma(k,\om)$ as
\begin{equation}
 {\cal A}(k,\om) = -\frac{\mbox{Im}\,\Sigma(k,\om)}{
 [\om-\frac{k^2}{2m}-\mbox{Re}\,\Sigma(k,\om)]^2+
 [\mbox{Im}\,\Sigma(k,\om)]^2} \:.
\label{e:SF2}
\end{equation}

The computational implementation of the SCGF method requires:
(1) to calculate the effective interaction ($T$-matrix) describing the
in-medium scattering of two nucleons,
(2) to extract the self-energy $\Sigma(k,\om)$ and
(3) to obtain the one-body propagator ${\cal G}(k,\om)$ by solving,
with the self-energy, the Dyson equation,
which is then inserted in the scattering equation,
repeating these three steps until a self-consistent solution is achieved.

Also for the SCGF method the inclusion of TBFs is essential.
So far TBFs were not included in the ladder approximation,
however a method has been developed recently in \cite{Carbone13a},
and applied to SNM using chiral nuclear interactions.
TBFs are included via effective one-body and two-body interactions,
and are found to improve substantially the saturation point \cite{Carbone13b}.

One has to note that because of the well-known
Cooper instability \cite{Cooper56},
through which a fermionic many-body system with an attractive interaction
tends to form pairs at the Fermi surface,
low-temperature nuclear matter is unstable with respect to the formation of
a superfluid or superconducting state.
The Cooper instability shows up as a pole in the $T$-matrix
when the temperature falls below the critical temperature $T_c$
for the transition to the superfluid/superconducting state.
Therefore current calculations are often performed at temperatures
above $T_c$ and extrapolated to zero temperature, see e.g.~\cite{Frick04}.
Further details on the SCGF method and on its applications
to nuclear problems can be found,
e.g.~in \cite{Muether00,Dickhoff04,Rios20}.

\subsubsection{Chiral effective field theory ($\chi$EFT) approach}
\label{s:EFT}

High-precision nuclear potentials based on the ChPE \cite{Entem03}
are nowadays very popular,
since ChPE allows to link nuclear physics with QCD.
In particular, the treatment of many-nucleon forces is very important,
those being mostly relevant in nuclear matter.
In the last years some effort has been devoted to the extension of chiral
perturbation theory to nuclear matter calculations,
i.e.~developing an effective field theory (EFT) for nuclear matter.
In this case another scale appears, $k_F/M$,
being $M$ the nucleon mass and $k_F$ the Fermi momentum,
approximately given as $k_F \approx 263$ MeV at saturation
which is smaller than a typical hadron scale.
In the chiral limit it is then natural to expand in $k_F/M$,
and this expansion can be obtained from the vacuum ChPE \cite{Kaiser02}.
For nuclear matter the correction thus obtained with respect to the
vacuum diagrams gives a direct contribution to the EoS of nuclear matter,
and this correction is clearly proportional to a power of $k_F/M$.
In this case a cut-off must be introduced,
and its tuning allows to obtain a saturation point and compressibility
in fair agreement with phenomenology.
Along the same lines more sophisticated approaches can be developed,
where the many-nucleon interactions built in vacuum
are directly used in nuclear matter calculations.
In this case the ChPE is used in conjunction with the EFT scheme ($\chi$EFT).

In recent years,
$\chi$EFT has been used for studying nuclear matter within various
theoretical frameworks like
many-body perturbation theory
\cite{Hebeler11,Wellenhofer14,Coraggio14,Drischler16},
SCGF framework \cite{Carbone13b},
in-medium chiral perturbation theory \cite{Holt13},
the BHF approach \cite{Kohno13,Li12}
and quantum Monte Carlo methods \cite{Gezerlis13,Roggero14,Lynn16}.
Several reliable calculations have been performed
up to twice the saturation density $n_0$,
beyond which uncertainties were estimated by analysing the
order-by-order convergence in the chiral expansion
and the many-body perturbation theory \cite{Holt17}.
Variations in the resolution scale \cite{Bogner05} and low-energy constants
appearing in the two-nucleon and three-nucleon forces
were systematically explored \cite{Hebeler10},
finding that the theoretical uncertainty band grows rapidly
with the density beyond $n_0$,
due to the missing third-order terms at low densities
and higher-order contributions in the chiral expansion.
This has consequences not only for the EoS,
but also for the symmetry energy at saturation density, $S_0$,
and the slope parameter $L$, as discussed in \cite{Baldo16}.

\subsubsection{Quantum Monte Carlo methods}
\label{s:QMC}

Quantum Monte Carlo (QMC) methods are very successful in solving the
nuclear many-body problem non-perturbatively and with controlled approximations,
thus making QMC methods quasi-exact in describing the ground state
of fermionic systems.
Among the most successfully described,
we mention the liquid $^3$He,
but also bosons, like atomic liquid $^4$He.
Moreover they have been applied in studies of nuclear matter and light nuclei
\cite{Carlson15,Lynn19}.
Several implementations of QMC methods have been developed over the years,
e.g.~Green's function Monte Carlo (GFMC) \cite{Carlson03},
or auxiliary-field diffusion Monte Carlo (AFDMC) \cite{Gandolfi09},
which differ in the treatment of the spin and isospin degrees of freedom.
In particular, the AFDMC method has been used extensively
to study nuclear matter for astrophysical applications
\cite{Lynn16,Gandolfi09,Lonardoni20}.

The main idea of Quantum Monte Carlo methods is to stochastically solve the
many-body Schr\"odinger equation to extract the ground state of a system,
by evolving a given trial wave function of the many-body system,
$\Psi_V$, in imaginary time $\tau = it$.
Monte Carlo sampling is then used to evaluate all possible configurations
until convergence is reached.
However, the accuracy of the different QMC versions is limited
by the so-called fermion sign problem \cite{Schmidt87},
for which different approximations are adopted \cite{Carlson12}.
In fact, for fermionic systems the wave function is antisymmetric
and contains several changes in sign.
Hence, the integrands in the QMC integrals are highly oscillatory thus producing
very large statistical uncertainties,
so that no information can be obtained from the calculation.
One possible way to cure the problem consists in splitting
the wave function space into regions of positive and negative wave functions,
defining a nodal surface at which the wave function changes sign.
Generally, configurations that cross the nodal surface
are removed from the evolution.
This approximation, called fixed-node approximation \cite{Schmidt87},
has been generalized to the constrained-path method \cite{Zhang95,Zhang97}
for complex wave functions.

When calculating nuclear matter, one typically simulates $N$ particles
in a cubic box with size $L$,
where $L$ is determined in such a way that the number density $n$ in the box
reflects a chosen value, $L = (N/n)^{1/3}$.
To probe the thermodynamic limit,
the particle number $N$ has to be chosen sufficiently large.
However, due to growing computational costs associated
with large particle numbers,
for neutron matter one typically chooses $N=66$
(33 spin up and 33 spin down neutrons),
thus finding results close to the thermodynamic limit.
Those approximations seriously limit the potentiality of the QMC methods.
Moreover, we notice that in spite of its recent progress,
it is not yet possible to perform GFMC and AFDMC calculations
with the Argonne $V_{18}$ potential,
mainly due to technical problems associated with the spin-orbit structure
of the interaction and the trial wave function,
which induce very large statistical errors.
In order to overcome this problem,
the full operatorial structure of current high-quality NN potentials
has been simplified and more manageable NN potentials have been developed
containing less operators with readjusted parameters.
In particular, we mention the $V'_8$, $V'_6$ and $V'_4$ potentials
\cite{Pudliner97,Wiringa02},
eventually supplemented with the Urbana TBFs.
Recently, a local chiral potential has been developed~\cite{Gezerlis13},
which is well suited for QMC techniques.

Modern computer technology has allowed the extension of the QMC method
to nuclear systems,
which have more complicated interactions and correlation structures.
It has to be noticed that the computing time increases exponentially
with the number of particles,
which limits the number of nucleons considered by GFMC up to 16 neutrons.
The largest nucleus considered is $^{12}$C.
A recent comparison has demonstrated that both methods give very close results
for neutron drops with $N\leq16$ \cite{Gandolfi11}.

\subsection{Phenomenological approaches}
\label{s:models-phenom}

\subsubsection{Non-relativistic energy density functionals}
\label{s:nrel}

Phenomenological approaches make use of effective interactions instead of
bare ones to treat dense matter,
and mostly rely on the energy density functional (EDF) theory,
which is able to recast the complex many-body problem of interacting particles
into an effective independent particle approach \cite{Duguet14}.
In the EDF theory, the total energy of the
system is usually expressed as a functional of the nucleon number densities,
the kinetic energy densities, and the spin-current densities,
but the exact form of the functional itself is not known a priori.
Therefore, one has to rely on phenomenological functionals,
which depend on a certain number of parameters fitted to reproduce some
properties of known nuclei and nuclear matter,
as well as ab-initio calculations of infinite nuclear matter.
A different approach to construct a phenomenological EoS is to use a
purely parametrized EoS,
without considering the main properties of the NN interaction.
An example is given by a metamodel for the nucleonic EoS,
which consists in a Taylor expansion around the saturation density of SNM
in terms of the empirical parameters \cite{Margueron18a},
and has been applied to study NS global properties \cite{Margueron18b}.

Non-relativistic approaches usually start from an Hamiltonian
$\hat{H} = \hat{T} + \hat{V}$
for the many-body system, where
$\hat{T} = \sum_i \hat{p}^2/2m_i$
is the kinetic term
($\hat{p}$ being the momentum operator and $m_i$ the mass of the species $i$)
and $\hat{V}$ is the potential term.
The latter accounts for the two-body (pseudo)potential,
that allows one to incorporate physical properties like effective masses,
but TBFs can also be included explicitely \cite{Vautherin72}.
A very popular scheme is based on the Skyrme-type effective interactions,
which are zero-range density-dependent interactions and allow for fast numerical
computations.
Since the pioneer work of Skyrme~\cite{Skyrme56},
several extensions have been proposed,
which accurately reproduce experimentally
measured properties of finite nuclei
\cite{Bender09,Chamel09,Margueron09,Goriely10,Chamel15}
and are applied to NSs.
We mention that many Skyrme models have been tested against
several nuclear-matter constraints in Ref.~\cite{Dutra12},
but most of the constraints are known with large error bars,
particularly the symmetry energy coefficients,
and therefore it is not possible to rule out some models on this basis.
Besides Skyrme models,
finite-range (density-dependent) interactions have been
derived from the Gogny interaction \cite{Decharge80},
which are less widely used in astrophysics because of
the larger numerical efforts
\cite{Goriely16,Sellahewa14}.

In addition to the Skyrme and Gogny effective interactions,
new EDFs have been recently constructed on the basis of the
Kohn-Sham density functional theory \cite{Baldo10, Baldo13}.
These Barcelona-Catania-Paris(-Madrid) (BCP and BCPM) EDFs have been derived
by introducing in the functional results from microscopic nuclear
and neutron-matter BHF calculations,
thus yielding a very good description of properties of finite nuclei and
NS masses with a reduced number of parameters.

\subsubsection{Relativistic mean-field (RMF) models}
\label{s:rmf}

Relativistic mean-field (RMF) models have been constructed on the basis of the
quantum hadrodynamics (QHD) framework \cite{Serot92},
a field-theoretical formalism where nucleons are represented
by four-component Dirac spinors,
and the NN interaction is modeled by exchange of mesons.
A nucleus is thus described as a system of Dirac nucleons
whose motion is governed by the Dirac equation.
RMF models have been successfully employed in nuclear structure,
to describe both nuclei close to the valley of stability and exotic nuclei
\cite{Niksic11}.

The common starting point in RMF models is an effective Lagrangian
$\mathcal{L} =
\mathcal{L}_\text{nuc} + \mathcal{L}_\text{mes} + \mathcal{L}_\text{int}$,
where the different terms account for the nucleon, the free mesons
and the interaction contribution, respectively.
The isoscalar scalar $\sigma$-meson and the isoscalar vector $\om$-meson
mediate the long-range attraction and short-range repulsion of the nuclear
interaction, respectively, in SNM.
Isovector mesons
(like the isovector vector $\rho$-meson and the isovector scalar $\delta$-meson)
need to be included as well to treat neutron-proton asymmetric systems.
The interaction term depends on the nucleon-meson coupling constants
that are usually determined by fitting nuclei or nuclear-matter properties.
From the Lagrangian above, the field equations for nucleons and mesons
are derived, and they are solved self-consistently,
usually in the relativistic mean-field approximation.
However, the correct description of nuclear matter and finite nuclei
requires the extension of that simple Lagrangian in order to include
a medium-dependent effective interaction.
This effect can either be introduced by including non-linear (NL)
meson self-interaction terms in the Lagrangian,
or by assuming an explicit density dependence (DD)
for the meson-nucleon couplings,
which gives rise to the so-called rearrangement contributions
mostly important for the thermodynamic consistency of the model.

The former approach has been employed in constructing several phenomenological
RMF interactions,
like the widely used NL3 \cite{Lalazissis97}, PK1, PK1R \cite{Long04}
and FSUGold \cite{Todd05}.
In the second approach,
the functional form of the density dependence of the coupling constants
can be derived by comparing results with the microscopic DBHF calculations
or it can be fully phenomenological,
with parameters adjusted to experimental data
(DD-RMF models, see Refs.~\cite{Long04,Typel05,Rocamaza11}).
We notice that all those models do not explicitly take into account
the antisymmetrization of the many-body wave function.
This has been included in relativistic Hartree-Fock (RHF)
or Hartree-Fock-Bogoliubov models accounting for pairing,
see Refs.~\cite{Meng06,Long07,Long10}.

Due to the mean-field character of the model and the charge independence
of the strong interaction,
the extension of the model to the full baryon octet is slightly more difficult
than the simple nuclear-matter case.
In Ref.~\cite{Dutra14}, an ensemble of 263 RMF models was tested
against nuclear matter properties up to about 3 times the saturation density,
also making use of heavy-ion collision data.
Only a limited set was found to be consistent
with the constraints taken into account.
This same set has been further used in Ref.~\cite{Lourenco19},
where the Love number and the corresponding tidal deformabilities
show very good agreement with the data from the GW170817 merger event.

A further RMF model is the quark-meson-coupling model,
which incorporates the internal quark structure of baryons,
where nucleons are treated as bound states of three quarks
and interact via meson exchange, pions included.
This model has been applied to study NS properties,
e.g.~in \cite{Whittenbury14}.

Both Skyrme and RMF based approaches turn out to be very popular
in astrophysical applications,
i.e.~core-collapse supernovae and binary NS mergers simulations,
where a wide range of densities, temperatures and charge fractions,
describing both clustered and homogeneous matter, is needed.
This range is covered by the so-called ``general-purpose" EoSs,
which are described in Sec.~\ref{s:finiteT_EoS}.
However, we like to point out
that phenomenological models have to be considered with some care,
as discussed in Ref.~\cite{Kruger13},
where a comparison between some phenomenological EoSs with those derived within
the chiral EFT interactions up to next-to-next-to-next-to-leading order (N3LO)
shows some inconsistency with the chiral pure neutron matter (PNM) band,
although this comparison is limited to a density range up to $\rho=0.165\fm3$.
This point is further illustrated in Sec.~\ref{s:bhfv18eos}.

\subsection{Comparing ab-initio and phenomenological approaches}
\label{s:bhfv18eos}

\begin{figure}[t]
\centering
\vskip-2mm
\includegraphics[scale=0.84]{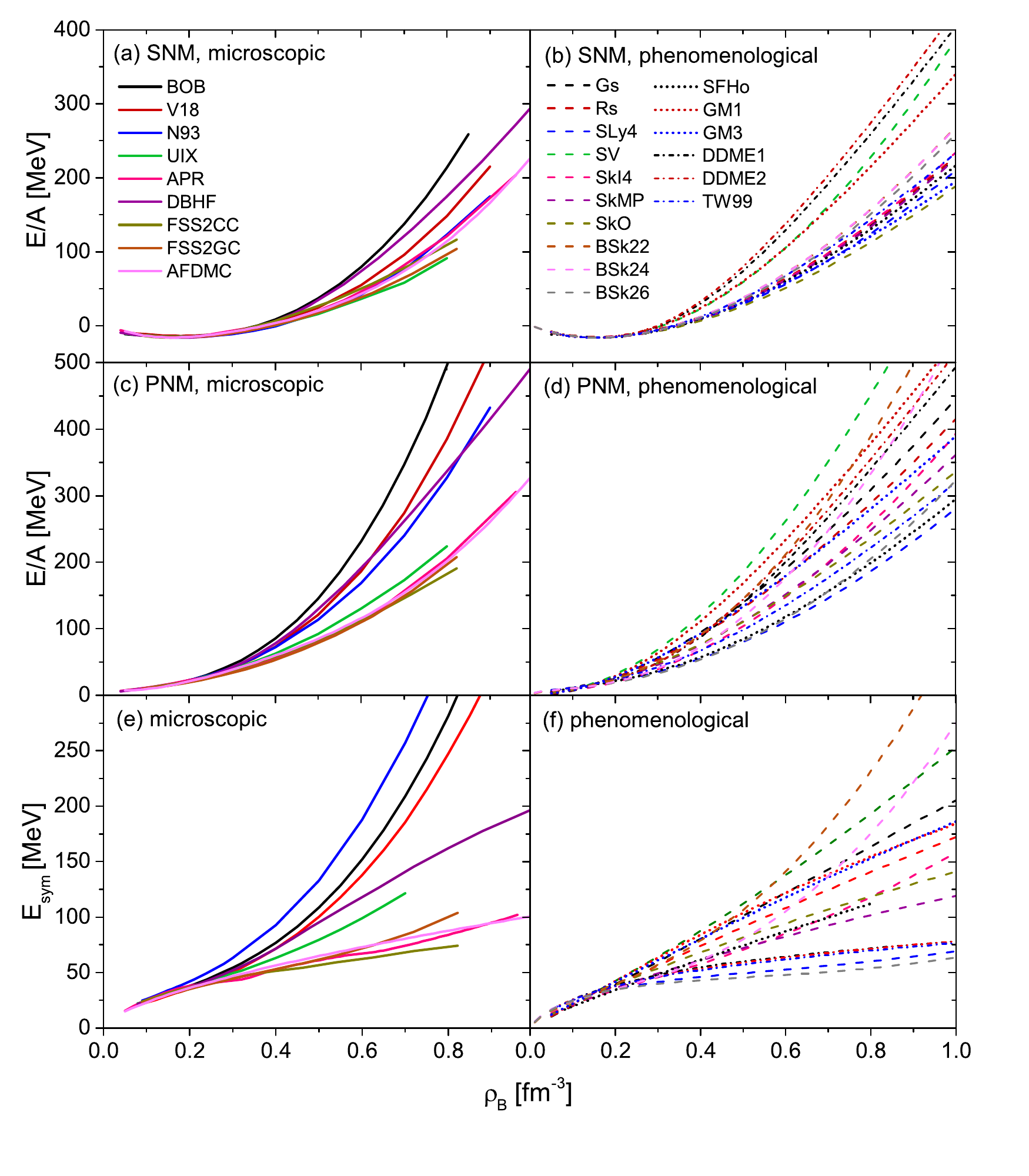}
\vskip-12mm
\caption{
Binding energy per nucleon
of symmetric matter (upper panels),
pure neutron matter (central panels)
and the symmetry energy (lower panels)
obtained with different EoSs.
See text for details.}
\label{fig:EOS}
\end{figure}

We now turn to discuss the main differences in the results obtained for
SNM and PNM,
employing microscopic and phenomenological EoS models.
The main results are summarized in Fig.~\ref{fig:EOS}.
In the upper (central) panels we display the binding energy per particle
as a function of the nucleon density obtained for SNM (PNM)
with a set of microscopic and phenomenological approaches and NN potentials,
plotted respectively in the left and right columns.
In the lower panels we display the symmetry energy for the same EoSs.

As far as microscopic approaches are concerned,
we adopt several BHF EoSs based on different NN potentials,
namely the Bonn B (BOB) \cite{Machleidt89},
the Nijmegen 93 (N93) \cite{Nagels78,Stoks94},
and the Argonne $V_{18}$ (V18) \cite{Wiringa95}.
In all those cases,
the two-body forces are supplemented by nucleonic TBFs,
either phenomenological or microscopic
\cite{Grange89,Baldo97,Zuo02a,Li08a,Li08}.
In the phenomenological case the corresponding EoS employs the Urbana model
for TBF and is labelled UIX.
Within the same theoretical framework,
we also studied an EoS based on a potential model which includes
explicitly the quark-gluon degrees of freedom,
i.e.~FSS2, which  reproduces correctly the saturation point of SNM
and the binding energy of few-nucleon system without the need of introducing TBF.
In the following we use two different EoS versions labelled respectively
as FSS2CC and FSS2GC.
Moreover, we compare these BHF EoSs with the often-used results
of the DBHF EoS \cite{Gross99}, which employs the Bonn~A potential,
the APR EoS \cite{Akmal98} based on the variational method
and the Argonne $V_{18}$ potential,
and a parametrization of a recent Auxiliary Field Diffusion Monte Carlo (AFDMC)
calculation \cite{Gandolfi10}.

As far as phenomenological approaches are concerned,
we show results obtained for a sample of seven Skyrme forces,
namely, GS and Rs \cite{Friedrich86},
SLy4 \cite{Chabanat95} of the Lyon group,
the old SV \cite{Beiner75},
SkI4 \cite{Reinhard95} of the SkI family,
SkMP \cite{Bennour89},
and SkO \cite{Reinhard99}.
We also consider two types of RMF models,
and in particular we adopt models with constant meson-baryon couplings
described by the Lagrangian density of the nonlinear Walecka model (NLWM),
and models with density-dependent couplings,
hereafter referred to as density-dependent models (DDM).
Within the first type,
we consider the models GM1 and GM3 \cite{Glendenning91a}.
For the DDM, we consider the models DDME1 and DDME2 \cite{Niksic02},
and TW99 \cite{Typel99}.
A further phenomenological RMF EoS, the SFHo EoS \cite{Steiner13},
has been used for comparison.
The latter belongs to the class of the so-called general-purpose EoSs,
and will be illustrated in Sec.~\ref{s:finiteT_EoS}.

Turning back to the discussion of Fig.~\ref{fig:EOS},
we see that in both PNM and SNM all approaches yield similar results
up to about twice the saturation density,
and diverge at larger density.
For the microscopic approaches of BHF type,
this is certainly due to the different high-density behaviour of TBFs,
either microscopic or phenomenological.
Also three-body correlations make a difference.
For phenomenological models, a similar spread at high density can be observed.
Indeed, these models are characterized by parameters which are fitted
on experimental data approximately known around saturation density,
therefore their behaviour at high density,
where no experimental data are avilable,
can produce large differences.
In the lower panels of Fig.~\ref{fig:EOS},
we display the symmetry energy as a function of the density.
We observe a monotonically increasing behaviour for all the considered EoSs,
and strong divergencies at large density,
which depend on either the nucleonic interaction or the many-body scheme;
those are important for determining structure and composition of NSs,
and also their cooling mechanism.

\begin{figure}[t]
\centering
\includegraphics[scale=0.8]{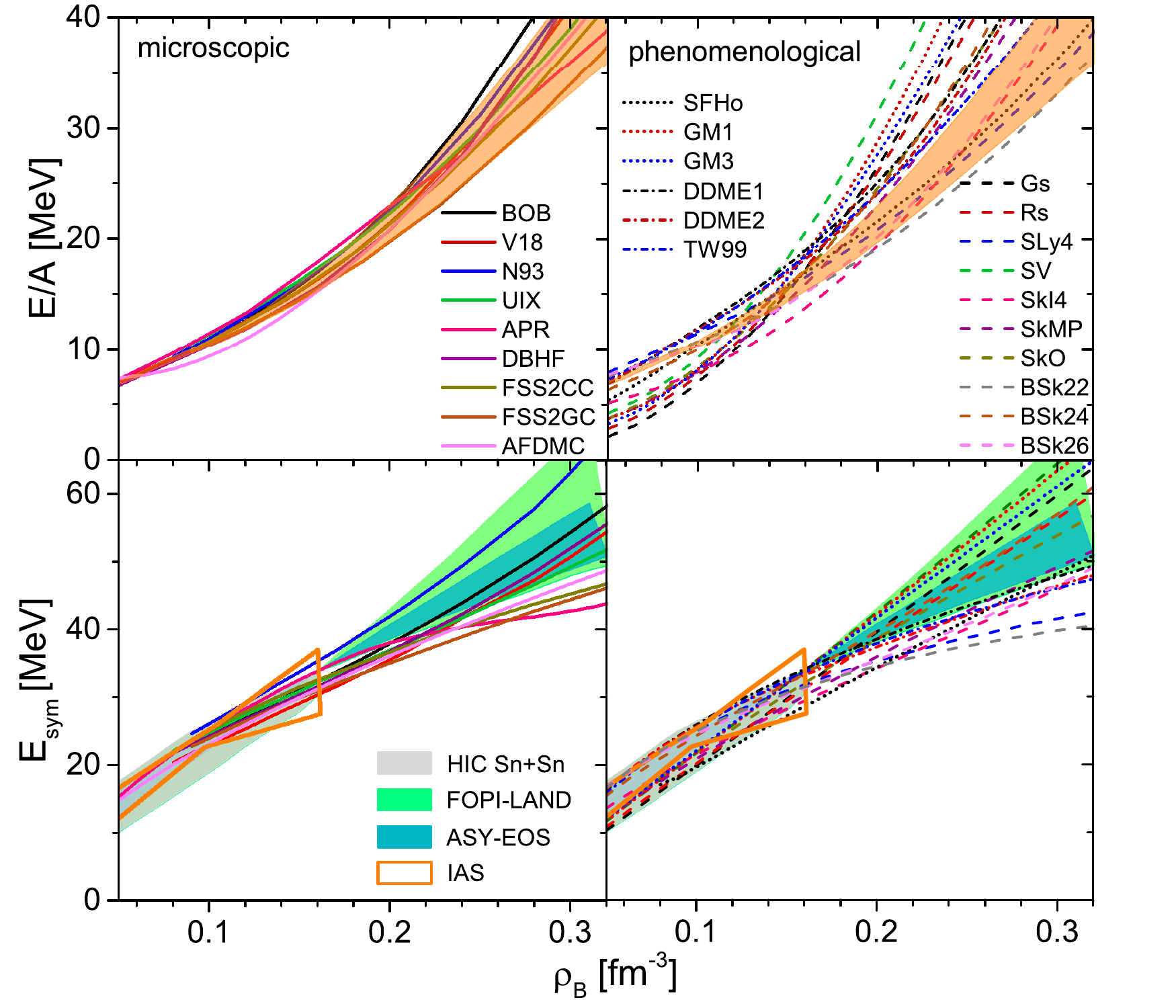}
\vskip-4mm
\caption{
The binding energy of PNM (upper panels)
for microscopic EoSs (left panels) and
phenomenological ones (right panels).
The light-orange band represents the predictions of $\chi$EFT theory
reported in Ref.~\cite{Drischler20}.
In the lower panels the symmetry energy is displayed,
along with currently available limits from heavy ion collisions
(green, blue and gray bands),
and the IAS calculations (orange contour).
See text for details.}
\label{fig:EOSconstr}
\end{figure}

It is worthwhile to perform a comparison of the above discussed EoSs
with the currently available constraints.
In Fig.~\ref{fig:EOSconstr} (upper panels)
we compare with the results obtained for PNM
within the chiral EFT interactions up to N3LO order \cite{Kruger13,Drischler20},
and shown by the orange-shaded band.
Please notice that this comparison is limited to a density range up to
$\rho=0.34\fm3$.
Whereas the microscopic calculations fall well inside the chiral band,
some inconsistency is evident for the phenomenological EoSs,
as already found in Ref.~\cite{Kruger13}.
In particular, we find that the SLy4, BSk22, BSk24, BSk26 and SFHo are
in agreement with the chiral calculations;
TW99, DDME1 and DDME2 are not at all compatible,
and the remaining EoSs are only partially
compatible with the chiral PNM results.
We stress that neutron matter is an ideal laboratory for nuclear interactions
derived from chiral effective field theory since all contributions are predicted
up to N3LO in the chiral expansion,
including three-body forces \cite{Drischler16}.
By comparing with results obtained within the
self-consistent Green’s function theory (SCGF) using the same three-body forces,
predictions for the EoS of neutron matter at zero temperature
at density slightly above the saturation density have been improved,
including estimates of theoretical uncertainties for astrophysical applications.
A recently proposed Bayesian machine-learning method allowed to improve
results up to twice nuclear saturation density
for the binding energy and pressure of neutron matter,
as well as for the nuclear symmetry energy and its slope,
which turn out to be consistent with experimental constraints \cite{Drischler20}
at saturation density.

In the lower panels of Fig.~\ref{fig:EOSconstr},
we display the symmetry energy over the same density range.
Results are compared with available experimental data,
which are displayed by the shaded areas.
In particular, the grey area represents the diffusion data of HICs,
the green area includes the data obtained by the
FOPI-LAND collaboration \cite{Ritman95} on the collective flow,
and the blue area is the experimental region checked by the
ASY-EOS collaboration \cite{Russotto16}.
The full orange contour shows the results on the isobaric analog states (IAS),
obtained in Ref.~\cite{Danielewicz14}.
We see that most of the considered EOSs
are compatible with experimental data up to around saturation density,
whereas for larger densities some EOSs tend to predict smaller values
of the symmetry energy, below the experimental areas.
This is a clear sign of discrepancy,
which results in a much larger difference at larger values of the baryon density,
as the ones characterizing the inner core of a NS.
We notice that the inferred constraints are model dependent,
since the data interpretation requires theoretical simulations.
Those discrepancies give rise to different predictions for NS structure
and cooling behaviour, as will be shown later.

\begin{figure}[t]
\centering
\vskip-12mm
\includegraphics[scale=0.6]{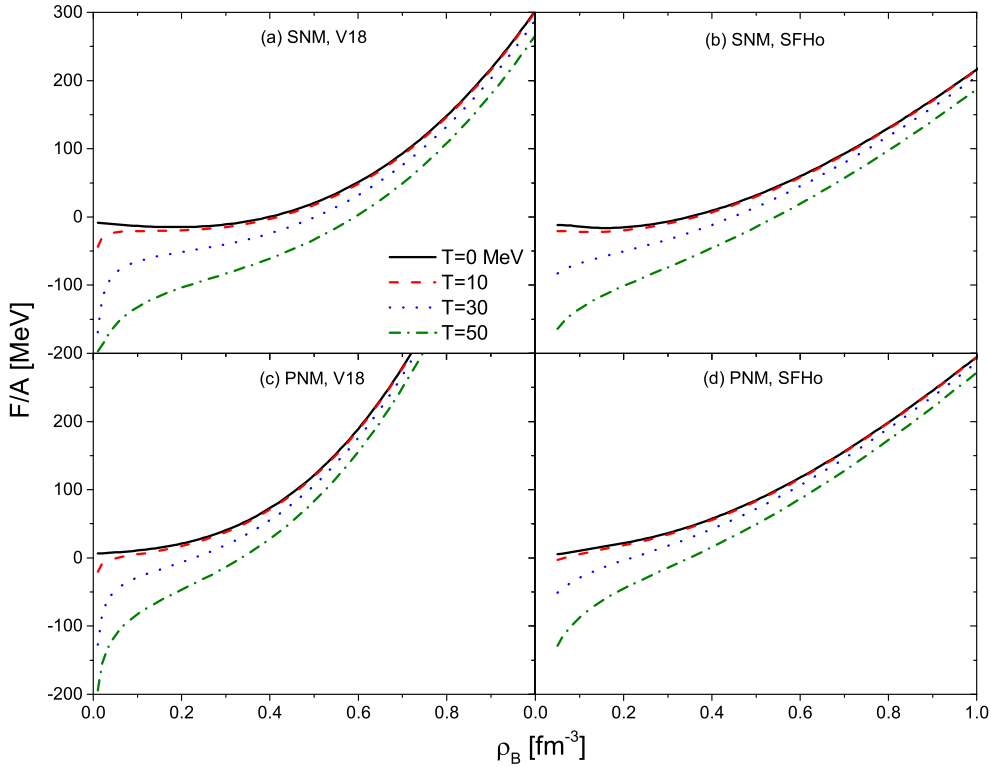}
\vskip-18mm
\caption{
Free energy per particle for several values of the temperature $T$,
for symmetric matter (upper panels)
and pure neutron matter (lower panels).
The V18 (left side) and SFHo (right side) EoSs are compared.
See text for details.}
\label{fig:EOS_T}
\end{figure}

\subsection{Finite-temperature EoSs}
\label{s:finiteT_EoS}

The EoS at finite temperature of asymmetric nuclear matter
plays a major role in the latest stage of the SN collapse
\cite{Burrows86,Prakash97,Pons99}
and binary NS mergers \cite{Baiotti17,Paschalidis17},
because it determines the final evolution of a transitory state
to either collapse to a black hole or to the formation of a NS.
A few finite-temperature nuclear EoSs for astrophysical simulations
are now available
\cite{Burgio10,Lu19,Lattimer91b,Shen98,Hempel10,Typel99,Steiner13,Togashi17},
and the predictions for the effects of temperature on stellar stability
are sometimes conflicting:
RMF models usually predict increasing stability (maximum mass) with temperature,
whereas BHF results indicate a slight reduction of the maximum mass.
There is therefore a possible mass range of the hot rotating
transitory metastable object that depends on the finite-temperature EoS.

As far as microscopic calculations of the nuclear EoS at finite temperature
are concerned,
we mention in particular the non-relativistic Brueckner-like calculations
\cite{Baldo99},
where the formalism by Bloch and De Dominicis \cite{Bloch58} was followed.
Those calculations confirmed that the resulting EoS for SNM
shows a typical Van der Waals behaviour,
which entails a liquid-gas phase transition,
with a definite critical temperature found to be around $T_c = 18 - 20\,$MeV,
very similar to the Friedman and Pandharipande findings \cite{Friedman81}
based on the variational method.
The main drawback in this approach is that the Hugenholtz-Van Hove (HVH) theorem
is not satisfied.
In other words,
the pressure $p$ calculated from the thermodynamic relation $p=-f+\mu\rho$,
being $f$ the free energy density, $\mu$ the chemical potential
and $\rho$ the number density,
does not coincide with the pressure calculated from
$p = -\Omega/V$, $\Omega$ being the grand potential and $V$ the volume.
In order to overcome this problem,
in Ref.~\cite{Baldo99} a procedure was proposed in which
the pressure is calculated from the derivative of the free energy per particle;
in this way, the HVH theorem is automatically satisfied.
The same drawback is present also in the relativistic DBHF.
On the contrary,
the HVH theorem is strictly fulfilled within the SCGF method \cite{Rios08}.
If only two-body forces are used,
the results at the two-body correlation level in some cases
are similar to the Brueckner ones,
in some others they differ appreciably according to the forces used.
The main difference with the Brueckner scheme is the introduction
of the hole-hole propagation in the ladder summation,
which gives a repulsive contribution.
As a result, the critical temperature in the SCGF approach
with Argonne $V_{18}$ potential is found to be about
$T_c \approx 11.6\,$MeV \cite{Rios08},
depending on the adopted NN interaction.
We remind that the values of the critical temperature depend both
on the theoretical approach and on the NN interaction employed.
For example, it turns out that the critical temperature within the DBHF scheme
is definitely smaller than in the non-relativistic scheme,
about 10~MeV \cite{Terhaar86,Huber98a}.
This cannot be due to relativistic effects,
since the critical density is about $1/3$ of the saturation density,
but to a different behaviour of the Dirac-Brueckner EoS at low density.
This point remains to be clarified.
However, there are experimental data from heavy-ion reactions that point towards
a value of $T_c > 15\,$MeV \cite{Borderie08}.
For completeness, we also mention an empirical extension of a
variational microscopic model based on the APR EoS \cite{Schneider19}.

The number of phenomenological EoSs available at finite temperature
is also quite small.
These are mostly based on density-functional theory (DFT),
either on a relativistic version with various parameterizations
or a non-relativistic one based on Skyrme functionals.
Those models are more extensively discussed in Sec.~\ref{s:genpurp} below.

In closing this section,
we show in Fig.~\ref{fig:EOS_T} a comparison between
the microscopic BHF V18 EoS (left panels)
and the general-purpose SFHo EoS (right panels).
In particular we display the free energy per particle
at different values of the temperature,
for SNM (upper panels) and PNM (lower panels).
In both cases we notice the typical Van der Waals behaviour
of the isothermal curves,
which are quite similar up to about twice the saturation density,
but diverge at large density, being SFHo softer than V18.
\footnote{
The original routine for the SFHo EoS is available at the CompOSE repository
https://compose.obspm.fr}
Recently, EoSs at finite temperature incorporating hyperonic degrees of freedom
have been also developed
\cite{Ishizuka08,Oertel12,Shen11a,Bhb14,Oertel16,Oertel17},
as well as those including a phase transition to quark matter
\cite{Nakazato08,Nakazato13,Sagert09,Sagert10,Fischer11,Fischer14b}.

\subsection{General-purpose EoSs}
\label{s:genpurp}

Besides the microscopic approaches,
the so-called ``general-purpose'' EoSs are able to cover a wide range
of densities, temperatures, and charge fractions,
describing both clustered and homogeneous matter.
These EoSs are therefore suitable for applications to SNe and mergers.
However, at present, only a few of them are available and directly applicable
to simulations;
we list below the general-purpose EoSs with only nucleonic degrees of freedom
currently used in astrophysical applications.

\begin{itemize}

\item The Hillebrandt and Wolff (H\&W) EoS \cite{Hillebrandt84}
has been calculated using a Nuclear Statistical Equilibrium (NSE) network
based on the model of \cite{Eleid80}
in the density range $10^9-3\times10^{12}\gc3$.
At higher densities, the EoS is computed in the single-nucleus approximation,
and the nuclear interaction employed is the Skyrme interaction.
This EoS is still used in recent numerical simulations \cite{Janka12}.

\item The Lattimer and Swesty (LS) EoS \cite{Lattimer91} is a very widely used
EoS in numerical simulations.
It models matter as a mixture of heavy nuclei
treated in the single-nucleus approximation (SNA),
$\al$ particles, free neutrons and protons,
immersed in a uniform gas of leptons and photons.
Nuclei are described within a medium-dependent liquid-drop model,
and a simplified NN interaction of Skyrme type is employed for nucleons.
With increasing density,
shape deformations of nuclei (non-spherical nuclei and bubble phases)
are taken into account by modifying the Coulomb and surface energies,
and the transition to uniform matter is described by a Maxwell construction.

\item The Shen et al.~(STOS) EoS \cite{Shen98a} is also widely used.
As the LS EoS,
matter is described as a mixture of heavy nuclei (treated in SNA),
$\al$ particles, and free neutrons and protons,
immersed in a homogeneous lepton gas.
For nucleons, a RMF model with the TM1 interaction \cite{Sugahara94} is used;
$\al$ particles are described as an ideal Boltzmann gas
with excluded-volume corrections.
The properties of the heavy nucleus are determined by WS-cell calculations
within the TF approach employing parameterized density distributions
of nucleons and $\al$ particles.

\item The Furusawa EoS of \cite{Furusawa11} is based on a NSE model,
including light and heavy nuclei.
For nuclei, the liquid-drop model is employed, and
the nuclear interaction is the RMF parameterization TM1 \cite{Sugahara94}.
This EoS has been applied in CCSN simulations to study the effect
of light nuclei in \cite{Furusawa13},
and the dependence of weak-interaction rates on the nuclear composition
during stellar core collapse in \cite{Furusawa17b}.

\item The SFHo and the SFHx EoSs \cite{Steiner13} are based on the
Hempel and Schaffner-Bielich (HS) \cite{Hempel10} approach,
which is inspired by the extended NSE model,
that takes into account an ensemble of nuclei and interacting nucleons.
Nuclei are described as classical Maxwell-Boltzmann particles,
and nucleons are described within the RMF model employing
new parameterizations fitted to some NS radius determinations.
Binding energies are taken from experimental data or from
theoretical nuclear mass tables.
Excluded-volume effects are implemented in a thermodynamically consistent way
so that it is possible to describe the transition to uniform matter.
We remind that NSE models have been employed for typical conditions
of core-collapse supernova \cite{Botvina10,Raduta10,Blinnikov11};
for a comparison among the different methods,
please refer to Ref.~\cite{Buyuk13}.

\item The EoSs SHT \cite{Shen11a} and SHO \cite{Shen11b},
are computed using different methods in different density-temperature domains.
At high densities, uniform matter is described within a RMF model.
For non-uniform matter at intermediate densities,
calculations are performed in the (spherical) WS approximation,
incorporating nuclear shell effects \cite{Shen10a}.
In this regime,
matter is modelled as a mixture of one average nucleus and nucleons,
but no $\al$ particles \cite{Shen10a}.

\item The GRDF (generalized relativistic density functional) is based on a
relativistic mean-field model of nuclear matter
with density-dependent nucleon-meson couplings
using the functional dependence introduced in Ref.~\cite{Typel10}.
Besides nucleons, electrons and muons with experimental masses,
photons and nuclei are included as degrees of freedom.
Two-nucleon correlations in the continuum are considered as effective resonances,
and the dissolution of nuclei is described with the help of
medium-dependent mass shifts.
Masses of nuclei are taken from \cite{Wang12}.

\item The EoS by Schneider, Roberts, and Ott (SRO) \cite{Schneider17}
is computed using the SLy4 parametrization \cite{Chabanat98}.
The model includes nucleons, which are treated as non-relativistic particles;
$\al$ particles, and photons, electrons and positrons,
all treated as thermally equilibrated non-interacting relativistic gases.
At low densities and temperatures nucleons may cluster into heavy nuclei
computed within the SNA.
Additional information on the open-source SRO EoS code can be found in
https://stellarcollapse.org/SROEOS.

\item The RG(SLy4) EoS corresponds to the extended NSE model proposed in
Refs.~\cite{Gulrad15,Raduta19},
where excluded volume effects between nuclear clusters and unbound nucleons
are implemented via energy shifts of clusters binding energies.
For nuclei for which experimental masses are known,
the mass tables \cite{Wang12} are used.
Then, up to the drip lines,
evaluated masses model by Duflo and Zuker \cite{Duflo95} are employed.
Beyond drip lines,
nuclear binding energies are described according to the liquid-drop-model-like
parametrization of \cite{Danielewicz09},
corresponding to the SLy4 parameterization.

\item The TNTYST EoS \cite{Togashi17} is based on the variational many-body
theory with realistic nuclear forces.
For uniform matter, the EoS is constructed with the cluster variational method
starting from the Argonne V18 two-body nuclear potential
and the Urbana IX three-body nuclear potential.
Non-uniform nuclear matter is treated in the Thomas-Fermi approximation.
In the FT version, the variational approach is combined
with the quantum approach for $d$, $t$, $h$ and $\al$,
as well as the liquid-drop model for the other nuclei
under the NSE assumption \cite{Furusawa11,Furusawa13}.

\end{itemize}

The interested reader can refer to the CompOSE repository,
https://compose.obspm.fr/,
for additional information regarding the practical use of the above listed EoS.
Furthermore, we'd like to mention a few hybrid general purpose EoSs
which are present in the repository, namely

\begin{itemize}

\item
The STOS EoS has been used for describing the hadron phase,
with the inclusion of a transition to quark matter \cite{Sagert09}
using a non-linear RMF model with the TM1 parametrization
of the effective interaction
and a bag model for the quark phase with three different values
of the bag constant and the QCD coupling constant $\al_s$.
The transition from the hadronic to the quark phase is done
via a Gibbs construction.
Non-uniform nuclear matter is calculated in the single-nucleus
Thomas-Fermi approximation with parametrized density distributions
in spherical Wigner-Seitz cells.
Only neutrons, protons, alpha particles and a single heavy nucleus
are considered.

\item The relativistic density functional (RDF) formalism is used to describe
homogeneous quark and hadron matter in a mean-field approximation
\cite{Bastian21},
assuming a first-order phase transition from hadron to quark matter.
The model was applied to NS configurations,
and its current extension to finite temperature and arbitrary charge fractions
has been employed to core-collapse supernova simulations \cite{Fischer18}
and binary NS merger simulations \cite{Bauswein20}.

\end{itemize}

\section{Structure and EoS of the neutron star crust}
\label{s:crust}

Although the crust is only a small fraction of the star mass and radius,
it plays an important role in various observed astrophysical phenomena such as
pulsar glitches,
quasiperiodic oscillations in soft gamma-ray repeaters (SGR)
and thermal relaxation in soft x-ray transients (SXT)
\cite{Haensel07,Chamel08,Strohmayer06,Sotani12,Newton13,Piekarewicz14},
which depend on the departure of the star from the picture of a homogeneous
fluid.

In modelling the NS crust,
it is usually assumed that it has the structure of a regular lattice,
in which nucleons can be either treated as a uniform system
of interacting particles,
or distributed within a defined shaped and sized cell.
In the latter case,
often the Wigner-Seitz (WS) approximation is used:
matter is distributed in charge-neutral cells,
which at lower densities are usually assumed spherical,
centered around the positive-charged ion surrounded by a
uniform electron and eventually free neutron gas,
whereas at higher densities nuclei can be non-spherical and other
geometries of the cell are considered.
The minimization of the (free) energy of the system with respect
to the variational variables,
e.g.~the nucleus atomic and mass number,
the radius of the cell,
and the free nucleon densities,
under baryon number and charge conservation,
allows the calculation of the EoS for the crust.
If ``pasta'' phases are included,
the minimisation is also performed on the shape of the cell.
In the following we discuss the main features of the structure and crust EoS,
and we refer the interested reader to Ref.~\cite{Blaschke18}
for a detailed treatment.

Current ab-initio many-body calculations with realistic interactions are
not affordable to describe clusterized matter,
and therefore for the crust we have to rely on phenomenological models.
As far as the outer crust is concerned,
the classical way to determine the EoS is to use the so-called
Baym-Pethick-Sutherland (BPS) model \cite{Baym71}.
In this model, the outer crust is supposed to be made of fully ionised atoms
arranged in a body-centered cubic lattice at $T=0$
and to contain homogeneous crystalline structures made of one type of nuclides,
coexisting with a degenerate electron gas.
The EoS in each layer of pressure $P$ is found by minimising the
Gibbs free energy per nucleon,
the only microscopic input being nuclear masses \cite{Haensel07}.
The BPS model has been updated by using modern nuclear data
and theoretical mass tables \cite{Ruster06},
which were compared to check their differences concerning the neutron drip line,
magic neutron numbers,
the EoS and the sequence of neutron-rich nuclei
up to the drip line in the outer crust.
Pairing and deformation effects were also included.

A very popular model to describe the inner crust was first introduced
by Baym, Bethe and Pethick (BBP) \cite{Baym71},
based on the compressible liquid drop model (CLDM) to take into account
the effect of the dripped neutrons.
Liquid-drop models were among the earliest to be used in astrophysical
applications to treat non-uniform matter at zero and finite temperature,
because of their applicability and limited computational efforts
\cite{Baym71,Lattimer81,Lattimer85,Lattimer91b,Lorenz93,Watanabe00,Douchin00,Douchin01,Oyamatsu07}).
Those models parameterize the energy of the system in terms of global properties
such as volume, asymmetry, surface, and Coulomb energy;
their parameters are fitted phenomenologically.
Moreover nucleons inside neutron-proton clusters and
free neutrons outside are treated separately,
and assumed to be uniformly distributed.
Clusters have a sharp surface, and quantum-shell effects are neglected.
A partially phenomenological approach, based on the CLDM,
was developed by Lattimer and Swesty (LS) \cite{Lattimer91},
who derived the EoS from a Skyrme nuclear effective force.
Another EoS was developed by Shen et al.~\cite{Shen98a,Shen98b}
based on a nuclear RMF model.
The crust was described in the Thomas-Fermi (TF) scheme
using the variational method
with trial profiles for the nucleon densities.
The LS and Shen EoSs are widely used in astrophysical calculations
for both NSs and supernova simulations due to their numerical simplicity
and the large range of tabulated densities and temperatures.

Douchin and Haensel (DH) \cite{Douchin01} formulated a unified EoS for NS
on the basis of the SLy4 Skyrme nuclear effective force \cite{Chabanat98},
where some parameters of the Skyrme interaction were adjusted to reproduce
the Wiringa et al.~microscopic calculation of neutron matter
\cite{Wiringa88} above saturation density.
Hence, the DH EoS contains certain microscopic input.
In the DH model the inner crust was treated in the CLDM approach.
More recently, unified EoSs for NSs have been derived
by the Brussels-Montreal group
\cite{Chamel11,Pearson12,Fantina13,Potekhin13}.
They are based on the BSk family of Skyrme nuclear effective forces
\cite{Goriely10}.
Each force is fitted to the known masses of nuclei and adjusted
among other constraints to reproduce a different microscopic EoS
of neutron matter with different stiffness at high density.
The inner crust is treated in the extended Thomas-Fermi approach
with trial nucleon density profiles including perturbative shell corrections
for protons via the Strutinsky integral method.
Analytical fits of these NS EoSs have been constructed in order to
facilitate their inclusion in astrophysical simulations \cite{Potekhin13}.
More recently, in \cite{Sharma15} has been proposed a unified EoS (BCPM)
from the outer crust to the core based on modern microscopic calculations
using the Argonne $V_{18}$ potential plus TBFs computed with the Urbana model.
To deal with the inhomogeneous structures of matter in the NS crust,
they used a nuclear energy density functional that is directly based on the
same microscopic calculations,
and which is able to reproduce the ground-state properties of nuclei
along the periodic table.
The EoS of the outer crust requires the masses of neutron-rich nuclei,
which are obtained through Hartree-Fock-Bogoliubov calculations
with the new functional when they are unknown experimentally.
To compute the inner crust, Thomas-Fermi calculations in Wigner-Seitz cells
are performed with the same functional.
We notice that the DH, BSk and BCPM approaches are the only ones which
construct a unified theory able to describe on a microscopic level the
complete structure of NS,
from the outer crust to the inner core within the same theoretical approach.

Quantal Hartree-Fock models have been also used for the NS crust,
in which pairing and shell effects are properly taken into account
\cite{Negele73,Bonche82,Baldo07b,Magierski03,Papa13,Pais14}.
Among those,
we mention the Negele-Vautherin (NV) EoS as the mostly used one of this class.
The main drawback of those calculations,
which employ non-relativistic interactions,
is that they are computationally very expensive.

\begin{figure}[t]
\centering
\vspace{-8mm}
\includegraphics[scale=0.36,clip]{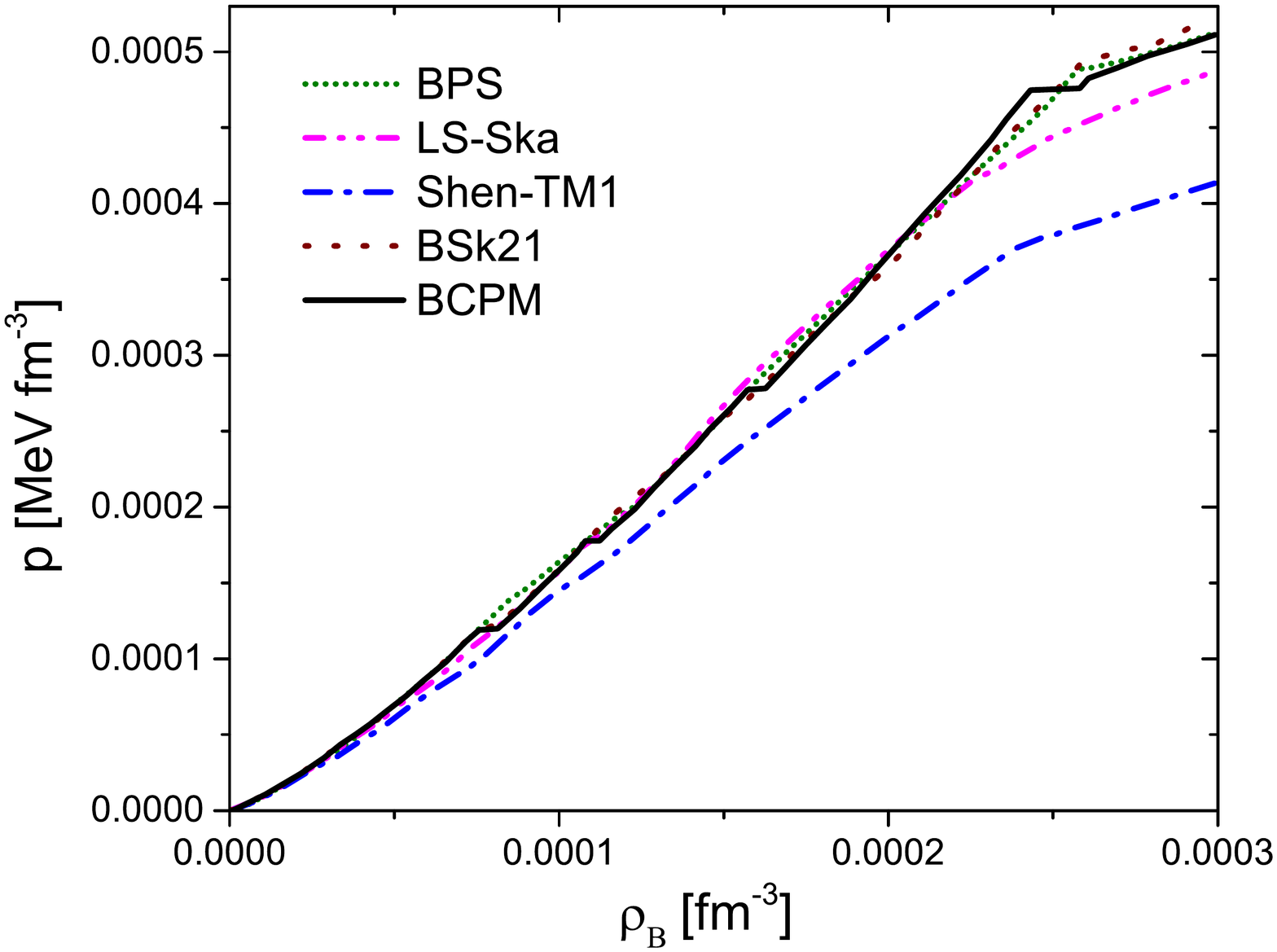}
\hspace{-19mm}
\includegraphics[scale=0.36,clip]{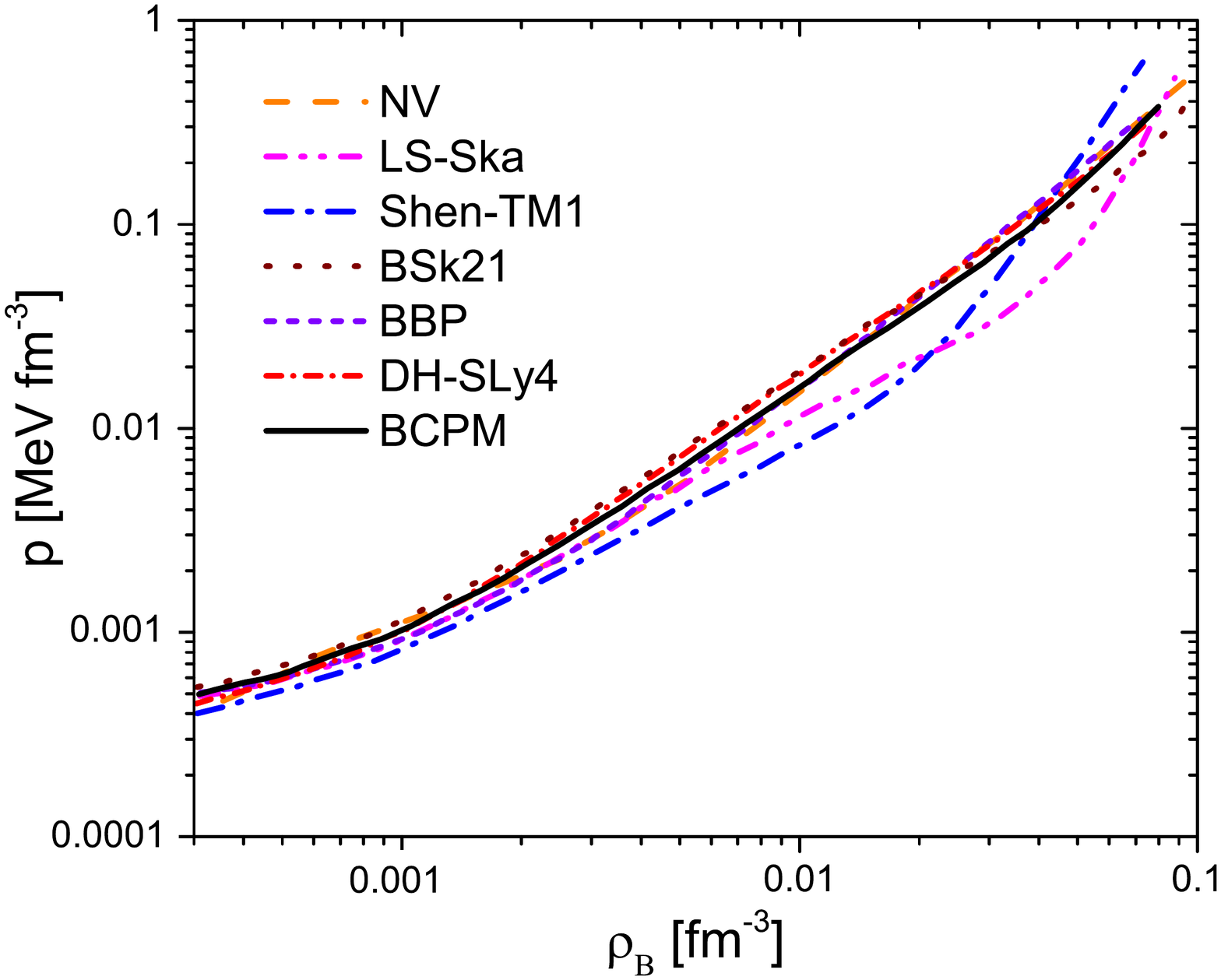}
\vspace{-10mm}
\caption{
Pressure $p$ as a function of the nucleon density
for the outer crust (left)
and the inner crust (right).
Note that the x-axis scale on the right figure starts at
$\rho = 3\times10^{-4}\fm3$,
where the left figure ends.
See text for details.
Adapted from Ref.~\cite{Sharma15}.}
\label{f:crust}
\end{figure}

In Fig.~\ref{f:crust},
we display in the left panel an EoS sample for the outer crust,
and in the right panel the one for the inner crust.
The initial baryon density in the right panel corresponds to the last density
shown in the left panel,
where the EoS of the outer crust is plotted.
The LS EoS \cite{Lattimer91},
taken here in its Ska version (LS-Ska),
and the Shen-TM1 EoS \cite{Shen98a,Shen98b} were computed with the
Skyrme force Ska and the relativistic mean-field model TM1, respectively.
In the two cases the calculations of masses are of semiclassical type and
$A$ and $Z$ vary in a continuous way.
Therefore, these EoSs do not present jumps at the densities associated
with a change from an equilibrium nucleus to another in the composition.
On the contrary those jumps are present for
the BCPM and the BSk21 Skyrme nuclear effective force
\cite{Pearson12,Fantina13,Potekhin13,Goriely10} tabulated in \cite{Potekhin13}.
The parameters of these forces were fitted to reproduce
with high accuracy almost all known nuclear masses,
and to various physical conditions including the neutron-matter EoS
from microscopic calculations.
We see that the BSk21 pressure is similar to the BCPM,
with just some displacement around the densities where the composition changes
from a nucleus to the next one.
In the seminal work of BPS \cite{Baym71} the nuclear masses for the outer crust
were provided by an early semi-empirical mass table.
In Fig.~\ref{f:crust} the corresponding EoS is seen to display a
similar pattern with the BCPM and BSk21 results.
The LS-Ska EoS shows good agreement with the previously discussed models,
with some departure from them in the transition to the inner crust.
The largest discrepancies in Fig.~\ref{f:crust} are observed with the
Shen-TM1 EoS that in this region predicts a softer crustal pressure
with density than the other models.

The crustal pressure is an essential ingredient entering the
Tolman-Oppenheimer-Volkoff (TOV) equations because of its implications for
astrophysical phenomena such as pulsar glitches \cite{Piekarewicz14}.
The pressure in the inner crust is provided by the free gases of the electrons
and of the interacting dripped neutrons
(aside from a correction from Coulomb exchange).
We note, however, that the pressure obtained in a WS cell in the inner crust
differs from the value in homogeneous $npe$ matter
in beta-equilibrium at the same average density owing to the lattice effects,
which influence the electron and neutron gases.
The lattice effects take into account the presence of nuclear structures in
the crust and are automatically included in the self-consistent TF calculation.
In the inner crust, the pressure from the BCPM functional is comparable in
general to the results of the NV, BBP, DH-SLy4 and BSk21 calculations.
Particular agreement is observed in the inner crust regime
between the BCPM and BSk21 pressures up to relatively high crustal densities.
In contrast, large differences are found when the BCPM pressure
is compared with the values from LS-Ska and Shen-TM1 models.
As the crust-core transition is approached,
these differences can be as large as a factor of two,
which may have an influence on the predictions of the
mass-radius relationship of NSs, particularly in small-mass stars.

All the approaches discussed above use the so-called single-nucleus
approximation,
i.e.~the composition of matter is assumed to be made of one
representative heavy nucleus,
usually the one most energetically favoured,
along with alpha particles and unbound nucleons.
It has been shown that this approximation has a small impact on
thermodynamic quantities \cite{Burrows84} with respect to the use of a
full distribution of nuclei,
though important differences might show up if the composition is dominated
by light nuclei,
or in the treatment of nuclear processes like electron captures in CCSNe.
Indeed, the nucleus that is energetically favoured from
thermodynamic arguments might not be the one with the highest reaction rate.

At finite temperature,
different configurations are expected and they can be calculated by using
time-dependent Hartree-Fock models based on a
wavelet representation \cite{Schuetrumpf13,Sebille11}.
Further methods have also been developed,
i.e.~the classical and quantum molecular dynamics,
where nucleons are represented respectively by point-like particles or wave
packets \cite{Dorso12,Maruyama12,Lopez14,Schneider14,Horowitz15},
but these approaches are computationally very demanding.

On the other hand, one can assume that the system is in thermodynamic
equilibrium and use NSE models,
where cluster degrees of freedom are introduced explicitly.
These models suppose that the system is composed of
a statistical ensemble of different nuclear species and nucleons in
thermal, mechanical and chemical equilibrium.
In the simplest version,
NSE approaches treat the matter constituents as a mixture of
non-interacting ideal gases,
assuming they obey a Maxwell-Boltzmann statistics,
although the Fermi-Dirac distribution is often adopted just for nucleons.
NSE calculations require as input the nuclear binding energies,
which can be either experimental,
whenever available \cite{Audi03,Audi12},
or theoretical (e.g.~CLDM or EDF-based mass models).
The main drawback of standard NSE-based models is that they neglect
interactions and in-medium effects,
that are essential for nuclear matter.
Therefore, homogeneous matter cannot be correctly described
by this kind of approaches.

A possible way out is given by the extended NSE models,
where the distribution of clusters is obtained self-consistently
under conditions of statistical equilibrium,
and interactions are properly included.
For example, in-medium corrections of nuclear binding energies have been
calculated for Skyrme interactions in \cite{Papa13,Aymard14}
within a local-density and extended TF approximation, respectively;
also the screening of the Coulomb interaction due to the electron background is
taken into account \cite{Radgul09,Raduta10}).
The interactions between the cluster and the surrounding gas are often treated
with an excluded-volume method (e.g.~\cite{Raduta10,Hempel12,Furusawa13}).
We notice that thermodynamic quantities are not very much affected by the
presence of the ensemble of nuclei with respect to the single-nucleus
approximation picture.
However, some quantitative differences arise in the matter composition,
in particular concerning the contribution of light and
intermediate mass nuclei (see, e.g.~\cite{Gulrad15,Raduta16}).
Besides the first applications of a NSE model for the EoS of SN cores at low
density,
NSE models have been subsequently employed for conditions encountered in CCSN,
e.g.~in
\cite{Botvina10,Hempel10,Raduta10,Hempel12,Furusawa13,Gulrad15,Furusawa17}.


\section{Structure and EoS of the core: Hyperons}
\label{s:hyp}

Since the pioneering work of Ambartsumyan and Saakyan \cite{Ambart60},
published in 1960, the presence of hyperons (baryons with a strangeness
content) in the NS interior and their effects on the properties of these
objects have been considered by many authors using either phenomenological
(see e.g.~Refs.~\cite{Balberg97,Balberg99,Glendenning82,Glendenning85,Glendenning87,Glendenning91a,Weber89,Knorren95,Schaffner96,Huber98})
or microscopic
(see e.g.~Refs.~\cite{Schulze95,Schulze98,Baldo98,Baldo00,Vidana00a,Vidana00b,Schulze06,Schulze11,Dapo10,Sammarruca09,Kohno10,Lonardoni14,Petschauer16}
approaches for the NS-matter EoS with hyperons.
Contrary to terrestrial
conditions, where hyperons are unstable and decay into nucleons through the
weak interaction, the equilibrium conditions in NSs can make the inverse
process happen. Hyperons may appear in the inner core of NSs at densities
of about $2-3 \rho_0$. At such densities, the nucleon chemical potential is
large enough to make the conversion of nucleons into hyperons energetically
favourable. This conversion relieves the Fermi pressure exerted by the
baryons and makes the EoS softer. Consequently, the mass of the star, and
in particular its maximum value, is reduced. How much the EoS is softened,
and how much the maximum mass is reduced depends on the attractive or
repulsive character of the YN and YY interactions.
In principle, attractive (repulsive) interactions cause an
earlier (later) onset and larger (smaller) concentration of hyperons and,
therefore, a stronger (more moderate) softening of the EoS and a larger
(smaller) reduction of the maximum mass.
However, it is well known
(see e.g.~Refs.~\cite{Schulze06,Schulze11}) that,
due to several compensation mechanisms,
hyperons equalize the effect of different nucleonic interactions:
a stiffer nucleonic EoS will lead to an earlier onset of hyperons,
thus enhancing the softening due to their presence.
Conversely, a later onset of a certain hyperon species will favour
the appearance of other species leading also to a softer EoS.

In the following we will review first some of the laboratory constraints of
the nuclear EoS with hyperons derived from hypernuclear research. Then we
will discuss the construction of YN and YY interactions on the basis of the
meson-exchange and chiral effective field theories. Recent developments
based on the so-called $V_\text{low\,k}$ approach and lattice quantum
chromodynamics will also be addressed. Finally, we go over some of the
effects of hyperons on the properties of neutron and PNSs with an emphasis
on the so-called ``hyperon puzzle''. We discuss some of the solutions
proposed to tackle this problem. We also re-examine the role of hyperons on
the cooling properties of newly born NSs and on the development of the
so-called r-mode instability.

\subsection{Laboratory constraints of the nuclear EoS with hyperons: hypernuclei}

The YN and YY interactions are still poorly constrained contrary to the NN one,
which is fairly
well known thanks to the large number of existing scattering data and known
properties, such as masses and radii, of more than 3000 isotopes.
Experimental difficulties associated with the short lifetime of hyperons in
addition to the low-intensity beam fluxes have limited the number of
$\la N$ and $\Sigma N$ scattering events to just a few hundred
\cite{Engelmann66,Alexander68,Sechi-Zor68,Kadyk71,Eisele71} and that of
$\Xi N$ events to very few.
In the case of the YY interaction the situation
is even worse because no scattering data exis,t at all.
In view of this limited amount of data,
which is not enough to fully constrain these interactions,
alternative information on them can be obtained from the study of hypernuclei,
bound systems composed of nucleons and one or more hyperons.
In the following we present a brief review of the different production
mechanisms of hypernuclei as well as a few aspects of their spectroscopy.

Hypernuclei were first observed in 1952 \cite{Danysz53}, when a
hyperfragment in a balloon-flown emulsion stack was found. Since then the
use of high-energy accelerators as well as modern electronic counters have
lead to the identification of more than 40 single $\la$-hypernuclei and
a few double-$\la$ and single-$\Xi$ ones.
The existence of single-$\Sigma$ hypernuclei
has not been experimentally confirmed yet without ambiguity,
suggesting that the $\Sigma$N interaction is most likely repulsive.

Single $\la$-hypernuclei can be produced by several mechanisms such as
$(K^-,\pi^-)$ {\em strangeness exchange reactions},
where a neutron of the
nucleus target hit by a $K^-$ is changed into a $\la$ emitting a
$\pi^-$. The analysis of these reactions showed many of the hypernuclear
characteristics such as, for example, the small spin-orbit strength of the
YN interaction.
The use of $\pi^+$ beams allowed $(\pi^+,K^+)$
{\em associated production reactions},
where an $s\bar{s}$ pair is created
from the vacuum and a $K^+$ and a $\la$ are produced in the final state.
The {\em electroproduction} of hypernuclei by means of the reaction
$(e,e'K^+)$ provides a high-precision tool for the study of hypernuclear
spectroscopy due to the excellent spatial and energy resolution of the
electron beams. Recently, the HypHI Collaboration at FAIR/GSI has proposed
a new way to produce hypernuclei by using stable and unstable heavy ion
beams \cite{Bianchin09}. The $\la$ and the $^3_\la$H and
$^4_\la$H hypernuclei have been observed in a first experiment using a
$^6$Li beam on a $^{12}$C target at 2 AGeV \cite{Rappold13}.

$\Sigma$ hypernuclei can also be produced in similar reactions. However, as we
said, there is not yet an unambiguous experimental confirmation of their
existence. The production of double-$\la$ hypernuclei, requires first
to create a $\Xi^-$ through reactions like
\begin{equation}
 p+\bar p \ra \Xi^-+\bar{\Xi}^+
\label{e:xi1}
\end{equation}
or
\begin{equation}
 K^-+p \ra \Xi^-+K^+ \:.
\label{e:xi2}
\end{equation}
Then it is necessary that the $\Xi^-$ be captured in an atomic orbit
and interacts with the nuclear core,
producing two $\la$'s by hitting one of the protons of the nuclei
\begin{equation}
 \Xi^-+p\ra \la+\la \:.
\end{equation}
This reaction releases about 30 MeV of energy that, in most of the cases,
is equally shared between the two $\la$'s, leading to the escape of one
or both hyperons from the nucleus. Single-$\Xi$ hypernuclei can be produced
by means of the reactions (\ref{e:xi1}) and (\ref{e:xi2}) and few of them
have been in fact identified. The analysis of the experimental data from
reactions such as $^{12}$C($K^-,K^+$)$^{12}_{\Xi^-}$Be \cite{Khaustov00}
indicates an attractive $\Xi$-nucleus interaction of about
$-14$ MeV. We should mention here the recent observation \cite{Nakazawa15}
of a deeply bound state of the $\Xi^-$-$^{14}$N system with a binding
energy of $3.87 \pm 0.21$ MeV \cite{Hiyama18}.
This event provides the first clear evidence
of a deeply bound state of this system by an attractive $\Xi$N
interaction.
Future $\Xi$ hypernuclei experiments are being planned at J-PARC.

Double-strange hypernuclei are currently the best systems to investigate
the properties of the baryon-baryon interaction in the strangeness $S=-2$
sector. The $\la\la$ bound energy $\Delta B_{\la\la}$ in
double-$\la$ hypernuclei can be determined experimentally from the
measurement of the binding energies of double and single-$\la$ hypernuclei as
\begin{equation}
 \Delta B_{\la\la} = B_{\la\la}(^A_{\la\la}Z) - 2B_{\la}(^{A-1}_{\la}Z) \:.
\end{equation}
Emulsion experiments have reported the formation of a few double-$\la$
hypernuclei: $^6_{\la\la}$He, $^{10}_{\la\la}$Be and
$^{13}_{\la\la}$B. From the subsequent analysis of these emulsion
experiments, a quite large $\la\la$ bound energy of around 4 to 5
MeV was deduced. We should also note that the identification of some of
these double-$\la$ hypernuclei was ambiguous. Therefore, careful
attention should be paid when using the data from this old analysis to put
any kind of constraint on the $\la\la$ interaction. However, a new
$^6_{\la\la}$He candidate with a $\la\la$ bound energy
$\Delta B_{\la\la}=1.01 \pm 0.2^{+0.18}_{-0.11}$ MeV (recently
corrected to $\Delta B_{\la\la}=0.67 \pm 0.17$) was unambigously
observed in 2001 at KEK \cite{Takahashi01}.

Hypernuclei can be produced in excited states if a nucleon in a $p$ or a
higher shell is replaced by a hyperon. The energy of these excited states
can be released either by emitting nucleons, or, sometimes, when the
hyperon moves to lower energy states, by the emission of $\gamma$-rays. The
detection of $\gamma$-ray transitions in $\la$ hypernuclei has allowed
the analysis of hypernuclear excited states with very good energy
resolution. However, there have been some technical difficulties in the
application of $\gamma$-ray spectroscopy to hypernuclei mainly related with
the detection efficiency of $\gamma$-ray measurements and with the
necessity of covering a large solid angle with $\gamma$-ray detectors. The
construction of large-acceptance germanium detectors, dedicated to
hypernuclear $\gamma$-ray spectroscopy, has allowed one to solve these
issues somehow. There exist still, however, several weak points in
hypernuclear $\gamma$-ray spectroscopy. A number of single-particle
$\la$ orbits are bound in heavy $\la$ hypernuclei with a potential
depth of around 30 MeV but the energy levels of many single-particle orbits
are above the neutron and proton emission thresholds. Therefore, the
observation of $\gamma$-rays is limited to the low excitation region, maybe
up to the $\la$ $p$-shell. The fact that the $\gamma$-ray transition
only measures the energy difference between two states is clearly another
weak point, since single energy information is not enough to fully identify
the two levels. The measurement of two $\gamma$-rays in coincidence might
help to resolve it. Systematic spectroscopic studies of single $\la$
hypernuclei indicate that the $\la$N interaction is attractive
\cite{Hashimoto06}.

Before closing this section,
we would like to note that the YN and YY interactions can be tested
experimentally by measuring the correlation in momentum space
for Yp or YY pairs produced at colliders
exploiting the so-called femtoscopy technique
\cite{Adamczewski16,Mihaylov18,Acharya19a,Acharya19b,Acharya19c,Acharya20,Fabbietti20,Tolos20}.
The correlation function between two hadrons is defined as the ratio
of the distribution of relative momenta
$k^*=|\pv_1-\pv_2|/2$
for correlated and uncorrelated baryon pairs.
In the case of an attractive interaction,
the measured correlation will be larger than one,
whereas if the interaction is repulsive or there exists a bound state,
it will take values between zero and one.
Theoretically, the correlation function can be expressed as \cite{Pratt86,Lisa05}
\begin{equation}
 C(k^*) = \int d^3\rv S(r)|\Psi(k^*,r)|^2 \:,
\label{e:cf}
\end{equation}
where $S(r)$ is the distribution at the distance $r$
at which particles are emitted (source) and
$\Psi(k^*,r)$ represents the relative wave function
of the hadron pair of interest.
The comparison of the measured correlation function with the theoretical one
allows to test and improve the existing baryon-baryon potentials.

\subsection{The hyperon puzzle}

Although, energetically, the presence of hyperons in NSs seems to be
unavoidable, the softening that their presence induces on the EoS leads to
maximum masses which are incompatible with observations.
The solution of this problem,
commonly known in the literature as the ``hyperon puzzle,''
is not easy.
It has become a subject of very active research
(see e.g.~Ref.~\cite{Chatterjee16} and references therein
for a comprehensive review),
since the measurements of unusually high masses of the millisecond pulsars
PSR J1903+0327 ($1.667 \pm 0.021\ms$) \cite{Champion08},
PSR J1614-2230 ($1.928 \pm 0.017\ms$) \cite{Demorest10},
PSR J0348+0432 ($2.01 \pm 0.04\ms$) \cite{Antoniadis13}
and the most recent
PSR J0740+6620 ($2.14^{+0.10}_{-0.09}\ms$) \cite{Cromartie20},
which rule out almost all currently proposed EoSs with hyperons
(both microscopic and phenomenological).
To solve this problem, a mechanism is necessary
that could provide the additional repulsion needed to make the EoS
stiffer and therefore
the maximum mass compatible with the current observational limits.
Three of the mechanisms proposed that could provide
such additional repulsion are:
(i) the inclusion of a repulsive
YY interaction through the exchange of vector mesons
\cite{Bednarek12,Weissenborn12,Oertel15,Maslov15},
(ii) the inclusion of repulsive hyperonic TBFs
\cite{Takatsuka02,Takatsuka08,Vidana11,Yamamoto13,Yamamoto14,Lonardoni15,Haidenbauer17,Kohno18}, or
(iii) the possibility of a
chiral phase transition and/or a phase transition
within color superconducting states
at densities below the hyperon threshold
\cite{Alford07,Weissenborn11,Ozel10,Schramm12,Zdunik13,Klahn13,Bonanno12,Lastowiecki12}.
The possible appearance of other hadronic degrees of freedom,
such as the $\Delta$ isobar or meson condensates,
that can push the onset of hyperons to higher densities
has also been considered.
An interesting possibility to circumvent the problem is the so-called
two-families scenario proposed in \cite{Drago16,Drago16b},
in which stars made of hadrons are stable only up to ($1.5-1.6$)$\ms$,
while most massive compact stars are entirely made of strange quark matter.
In the following, we briefly review some of these possible solutions.

\subsubsection{Hyperon-hyperon repulsion}

This solution, mainly explored in the context of RMF models
(see e.g.~Refs.~\cite{Bednarek12,Weissenborn12,Oertel15,Maslov15}),
is based on the known fact that in meson-exchange models of nuclear forces,
vector mesons are responsible for the repulsion felt by nucleons at short
distances,
while the $\sigma$ meson generates the intermediate-range attraction.
Therefore, it has been argued that if the interaction of two hyperons,
mediated through the exchange of a vector meson,
is repulsive enough or the attraction driven by the $\sigma$ meson
is weak enough,
then it can be possible to reconcile the current high pulsar mass observations
with the existence of hyperons in NSs.
The existence of hypernuclei, however,
shows that at least the $\la$N interaction is attractive \cite{Hashimoto06}.
Consequently, to be consistent with hypernuclear data,
repulsion in these models is included only through the exchange of the
hidden strangeness $\phi$ vector meson that couples only to the hyperons.
In such a way, the onset of hyperons is shifted to higher densities,
and it is possible to obtain NSs with maximum masses larger than $2\ms$
and with a non-negligible fraction of hyperons in their interior.
A question, however, immediately arises:
how does this additional YY repulsion affect the binding
energy of two $\la$s in double-$\la$ hypernuclei ?
Although the answer is not simple,
it has been shown recently \cite{Fortin17} that with
currently available hypernuclear experimental data and due to the lack of
stronger constraints on the asymmetric nuclear-matter EoS,
in RMF models it is still possible to find a wide range of values
of the $\la\phi$ coupling that predict a $\la\la$ binding energy compatible with
that derived from experimental data on $^6_{\la\la}$He,
and that simultaneously lead to the prediction of NS maximum masses
compatible with the $2\ms$ constraint.
The interested reader is referred to any of the works
(see e.g.~Ref.~\cite{Chatterjee16} and references therein)
that have explored this solution over the last years for further information.

\subsubsection{Hyperonic three-body forces}

It is well known from conventional nuclear physics that TBFs are
fundamental to reproduce properly the properties of few-nucleon systems as
well as the empirical saturation point of SNM.
Then, it is natural to think that TBFs involving one or more hyperons
(i.e.~NNY, NYY and YYY)
could also play an important role in the determination of the
NS-matter EoS and contribute to the solution of the hyperon puzzle.
Indeed if hyperonic TBFs are repulsive enough then they can make the EoS stiffer
at higher densities and, therefore, make the maximum mass of the star
compatible with observations.
This solution was suggested years ago before
the observation of NSs with $\sim 2\ms$ \cite{Takatsuka02,Takatsuka08}
and it has been considered later by other authors
\cite{Vidana11,Yamamoto13,Yamamoto14,Lonardoni15,Haidenbauer17,Kohno18}.
A general consensus regarding the role of these forces on the solution of the
hyperon puzzle, however, does not exist yet.
In Refs.~\cite{Yamamoto13,Yamamoto14}, for instance, it was found that the
inclusion of hyperonic TBFs is enough to obtain hyperon stars with $\sim2\ms$.
However, in Ref.~\cite{Vidana11} it was shown that the largest
maximum mass that they can support is $\sim 1.6\ms$.
Finally, the results of Ref.~\cite{Lonardoni15} were not conclusive enough
because they depend strongly on the $\la NN$ interaction employed.
It seems, therefore, that hyperonic TBFs are not the full solution
of the hyperon puzzle,
but they can probably contribute to it in an important way.
We want to mention here that the effect of NNY forces,
derived in the context of chiral effective field theory,
on the properties of hyperonic matter has been recently studied in
\cite{Haidenbauer17} and \cite{Kohno18}.
The interested reader is referred to all the works quoted,
Refs.~\cite{Takatsuka02,Takatsuka08,Vidana11,Yamamoto13,Yamamoto14,Lonardoni15,Haidenbauer17,Kohno18},
for the specific details of the calculations.

\subsubsection{Quarks in neutron stars}

It has been suggested that an early chiral phase transition
and/or a phase transition within color superconducting states
at densities below the hyperon threshold
could provide a solution to the hyperon puzzle.
Compact stars which possess a quark-matter core,
either as a mixed phase of deconfined quarks and hadrons,
or as a pure quark matter phase,
surrounded by a shell of hadronic matter are called hybrid stars.
Massive stars could therefore actually be hybrid stars
with a stiff quark-matter core.
The question that arises then is whether quarks can provide sufficient
repulsion to support a $2\ms$ NS.
A brief summary of some important conclusions drawn in the last years
using the massive NS constraint is given in the following.
This summary does not pretend to be exhaustive
and therefore we apologize in advance to those groups
whose results are not mentioned here.

Following the discovery of the massive NS EXO 0748-676,
it was argued \cite{Alford07} that hybrid or strange stars
(i.e.~compact stars made of deconfined $u,d,s$ quark matter)
can reach a mass of $2\ms$,
as formerly demonstrated within the framework of the MIT bag model,
perturbative QCD as well as the Nambu--Jona--Lasinio (NJL) model.
A systematic study of the consequences of the NS mass limit on the
properties of quark stars and hybrid stars with a pure quark core was
performed in \cite{Weissenborn11}.
Using an extended quark bag model,
the authors concluded that massive strange stars require
strong QCD corrections and large contribution from color superconductivity.
For the case of hybrid stars,
the EoSs with quark-hadron phase transition are compatible
with the $2\ms$ mass constraint,
provided the quarks are strongly interacting and in a color-superconducting phase
\cite{Ozel10,Weissenborn11}.
Ref.~\cite{Schramm12} defined a hadronic flavor-SU(3) model
to combine hadronic and quark degrees of freedom in a unified way.
They obtained a maximum mass of $2.06\ms$ for models with hyperons only,
and reported a mass reduction of $10\%$
on inclusion of quark degrees of freedom.
A further increase of about $20\%$ in the maximum mass was achieved
by considering rotation.
Ref.~\cite{Zdunik13} used the heavy NS mass observation to put
general constraints on the two-flavor color-superconducting (2SC) and
color-flavor-locked (CFL) phases of quark matter in the core of hybrid stars.
They deduced that for thermodynamic stability, both a stiff hyperon
repulsion at high baryon density accompanied by a stiff quark matter EoS
are required to reproduce this result.
Ref.~\cite{Klahn13} described a large number of possible parametrizations
based on the NJL model that can reach masses up to $2\ms$.
They demonstrated that to reconcile NS observations and heavy-ion flow data,
large values of the diquark and scalar couplings are necessary,
as well as repulsion arising from the vector-meson interaction.
Ref.~\cite{Bonanno12} succeeded in constructing stable NS
configurations with mass $\geq 2\ms$ using an EoS exploiting features of
phenomenological RMF density functionals at low densities, and the NJL
model of superconducting quark matter at high densities. The constraints
were satisfied subject to the condition that the nuclear EoS at
post-saturation density is sufficiently stiff,
followed by a transition to quark matter at a few times saturation density,
with substantial repulsive vector interactions in quark matter.
Ref.~\cite{Lastowiecki12} studied the possible appearance of hyperons
and strange quark matter in NSs subject to the constraints of observations
of heavy compact stars and flow constraints from heavy ion collisions (HICs).
Applying a color superconducting three-flavor NJL model for the quark sector,
and a density-dependent hadron field model
and the DBHF theory for the hadronic sector,
they showed that it is possible to have deconfined quark matter
in the core of massive stars.

\subsubsection{$\Delta$ isobar and meson condensation in neutron stars}

The appearance of other hadronic degrees of freedom such as for instance
the $\Delta$ isobar or meson condensates
that can push the onset of hyperons to higher densities
has also been considered as a possible solution to the hyperon puzzle.
The presence of the $\Delta$ isobar in NSs has been usually neglected
because its threshold density was found to be larger
than the typical densities prevalent in the NS core.
However, it has been recently shown \cite{Drago14} that the onset density
of the $\Delta$ isobar depends crucially on the slope parameter $L$
of the nuclear symmetry energy.
Using recent experimental constraints and a state-of-the-art EoS,
these authors have shown that the $\Delta$ isobar could actually appear
at densities below the hyperon threshold.
However, they found that as soon as the $\Delta$ appears,
the EoS becomes also considerably soft and consequently the maximum mass
is reduced to values below the current observational limit,
giving rise to what they have named ``the $\Delta$ puzzle.''
The possibility that pions or kaons form Bose--Einstein condensates
in the inner core of NSs has also been extensively considered in the literature
\cite{Kaplan86,Kaplan86b,Brown94,Thorsson94,Lee96,Glendenning98}.
However, as in the case of the hyperons or the $\Delta$ isobar,
the appearance of a meson condensate induces also a strong softening
of the EoS and therefore leads to a reduction of the maximum NS mass
to values also below the current observational limits.

\begin{figure}[t]
\centering
\vskip-9mm
\includegraphics[scale=0.635,clip]{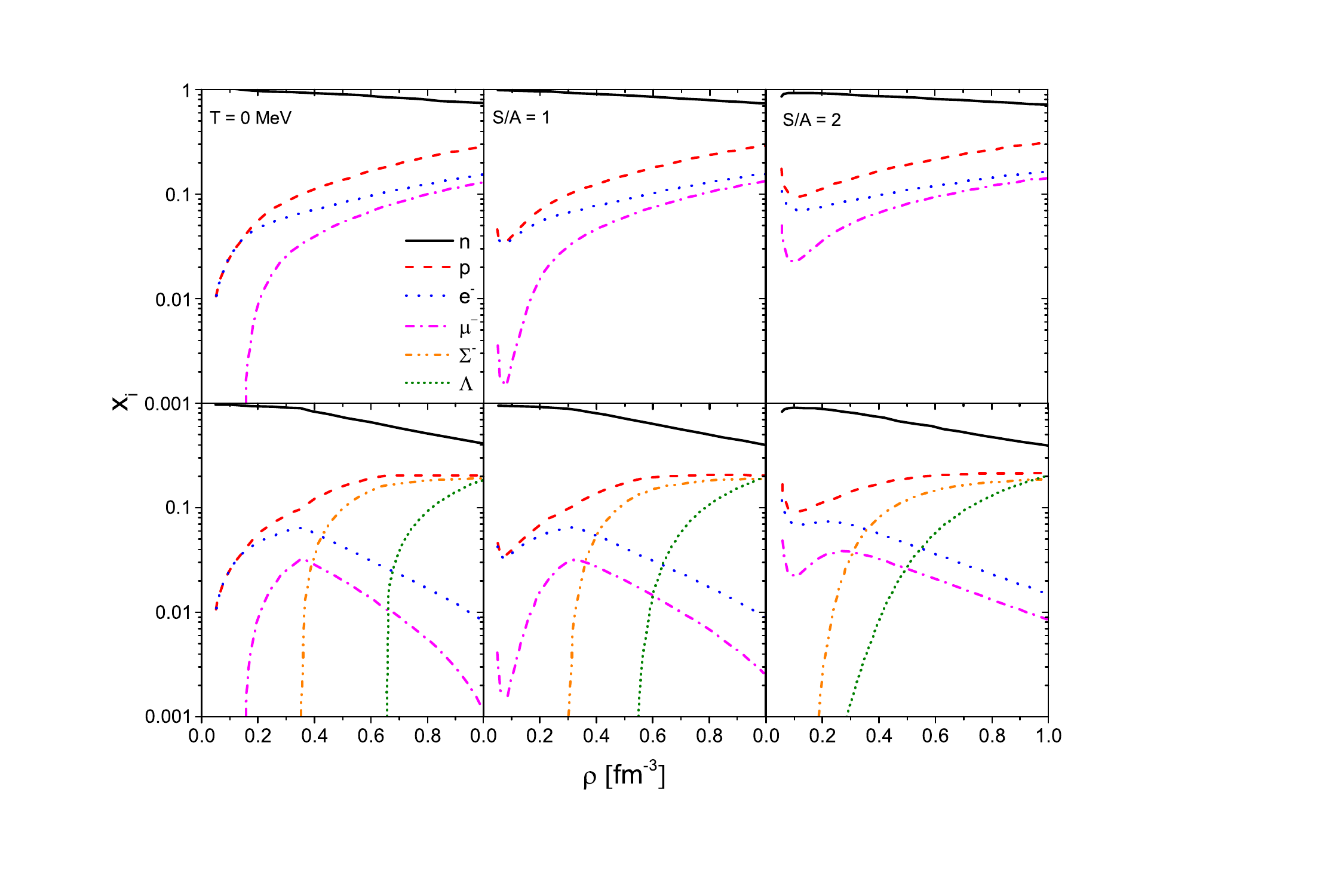}
\vskip-17mm
\includegraphics[scale=0.635,clip]{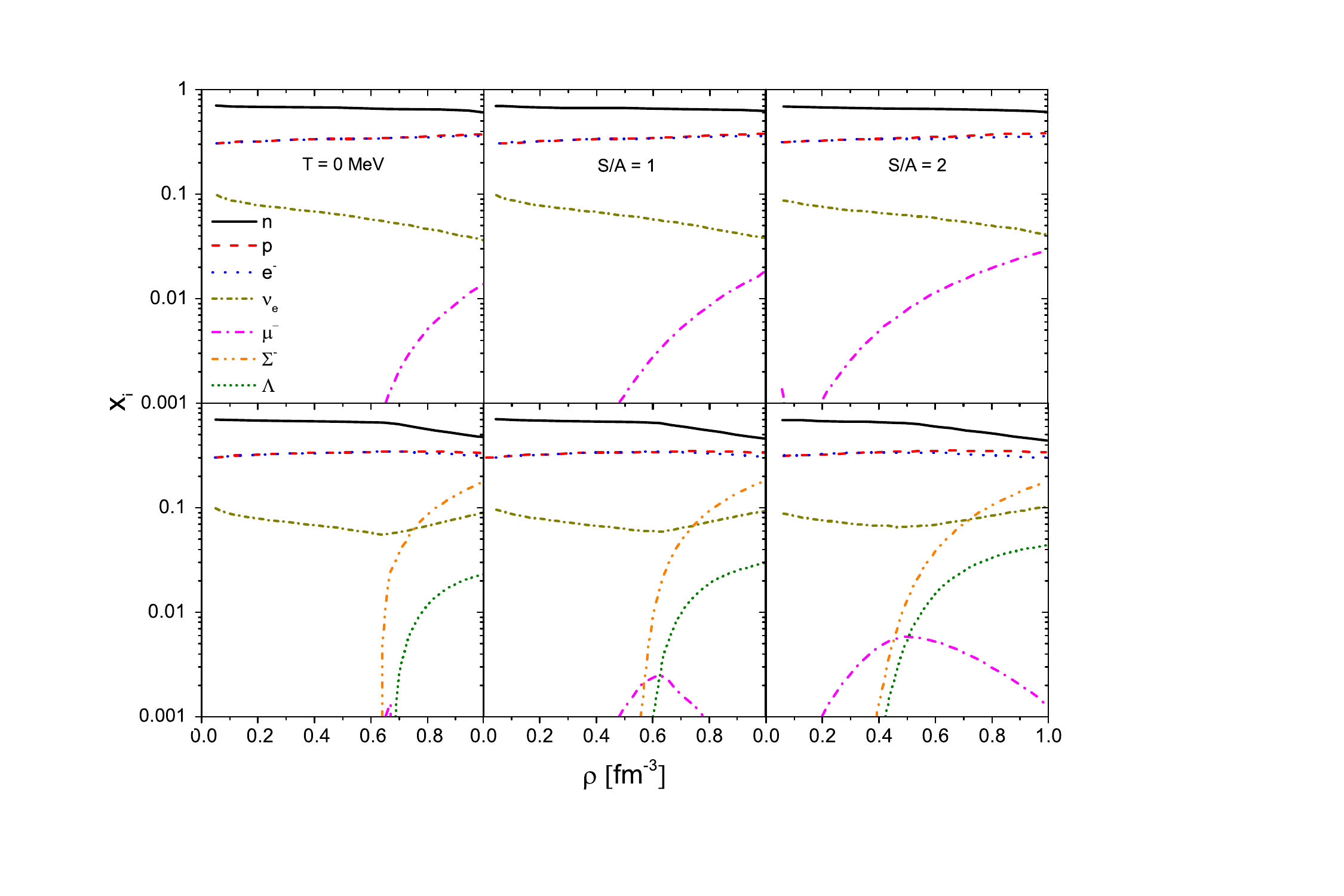}
\vskip-14mm
\caption{
Composition for neutrino-free (upper plot)
and neutrino-trapped (lower plot) matter
for several fixed values of the entropy per particle $S/A=0,1,2$
with (lower rows)
and without (upper rows) hyperons.
Figure adapted from Ref.~\cite{Burgio11}.
}
\label{f:COMPO_EOS}
\end{figure}

\begin{figure}[t]
\centering
\vskip-10mm
\includegraphics[scale=0.5,clip]{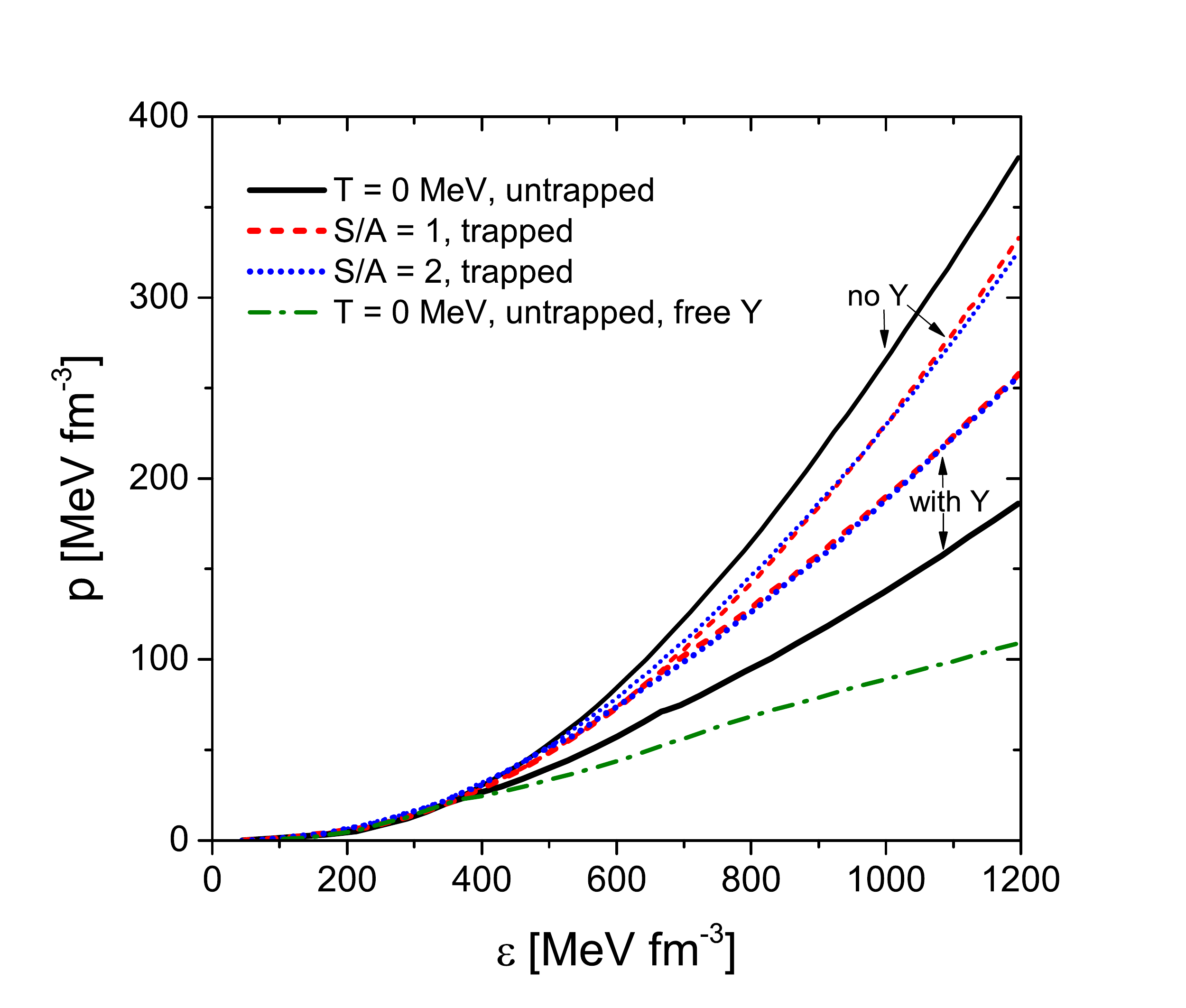}
\vskip-6mm
\caption{
EoS for neutrino-free and neutrino-trapped matter
for several fixed values of the entropy per particle $S/A=0,1,2$
with and without hyperons.
Figure adapted from Ref.~\cite{Burgio11}.
}
\label{f:COMPO_EOS2}
\end{figure}

\begin{figure}[t]
\centering
\vskip-5mm
\includegraphics[scale=0.4,clip=true]{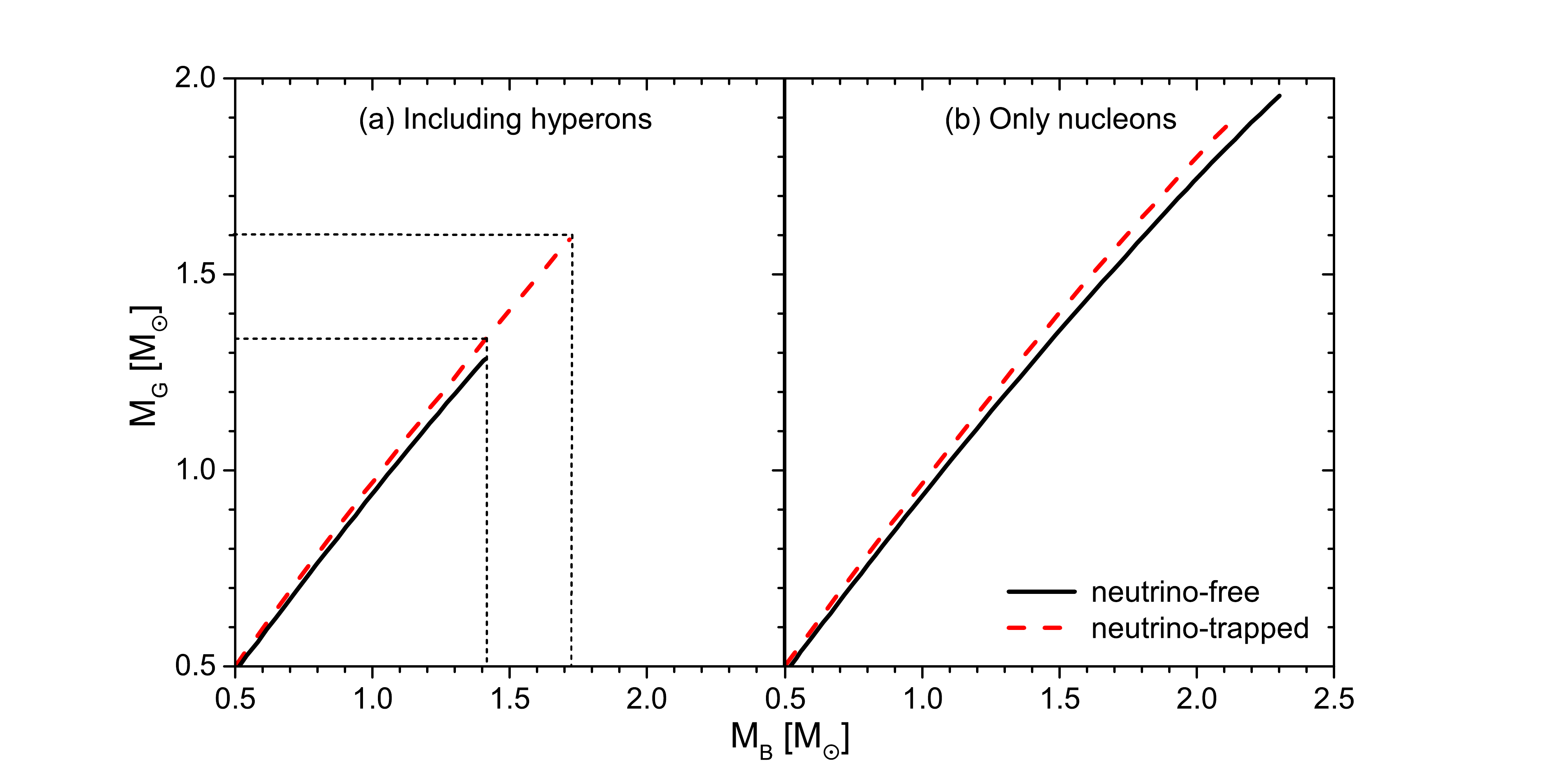}
\vskip-4mm
\caption{
Gravitational mass as a function of the baryonic mass
for neutrino-free (solid lines)
and neutrino-trapped (dashed lines) matter.
Panel (a) shows the results for matter including nucleons and hyperons,
whereas panel (b) shows the results for pure nucleonic matter.
Dotted horizontal and vertical lines show the window of metastability
in the gravitational and baryonic masses.
Figures adapted from Ref.~\cite{Vidana03}.}
\label{f:AA399}
\end{figure}

\subsection{Effect of hyperons on proto-NSs and neutron star cooling}

Thermal effects and neutrino trapping
characterise
a newly born proto-NS (PNS).
If hyperons are present,
their number is on average smaller
and their onset is shifted to higher densities
than in cold NSs.
Consequently, the EoS is stiffer in comparison with the
cool and neutrino-free case.
All these features are illustrated in
Figs.~\ref{f:COMPO_EOS} and \ref{f:COMPO_EOS2}
(adapted from Ref.~\cite{Burgio11}),
where the composition and EoS of neutrino-free and neutrino-trapped
($Y_e=0.4$) matter
with and without hyperons are shown for several fixed values of the
specific entropy.

An important consequence of neutrino trapping in NSs with hyperons
is the possible delayed
collapse of the hot newly born NS to a black hole,
as was already pointed out in \cite{Baumgarte96}.
This possibility is illustrated in Fig.~\ref{f:AA399},
which shows the gravitational mass $M_G$ of the star
as a function of its corresponding baryonic mass $M_B$
for the BHF calculation of Ref.~\cite{Vidana03}.
When hyperons are present in the star (left panel),
the deleptonization lowers the range of gravitational masses
that can be supported by the EoS from $1.59\ms$ to $1.28\ms$
(see dotted horizontal lines in the figure).
The NS baryonic mass can be considered to a good approximation
constant during the evolution from the initial PNS configuration
to the final neutrino-free one,
because most of the matter accretion on the forming NS
happens in the very early stages after its birth ($t \lesssim 1\,$s).
For the particular model calculation shown in the figure,
PNSs which at birth have a gravitational mass between $1.28\ms$ and $1.59\ms$
(a baryonic mass between $1.40\ms$ and $1.72\ms$)
will be stabilized by neutrino effects long enough
to carry out the nucleosynthesis accompanying a type-II supernova explosion.
Once neutrinos have left the star,
the EoS softens and it cannot any more support the star against its own gravity.
Thus the newborn NS collapses into a black hole
\cite{Keil96,Bombaci96,Prakash97}.
Conversely, when nucleons are considered to be the only relevant
baryonic degrees of freedom present in the NS core,
a black hole is unlikely to be formed during the deleptonization
since in this case (see right panel) the gravitational mass
increases during this stage which happens at (almost) constant baryonic mass.
If a black hole were to form from a star with only nucleons,
it is much more likely to form during the post-bounce accretion stage.

The presence of hyperons may also affect the cooling of NSs,
since it provides fast additional cooling mechanisms,
such as for example direct and modified hyperonic Urca processes,
\bal
 & Y \ra B+l+\bar\nu_l \ , \,\,\,
   B+l \ra Y +\nu_l \:,
\label{e:ydurca}
\\
 & B' + Y \ra B'+B+l+\bar\nu_l \ , \,\,\,
   B'+B+l \ra B'+ Y +\nu_l \:.
\label{e:ymurca}
\eal
Such additional cooling mechanisms, however,
can lead to predicted surface temperatures much lower than those observed
(see Sec.~\ref{s:cool}),
unless they are suppressed by hyperon pairing gaps.
Therefore, the study of hyperon superfluidity becomes of particular interest
since it could play a key role in the thermal history of NSs.
However, whereas the presence of superfluid neutrons in the inner crust of NSs
and superfluid neutrons together with superconducting protons
in their quantum fluid interior is well established
and has been the subject of many studies
(Sec.~\ref{s:gap}),
a quantitative estimation of the hyperon pairing gaps has not received
much attention and just a few calculations exist in the literature
\cite{Balberg98,Takatsuka99,Takatsuka00,Takatsuka02b,Vidana04,Zhou05,Wang10}.

\subsection{Hyperons and the r-mode instability of NSs}

It is well known that the Kepler frequency $\Omega_K$
is the absolute maximum rotational frequency of a NS,
above which matter is ejected from the star's equator
\cite{Lindblom86,Friedman86}.
Instabilities caused by several types of perturbations may prevent NSs
from reaching rotational frequencies as high as $\Omega_K$,
imposing therefore more stringent limits on their rotation \cite{Lindblom95}.
Among the different instabilities that can operate in a NS,
the so-called r-mode instability \cite{Anderson98,Friedman98},
a toroidal mode of oscillation whose restoring force is the Coriolis force,
is particularly interesting.
The r-mode instability leads to the emission of GWs
in hot and rapidly rotating NSs
through the Chandrasekhar--Friedman--Schutz mechanism
\cite{Chandrasekhar70,Friedman78,Friedman78b,Friedman78c}.
Gravitational radiation makes an r-mode grow, while viscosity stabilizes it.
Therefore, if the driving time of the gravitational radiation
is shorter than the damping time associated to viscous processes,
the r-mode becomes unstable.
In this case, a rapidly rotating NS could transfer a significant fraction
of its angular momentum and rotational energy to the emitted GWs,
whose detection could provide invaluable information on the internal structure
of the star and constraints on the EoS.

The main dissipation mechanisms of r- and other pulsation modes of NSs
are the bulk ($\xi$) and shear ($\eta$) viscosities.
Bulk viscosity is the dominant one at high temperatures ($T>10^9\,$K)
and therefore it is important for hot young NSs.
It is the result of the variations in pressure and density
induced by the pulsation mode,
variations that drive the star away from beta-equilibrium.
Energy is then dissipated as the weak interaction tries to reestablish
the equilibrium.
In the absence of hyperons or other exotic components,
the bulk viscosity of NS matter is mainly determined by the reactions
of direct and modified Urca processes.
However, as soon as hyperons appear,
new mechanisms such as weak non-leptonic hyperon reactions
\begin{equation}
 N+N \leftrightarrow N+Y \ , \,\,\, N+Y \leftrightarrow Y+Y \:,
\end{equation}
strong interactions
\begin{equation}
 Y+Y \leftrightarrow N+Y \ , \,\,\, N+\Xi \leftrightarrow Y+Y \ , \,\,\,
 Y+Y \leftrightarrow Y+Y \:,
\end{equation}
or direct and modified hyperonic Urca reactions,
Eqs.~(\ref{e:ydurca}) and (\ref{e:ymurca}),
contribute to the bulk viscosity and dominate it for densities
$\rho \geq 2-3 \rho_0$.
Hyperon bulk viscosity has been considered by several authors,
see e.g.~Refs.~\cite{Langer69,Jones71,Levin99,Jones01,Jones01b,Lindblom02,Haensel02,vanDalen04,Chatterjee06,Bondarescu07,Chatterjee08,Gusakov08,Shina09,Jha10}.

\begin{figure}[t]
\centering
\vskip-6mm
\includegraphics[scale=0.45,clip=true]{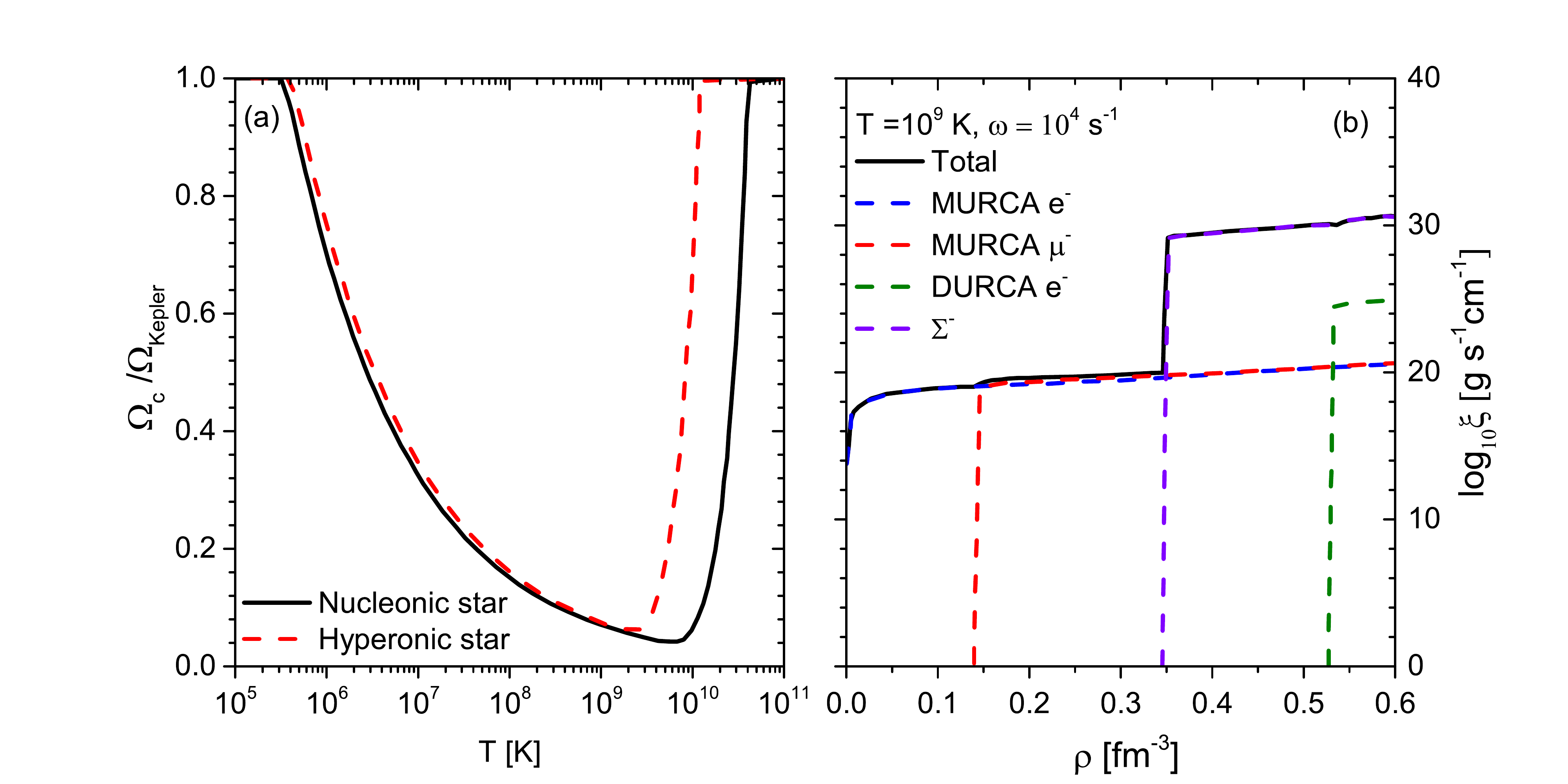}
\vskip-3mm
\caption{
Panel (a):
r-mode instability region for a pure nucleonic and a hyperonic star
with $1.27\ms$.
The frequency of the mode is taken as $\omega=10^4\,$s$^{-1}$.
Panel (b):
Bulk viscosity as a function of the density for $T=10^9\,$K
and $\omega=10^4\,$s$^{-1}$.
Contributions of direct and modified nucleonic Urca processes as well as
from the weak non-leptonic process
$n+n\leftrightarrow p+\Sigma^-$ are included.}
\label{f:RSPAFIG10}
\end{figure}

The time dependence of an r-mode oscillation is given by
$e^{i\omega t-t/\tau(\Omega,T)}$,
being $\omega$ the frequency of the mode and $\tau(\Omega,T)$ an
overall time scale describing both the exponential growth of the mode
due to GW emission as well as its decay due to viscous damping.
It can be written as
\begin{equation}
 \frac{1}{\tau(\Omega,T)} =
 -\frac{1}{\tau_{GW}(\Omega)} + \frac{1}{\tau_\xi(\Omega,T)}
 +\frac{1}{\tau_\eta(\Omega,T)} \:,
\end{equation}
where $\tau_{GW}, \tau_\xi$ and $\tau_\eta$
are the time scales associated to the GW emission
and the bulk and shear viscosity dampings, respectively.
If $\tau_{GW}$ is shorter than both $\tau_\xi$ and $\tau_\eta$,
the mode will exponentially grow,
while in the opposite case it will be quickly damped away.
For each star at a given temperature $T$ one can define a critical angular
velocity $\Omega_c$ as the smallest root of the equation
\begin{equation}
 \frac{1}{\tau(\Omega_c,T)}=0 \:.
\end{equation}
This equation defines the boundary of the so-called r-mode instability region.
A star will be stable against the r-mode instability
if its angular velocity is smaller than its corresponding $\Omega_c$.
On the contrary, a star with $\Omega > \Omega_c$ will develop an instability
that causes a rapid loss of angular momentum through gravitational radiation
until its angular velocity falls below the critical value.
As an example, in panel (a) of Fig.~\ref{f:RSPAFIG10} is presented
the r-mode instability region
for a pure nucleonic (black solid line)
and a hyperonic (red dashed line) star with $1.27\ms$.
The contributions to the bulk viscosity from
direct and modified nucleonic Urca processes
as well as from the weak non-leptonic process
$n+n\leftrightarrow p+\Sigma^-$
included in the calculation are shown in panel (b) of the figure.
Clearly the r-mode instability is smaller for the hyperonic star,
because of the increase of the bulk viscosity due to the presence of hyperons,
which makes the damping of the mode more efficient.

\section{Constraints on the EoS}
\label{s:constr}

\subsection{Nuclear experiment constraints of the EoS}

It is well known that around saturation density $\rho_0$
and isospin asymmetry
$\beta=(\rho_n-\rho_p)/(\rho_n+\rho_p)=0$,
the nuclear EoS can be characterized
by a set of a few isoscalar ($E_0, K_0, Q_0$)
and isovector ($S, L, K_\text{sym},Q_\text{sym}$) parameters.
$E_0$ is the energy per particle of SNM at $\rho_0$,
$K_0$ the incompressibility parameter,
$Q_0$ the so-called skewness,
$S_0$ the value of the nuclear symmetry energy at $\rho_0$,
$L$ the slope of the symmetry energy,
$K_\text{sym}$ the symmetry incompressibility
and $Q_\text{sym}$ the third derivative of the symmetry energy
with respect to the density.
These parameters can be constrained by nuclear experiments and are related to
the coefficients of a Taylor expansion of the energy per particle
of asymmetric nuclear matter in density and isospin asymmetry,
\begin{equation}
 \frac{E}{A}(\rho,\beta) = E_0 + \frac{1}{2}K_0x^2 + \frac{1}{6}Q_0x^3
 + \left(S_0+Lx+\frac{1}{2}K_\text{sym}x^2
 + \frac{1}{6}Q_\text{sym}x^3\right)\beta^2
 + \ldots \:,
\label{ec:expansion}
\end{equation}
where $x=(\rho-\rho_0)/3\rho_0$.

\begin{figure}[t]
\centering
\vskip-3mm
\includegraphics[scale=0.7]{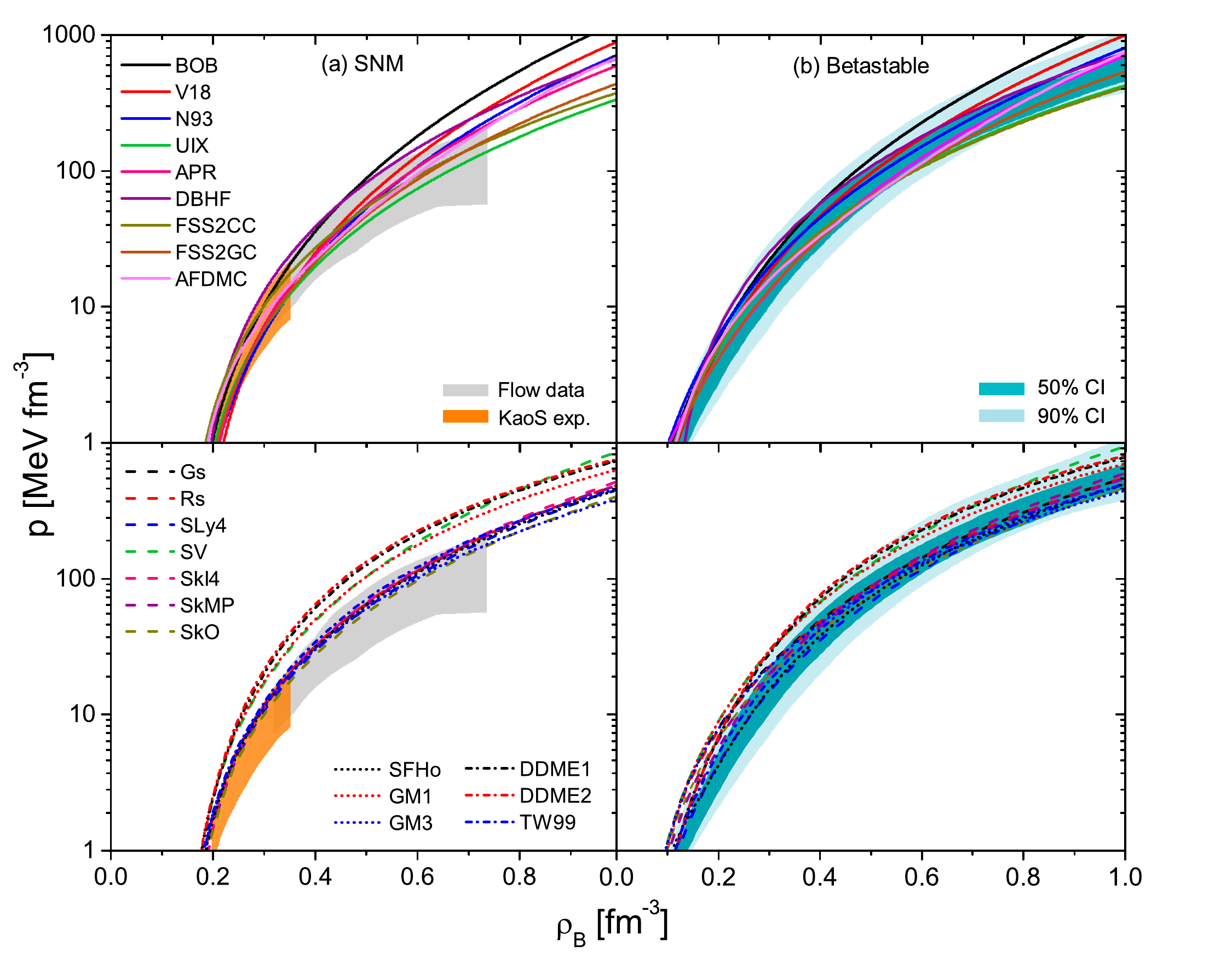}
\vskip-3mm
\caption{
The pressure as a function of the baryon density for symmetric (left panels)
and beta-stable matter (right panels).
Results for microscopic (phenomenological) models are shown
in the upper (lower) panels.
Constraints from HIC data of the KaoS experiment (orange bands)
and flow data (gray bands) are shown together
with the limits deduced by the GW170817 event
(blue bands in the right panels).}
\label{fig:EOS_data}
\end{figure}

\subsubsection{Constraints of the isoscalar parameters}

Measurements of density distributions \cite{Vries87}
and nuclear masses \cite{Audi03}
yield $\rho_0=0.15-0.16\fm3$ and $E_0=-16\pm 1\,$MeV, respectively.
Although its extraction is complicated and not unambiguous,
the value of $K_0$ can be determined experimentally
from the analysis of isoscalar giant monopole resonances in heavy nuclei.
Results of Ref.~\cite{Colo04}, for instance, suggest $K_0=240\pm 10\,$MeV,
whereas in Ref.~\cite{Piekarewicz04} a value of $K=248 \pm 8\,$MeV was reported,
and Ref.~\cite{Blaizot80} gives $K=210\pm 30\,$MeV.
Recently, in Ref.~\cite{Khan12} it was shown that
the third derivative $M$ of the energy density of SNM is constrained
by giant monopole resonance measurements not at saturation density
but rather around what has been called crossing density
$\rho\approx 0.11\fm3$.
The authors of this work found $M=1100\pm 70\,$MeV, which,
once extrapolated to $\rho_0$, implies $K_0=230\pm 40\,$MeV.
We note that the value of the skewness parameter $Q_0$
is more uncertain and not very well constrained yet,
being its estimate in the range $-500\leq Q_0\leq 300\,$MeV.

A rather ``soft'' EoS,
i.e.~a lower value of $K_0$ \cite{Fuchs01},
would be pointed out by HIC experiments.
Constraints on the nuclear EoS for SNM from HIC data
of the KaoS experiment \cite{Miskowiec94} and flow data \cite{Danielewicz02}
are shown in the left panels of Fig.~\ref{fig:EOS_data},
together with the predictions of some microscopic
(solid lines in the upper panels) and phenomenological
(broken lines in the lower panels) EoSs considered.
Any reliable model for the EoS should pass through the region
defined by the experimental data.
We notice that most of the EoSs considered are in general compatible
with these data,
with some exception in the case of the phenomenological ones,
which are too repulsive and some of the microscopic ones (BOB, DBHF, V18),
which are only marginally compatible.
We should point our that the constraints inferred from HICs
are however model dependent,
since the analysis of the measured data requires the use of transport models.
For completeness, in the right panels of the figure
we show the astrophysical constraints (see next section)
on the EoS for beta-stable matter imposed by the GW170817 event \cite{Abbott17}.

\begin{figure}[t]
\centering
\vskip-3mm
\includegraphics[scale=0.5]{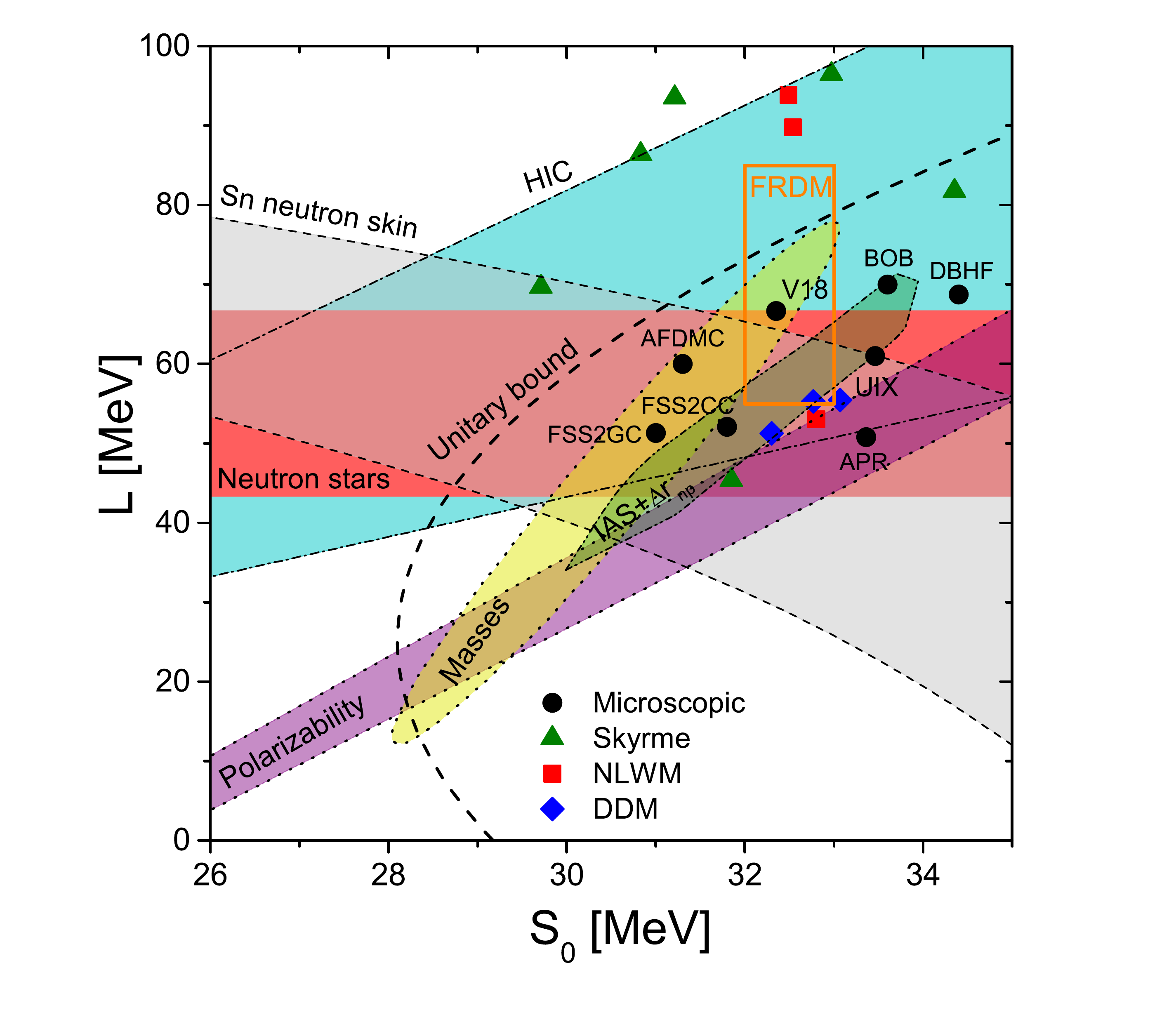}
\vskip-9mm
\caption{
Slope of the symmetry energy $L$ as a function of the symmetry energy
at saturation $S_0$.
Full symbols show the predictions from microscopic approaches (black circles),
the Skyrme EoS (green triangles) and the NLWM (red squares)
and DDM models (blue diamonds).
The shaded areas represent experimental bands
whereas the dashed line shows the unitary limit constraint
determined in Ref.~\cite{Tews17}.
}
\label{f:LvsS}
\end{figure}

\subsubsection{Constraints of the isovector parameters}

The isovector parameters of the nuclear EoS can be constrained experimentally
from isospin diffusion measurements \cite{Chen05},
the analysis of giant \cite{Garg07} and
pygmy \cite{Klimkiewicz07,ACarbone10} resonances,
isoscaling \cite{Shetty07},
isobaric analog states \cite{Danielewicz09},
pion \cite{BALi05} and kaon \cite{Fuchs06} production in HICs
or measurements of the neutron skin thickness in heavy nuclei
\cite{Brown00,Typel01,Horowitz01,Brown07,Centelles09,Warda10,Rocamaza11,Reed21}.
Astrophysical observations can also be used to constrain these parameters.
It has been shown, for instance,
that the slope parameter $L$ of the symmetry energy
is correlated with the radius \cite{Hu20}
and the tidal deformability \cite{Zhao18,Tsang19} of a $1.4\ms$ NS,
and that precise and independent measurements of the radius
and the tidal deformability from multiple observables of NSs
can potentially pin down the correlation between $K_\text{sym}$ and $L$ and
thus the high-density behaviour of the nuclear symmetry energy \cite{Li20}.
Nevertheless, whereas $S_0$ is more or less well established
($\approx 30\,$MeV),
the values of $L$,
and specially those of $K_\text{sym}$ and $Q_\text{sym}$,
are still uncertain and poorly constrained.
Combining different data, for instance,
the authors of Ref.~\cite{Lattimer13} give
$29.0 < S_0 < 32.7\,$MeV and $40.5<L<61.9\,$MeV,
while in Ref.~\cite{Danielewicz14}
it was suggested $30.2<S_0<33.7\,$MeV and $35<L<70\,$MeV.
Why the isovector part of the nuclear EoS is so uncertain
is still an open question whose answer is related to our limited knowledge
of the nuclear force and in particular to its spin and isospin dependence.

We show in Fig.~\ref{f:LvsS} the slope of the symmetry energy $L$
as a function of the symmetry energy at saturation $S_0$.
The different symbols show the predictions from
microscopic approaches (black circles),
the Skyrme EoS (green triangles)
and the NLWM (red squares)
and DDM models (blue diamonds).
Currently available experimental data from
neutron-skin thickness in Sn isotopes \cite{Chen10},
isospin diffusion in HICs \cite{Tsang09},
electric dipole polarizability \cite{RocaMaza15},
isobaric-analog-state (IAS) phenomenology
combined with the skin-width data \cite{Danielewicz14}
or the Bayesian analysis of mass and radius measurements in NSs \cite{Steiner13}
are also indicated in the figure.
The dashed line shows the unitary limit constraint determined
in Ref.~\cite{Tews17},
which combined with the experimental and observational data
indicates that only values of ($S_0,L$) to the right of this line are permitted.
As can be seen, all the microscopic and most of the phenomenological models
considered fulfill all these constraints.

\subsection{Astrophysical constraints}

The observation of NSs provides the main astrophysical constraints
on the nuclear EoS.
An enormous amount of data on different NS observables has been
collected after fifty years of observations.
From these observables it is possible to infer valuable information
on the internal structure of these objects
and therefore also on the nuclear EoS,
which is the only ingredient needed to solve the structure equations of NSs.
In the following we shortly review some of these observables.

\subsubsection{Masses}

NS masses can be inferred directly from observations of binary systems
and likely also from supernova explosions.
In any binary system,
there exist five orbital parameters
(usually known as Keplerian parameters),
which can be precisely measured.
These are:
the eccentricity of the orbit $e$,
the orbital period $P_b$,
the projection of the pulsar's semi-major axis on the line of sight
$x\equiv a_1 \mbox{sin}\, i/c$, where $i$ is the inclination of the orbit,
and the time $T_0$ and longitude $\omega_0$ of the periastron.
With the help of Kepler's third law,
these parameters can be related to the masses of the NS ($M_p$)
and its companion ($M_c$) through the so-called mass function
\begin{equation}
  f(M_p,M_c,i) = \frac{(M_c\,\mbox{sin}\,i)^3}{(M_p+M_c)^2}
  = \frac{P_bv_1^3}{2\pi G} \:,
\label{e:massfun}
\end{equation}
where $v_1=2\pi a_1\mbox{sin}\,i/P_b$
is the projection of the orbital velocity of the NS along the line of sight.
One cannot proceed further than Eq.~(\ref{e:massfun})
if only one mass function can be measured for a binary system,
unless additional assumptions are made.
Deviations from the Keplerian orbit due to general relativity effects
can be detected fortunately.
These relativistic corrections are parametrized in terms of one or more
post-Keplerian parameters.
The most significant ones are:
the advance of the periastron of the orbit ($\dot \omega$),
the combined effect of variations in the transverse Doppler shift
and gravitational redshift around an elliptical orbit ($\gamma$),
the orbital decay due to the emission of quadrupole gravitational radiation
($\dot P_b$),
and the range ($r$) and shape ($s$) parameters that characterize
the Shapiro time delay of the pulsar signal
as it propagates through the gravitational field of its companion.
These post-Keplerian parameters can be written in terms of measured quantities
and the masses of the star and its companion as \cite{Taylor92}:
\bal
 \dot\omega &= 3n^{5/3}T_\odot^{2/3}\frac{(M_p+M_c)^{2/3}}{1-e} \:,
\label{ec:ome}
\\
 \gamma &= eT_\odot^{2/3}\frac{M_c(M_p+2M_c)}{n^{1/3}(M_p+M_c)^{4/3}} \:,
\label{ec:gam}
\\
 \dot P_b &= -\frac{192\pi}{5}(nT_\odot)^{5/3}\left(1+\frac{73}{24}e^2+\frac{37}{96}e^4\right)
\frac{1}{(1-e^2)^{7/2}}\frac{M_pM_c }{(M_p+M_c)^{1/3}} \:,
\label{ec:pbdot}
\\
 r &= T_\odot M_c \:,
\label{ec:r}
\\
 s &= x\frac{n^{2/3}}{T_\odot^{1/3}}\frac{(M_p+M_c)^{2/3}}{M_c} \:,
\label{ec:s}
\eal
where $n=2\pi/P_b$ is the orbital angular frequency and
$T_\odot\equiv GM_\odot=4.925490947\times10^{-6}\,$s.
The measurement of any two of these post-Keplerian parameters
together with the mass function $f$ is sufficient to determine uniquely
the masses of the two components of the system.
An example of a high-precision mass measurement is that of the famous
Hulse--Taylor binary pulsar \cite{Hulse75} with measured masses
$M_p=1.4408\pm 0.0003\ms$ and $M_c=1.3873 \pm 0.0003\ms$.
Other examples are those of the millisecond pulsars
PSR J1614-2230 \cite{Demorest10},
PSR J0348+0432 \cite{Antoniadis13}
and the most recently observed PSR J0740+6620 \cite{Cromartie20}
with masses $M_p=1.928\pm 0.017\ms$,
$M_p=2.01\pm 0.04\ms$ and
$M_p=2.14^{+0.10}_{-0.09}\ms$, respectively.
These are binary systems formed by a NS and a white dwarf.
The measurement of these unusually high NS masses constitutes nowadays
one of the most stringent astrophysical constraints on the nuclear EoS.

\subsubsection{Radii}

NS radii are very difficult to measure,
mainly because NSs are very small objects and are very far away from us
(e.g.~the closest NS is probably the object RX J1856.5-3754
in the constellation Corona Australis,
which is about 400 light-years from earth).
Direct measurements of radii do not exist.
However, a possible way to determine them is to use the thermal emission
of low-mass x-ray binaries.
The observed x-ray flux $F$ and temperature $T$,
assumed to be originated from a uniform black body,
together with a determination of the distance $D$ of the star
can be used to obtained the radius through the following implicit relation
\begin{equation}
 R = \sqrt{\frac{FD^2}{\sigma T^4}\left(1-\frac{2GM}{R}\right)} \:,
\label{ec:radinf}
\end{equation}
where $\sigma$ is the Stefan--Boltzmann constant and $M$ the mass of the star.
The major uncertainties in the measurement of the radius through
Eq.~(\ref{ec:radinf}) originate from the determination of the temperature,
which requires the assumption of an atmospheric model,
and the estimation of the distance of the star.
The analysis of present observations from quiescent low-mass x-ray binaries
is still controversial.
Whereas the analysis of Ref.~\cite{Steiner13,Lattimer14}
indicates NS radii in the range of $10.4-12.9$ km,
that of Ref.~\cite{Guillot13,Guillot14} points towards smaller radii
of $\sim 10$ km or less.

We would like to note here that the simultaneous measurement
of both mass and radius of the same NS would provide the most definite
observational constraint on the nuclear EoS.
Very recently the NICER mission has reported a Bayesian parameter estimation
of the mass and equatorial radius of the millisecond pulsar PSR J0030+0451
\cite{Riley19,Miller19}.
The values inferred from the analysis of the collected data are, respectively,
$1.34^{+0.15}_{-0.16}\ms$ and $12.71^{+1.14}_{-1.19}\,$km.
We want to stress here that the measurement performed by the NICER mission
does not make use of Eq.~(\ref{ec:radinf})
and constitutes the first model-independent one,
since only the geometry of the hot spots of the NS
and general relativity effects enter in the determination of the star radius.

\subsubsection{Gravitational waves}

On August 17 2017, a GW signal from a binary NS merger was detected
for the first time by the Advanced LIGO and Advanced VIRGO collaborations
\cite{Abbott17}, inaugurating with this event (known as GW170817)
a new era in the observation of NSs.
GWs originated during the coalescence of two NSs or a black hole and a NS
constitute now a new and valuable source of information
on the EoS and internal structure of NSs.
In particular, the so-called tidal deformability $\lambda$,
or equivalently the tidal Love number
$k_2$ of a NS \cite{Hinderer08,Hinderer09,Hinderer10},
can provide priceless information and constraints on the related EoS,
because it depends strongly on the compactness $\beta\equiv M/R$ of the object.
The Love number, defined as
\begin{equation}
 k_2 = \frac{3}{2}\frac{\lambda}{R^5} = \frac{3}{2} \beta^5 \la
 = \frac{8}{5}\frac{\beta^5 z}{F} 
\label{e:l}
\end{equation}
with $\la = \lambda/M^5$ and
\bal\nonumber
 z &\equiv (1-2\beta)^2 [2-y_R+2\beta(y_R-1)] \:,
\\\nonumber
 F &\equiv 6\beta(2-y_R) + 6\beta^2(5y_R-8) + 4\beta^3(13-11y_R)
  + 4\beta^4(3y_R-2) + 8\beta^5(1+y_R) + 3z\ln(1-2\beta) \:,
\eal
can be obtained by solving the TOV equations
\bal
 \frac{dp}{dr} &= -\frac{\left(\eps+p\right)\left(m+4\pi r^3p\right)}
{r\left(r-2m\right)} \:,
\label{e:tov1}
\\
 \frac{dm}{dr} &= 4\pi r^2 \eps \:,
\label{e:tov2}
\eal
together with the additional equation \cite{Lattimer07}
\begin{equation}
 \frac{dy}{dr} = -\frac{y^2}{r} - \frac{y-6}{r-2m} - rQ \:,
\label{e:tov3}
\end{equation}
where
\begin{eqnarray}
  Q = 4\pi \frac{(5-y)\eps+(9+y)p+(\eps+p)(dp/d\eps)^{-1} }{1-2M/r}
  - \Bigg[ \frac{2(m+4\pi r^3 p)}{r(r-2m)} \Bigg]^2 \:.
\end{eqnarray}
We note that $y_R=y(R)$ is appearing in the definitions of $z$ and $F$
given above.
Equations~(\ref{e:tov1}), (\ref{e:tov2}) and (\ref{e:tov3})
are solved using the EoS $p(\eps)$ as the only input needed,
and imposing the following set of boundary conditions
\begin{equation}
 [p,m,y](r=0) = [p_c,0,2] \:.
\end{equation}

The average tidal deformability of an asymmetric binary NS system
with mass asymmetry $q=M_2/M_1$ and known chirp mass
$M_c+(M_1M_2)^{3/5}/(M_1+M_2)^{1/5}$,
characterizing the GW signal waveform, is given by
\begin{equation}
 \tilde{\la} = \frac{16}{13}
 \frac{(1+12q)\la_1 + (q+12)\la_2}{(1+q)^5} \:.
\label{e:lq}
\end{equation}

A value of $M_c=1.186{+0.001\atop-0.001}\ms$ was obtained
from the analysis of the GW170817 event \cite{Abbott17,Abbott18,Abbott19},
corresponding to $M_1=M_2=1.365\ms$ for a symmetric binary system,
$q=0.73-1$ and $\tilde{\la}<730$
from the phase-shift analysis of the observed signal.
We stress that the GW170817 observation put a strong
constraint on the radius of a $1.4\ms$ NS.
Requiring both NSs to have the same EoS,
this leads to the constraints $70 < \la_{1.4} < 580$ and
$10.5 < R_{1.4} < 13.3\,$km \cite{Abbott18}
for this radius.
In Ref.~\cite{Annala18} a general polytropic parametrization of the EoS
compatible with perturbative QCD at very high density was used,
and the constraint $\la_{1.4}<800$ yielded a similar upper limit
$R_{1.4}<13.4\,$km.

The high luminosity of the kilonova AT2017gfo following the NS merger event
imposes also a lower limit on the average tidal deformability,
$\tilde{\la}>400$,
which was deduced in order to justify the amount of ejected material
being heavier than $0.05\,\ms$.
This lower limit,
which was used in Refs.~\cite{Most18,Lim18,Malik18,Burgio18}
in order to constrain the EoS,
could indicate that $R_{1.4}\gtrsim 12\,$km.
However, this constraint of $\tilde{\la}$ has to be taken with great care and,
in fact, it has been recently revised to
$\tilde{\la} \geq 300$ \cite{Radice19},
but considered of limited significance in Ref.~\cite{Kiuchi19}.
We thus notice that the determination of the average tidal deformability
of the binary NS system GW170817
has imposed constraints on the NS radii,
to lie between about 12 and 13 kilometers \cite{Wei19b}.
This is complementary to the mass-radius measurement by NICER mentioned before,
and contributes to selecting the suitable EoS.

\begin{figure}[t]
\centering
\vskip-12mm
\includegraphics[scale=0.5]{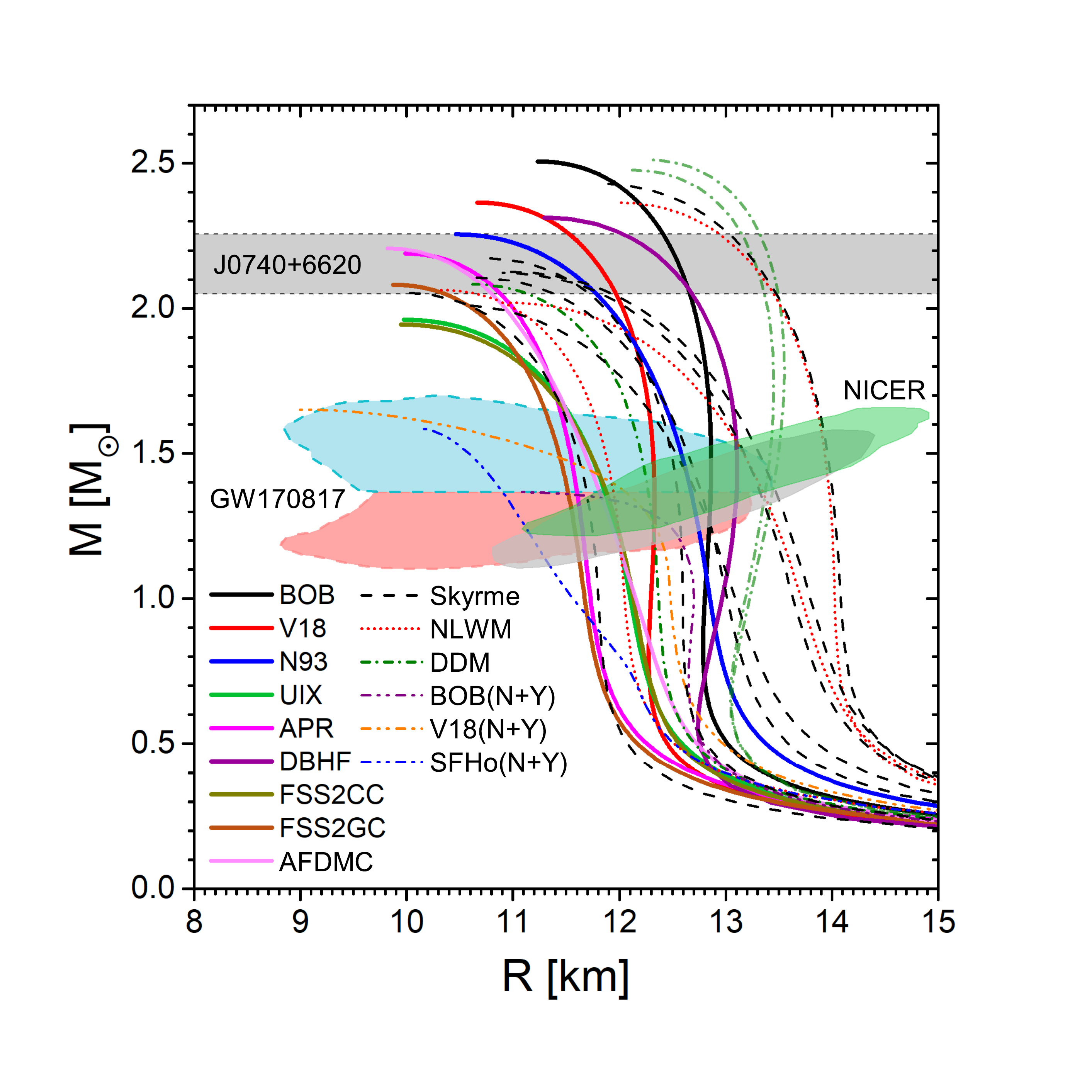}
\vskip-14mm
\caption{
Mass-radius relations obtained with different microscopic (solid lines)
and phenomenological (broken lines) EoSs.
The mass of the most heavy pulsar PSR J0740+6620 \cite{Cromartie20}
observed until now is also shown,
together with the constraints from the GW170817 event \cite{Abbott17}
and the mass-radius constraints of the NICER mission \cite{Riley19,Miller19}.}
\label{f:MR}
\end{figure}

Before discussing other astrophysical constraints,
we show in Fig.~\ref{f:MR} the mass-radius relations
obtained with different microscopic (solid lines)
and phenomenological (broken lines) EoSs.
Most of the models considered,
except the soft microscopic UIX and FSS2GC,
predict values of the maximum mass larger than $2\ms$,
therefore being compatible with current observational limits
\cite{Demorest10,Antoniadis13,Cromartie20}.
The mass of the most heavy pulsar PSR J0740+6620 \cite{Cromartie20}
observed until now is also shown in the figure,
together with the constraints of the NICER mission \cite{Riley19,Miller19}
mentioned above and those from the GW170817 event \cite{Abbott17}.
Some recent theoretical analyses of this event
indicate an upper limit of the maximum mass
of $\sim2.2-2.3\ms$ \cite{Shibata17,Margalit17,Rezzolla18},
with which several of the microscopic and phenomenological EoSs considered
would be compatible as well.

\subsubsection{Rotational periods}

The majority of the known NSs are observed as radio pulsars
which exhibit in most of the cases very stable rotational periods.
Highly accurate pulsar timing has allowed observers to measure the
rotational period $P$ and the first time derivative $\dot P$
and sometimes even the second one $\ddot P$ of many radio pulsars
(see e.g.~\cite{Lyne98,Lorimer01,Livingstone06} and references therein).
Two different classes of pulsars have been identified thanks to the observation
of their rotational periods:
the {\em normal} pulsars with rotational periods of the order of $\sim 1$s,
and the so-called {\em millisecond} pulsars with rotational periods
three orders of magnitude smaller.
The first millisecond pulsar was discovered in 1982 with the help of the
Arecibo radio telescope and since then more than 200 pulsars of this class
have been observed.
Nowadays, the fastest pulsar known until now,
with a rotational period of 1.39595482 ms,
is the object named PSR J1748-2446ad which was discovered in 2005
in the globular cluster Terzan 5 in the Sagittarius constellation.

\subsubsection{Surface temperatures}

The effective surface temperatures of NSs can be determined
from the detection of thermal photons from the stellar surface in isolated NSs
(see Sec.~\ref{s:lum})
by fitting the observed spectra to blackbody ones.
One should keep in mind, however, that NSs are not black bodies,
because the hydrogen and helium (or even carbon) in their atmospheres
modify the blackbody spectrum.
Furthermore, the surface emission can be modified by the presence
of strong magnetic fields.
In fact, when realistic atmosphere models are used in the fit
of the measured spectrum,
surface temperatures are reduced.
We note that any uncertainty in the determination of the surface temperature
changes the corresponding luminosity $L$ of the star by a large factor
according to the Stefan--Boltzmann law,
$L=4\pi R^2\sigma T^4$.
Consequently, it is not appropriate to use the surface temperature
when comparing with observational data, but the luminosity instead.

\subsubsection{Gravitational redshift}

A source of very valuable information on NS structure is provided
by the measurement of its gravitational redshift
\begin{equation}
  z = \left(1-\frac{2GM}{R}\right)^{-1/2}-1 \:,
\label{ec:z}
\end{equation}
which allows to constrain the $M/R$ ratio and therefore the nuclear EoS.
The interpretation of measured $\gamma$-ray bursts,
as gravitationally redshifted 511 keV $e^{\pm}$ annihilation lines
from the surface of NSs,
supports a NS redshift range of $0.2 \leq z \leq 0.5$
with the highest concentration in the narrower range
$0.25 \leq z \leq 0.35$ \cite{Liang86}.

\subsubsection{Quasiperiodic oscillations}

The observation and analysis of quasiperiodic x-ray oscillations (QPOs)
\cite{Ingram20} in x-ray binaries can provide very useful information to
understand better the innermost regions of accretion disks as well as to
put stringent constraints on the masses, radii and rotational periods of NSs.
They may also serve as a unique proof of strong-field general relativity.
QPOs measure the difference between the rotational frequency
of the NS and the Keplerian frequency of the innermost stable orbit of
matter elements in the accretion disk formed by the diffused material of
the companion in orbital motion around the star. However, their theoretical
interpretation is not simple and remains still controversial.

\subsubsection{Magnetic fields}

Since the pioneering work of Gold in 1968 \cite{Gold68}
pulsars are generally believed to be rapidly rotating NSs
with strong surface magnetic fields.
The strength of the surface field could be of the order of $10^8-10^9\;$G
in the case of millisecond pulsars,
about $10^{12}\;$G in normal pulsars,
or even $10^{14}-10^{15}\;$G in the so-called {\em magnetars}.
The observation of the pulsar rotational period $P$
and its first derivative $\dot P$
can be used to estimate the strength of the surface magnetic field of a pulsar,
e.g.~within the so-called magnetic dipole model \cite{Shapiro08}
through the relation
\begin{equation}
 P{\dot P} = \frac{2\pi^2 B^2 R^6 \sin^2\al}{3I} \:,
\label{e:ppdot}
\end{equation}
where $B$ is the strength of the magnetic field at the pole,
$R$ is the radius of the NS,
$\al$ is the angle between the magnetic and the rotation axis,
and $I$ is the moment of inertia of the star.

\subsubsection{Glitches}

Pulsars are observed to spin down gradually due to the transfer
of their rotational energy to the emitted electromagnetic radiation.
However, sudden jumps $\Delta\Omega$ of the rotational frequency $\Omega$
have been observed in several pulsars followed by a slow partial relaxation
that can last days, months or years.
These jumps, mainly observed in relatively young radio pulsars,
are known as {\em glitches}.
The relative increase of the rotational frequency $\Delta\Omega/\Omega$
varies from $\sim 10^{-10}$ to $\sim 5\times 10^{-6}$.
The first glitches were detected in the Crab and Vela pulsars
\cite{Boyton69,Radhakrishnan69,Reichley69}.
Nowadays we know more than 520 glitches in more than 180 pulsars.

\subsubsection{Timing noise}

The rotational stability is one of the most remarkable features of radio pulsars.
Some of them, however, show slow irregular or quasi-regular
variations of their pulses over time scales of months, years and longer
times which have been called {\em pulsar timing noise}.
These timing imperfections appear as random walks in the pulsar rotation
(with relative variations of the rotational period $\leq 10^{-10}-10^{-8}$),
the spin-down rate or the pulse phase,
which could stem from changes in the internal structure and/or
the magnetosphere of NSs,
although their nature is still uncertain and many hypotheses have been made,
see e.g.~Ref.~\cite{Cordes81}.


%
%
%
\section{Equation of state and neutron star cooling}
\label{s:cool}

We discuss in this chapter another physical phenomenon of NSs
that is strongly related to the nuclear interactions and the EoS,
namely the cooling properties of (isolated) NSs.
Nuclear forces determine via the EoS not only the NS structure,
but also the composition of NS cores,
as well as reaction matrix elements, effective masses,
and superfluid gaps of baryons,
which are crucial for the computation of
heat capacity and neutrino emission rates in NS cores.
These two quantities govern the cooling of middle-aged ($\lesssim 10^5\,$yr)
isolated NSs,
which can be confronted with astrophysical data on
luminosity and surface temperatures for NSs of known or estimated age.

The problem of NS cooling is a very rich field of physics that has been
covered in several specialized reviews
\cite{Yakovlev99,Yakovlev01,Yakovlev04,Page06,Page06b,Page09,Tsuruta09,
Potekhin15,Geppert17,Schmitt18,Sedrakian19}
and it has become impossible to give a complete overview.
We try to single out here only the essential aspects
related to the nuclear interactions and EoS.
In particular, we focus on the cooling of isolated NSs
and disregard the even more challenging problem of the cooling of
accreting NSs in X-ray transients in quiescence (qXRT)
\cite{Yakovlev04,Beznogov15a}.
NSs may also be reheated by various other mechanisms,
e.g.~\cite{Gonzalez10}.
Furthermore, only purely nucleonic NSs are considered,
disregarding scenarios comprising hyperons and quark matter,
and also the important effects of magnetic fields are not considered
in this brief chapter.

\subsection{Thermal evolution of a isolated neutron star}

We start by sketching the thermal evolution of a isolated NS
(see \cite{Yakovlev01,Yakovlev04,Page09} for a detailed review).
A proto-NS is formed in a supernova event with a high temperature
$T\sim {\cal O}(10^{11}\,$K).
The proto-NS becomes a NS when it gets transparent to the
neutrinos that are formed in its interior.
During $\sim 100$ years,
the crust stays hot due to its low thermal conductivity,
while the core cools by emission of neutrinos.
Therefore, the core and the crust cool independently
and the evolution of the surface temperature reflects the
thermal state of the crust and is sensitive to its physical properties
(see, e.g.~\cite{Fortin10,Carreau19,Fantina20,Potekhin19}
and references therein).
Then the thermal evolutions of core and crust couple.
The cooling wave from the core reaches the surface and the whole NS
cools by the emission of neutrinos, mainly from the core.
This is the neutrino cooling stage.
The evolution of the surface temperature depends mainly on the
physical properties of the core.
As the temperature in the interior of the NS continues to decrease,
the neutrino luminosity becomes comparable to the photon luminosity.
The NS then enters the photon cooling stage and the
evolution of the internal temperature is governed by the emission of
photons from the surface and is sensitive to the properties of the outer
parts of the star.

Currently, all isolated NSs for which the surface temperature has been measured
thanks to X-ray observations are at least $\sim 300$ years old
(Recently the ALAM radio telescope discovered a red blob
in the remnant of SN1987 which has a high possibility to be a NS.
If correct, this NS would be the youngest star observed with 33~yr
\cite{Beznogov16,Page20}.)
and they are all in the neutrino- and photon-cooling stages
(see, e.g.~Table~1 in \cite{Potekhin20}).
Therefore they have an isothermal interior,
i.e.~the redshifted internal temperature
$T_i(t)=T(t,r) e^\phi(r)$
is constant throughout the interior for densities larger than
$\rho\approx 10^{10}\gc3$.
Here $T(r,t)$ is the local temperature and $\phi(r)$ the metric function.

The cooling evolution of a spherically symmetric NS can be characterized
by two equations:
the equation of thermal balance and the equation of thermal transport.
For a spherical shell in the star,
the change of thermal energy is caused by the neutrino emission,
the heat flux passing through the surface,
and the possible heating sources,
for example, by converting magnetic or rotational energy into thermal energy
\cite{Page06}.
This is expressed by the thermal balance equation
\be
 \frac{\sqrt{1-2Gm/r}}{4\pi r^2 e^{2\phi}}
 \frac{\partial}{\partial r}(e^{2\phi}L) =
 Q_h - Q_\nu - \frac{C_v}{e^\phi}\frac{\partial T}{\partial t} \:,
\label{e:tb}
\ee
where
$Q_h$ stands for all relevant heating sources,
$Q_\nu$ is the neutrino emissivity,
$C_v$ the specific heat capacity, and
$m$ is the gravitational mass enclosed inside the sphere of radius $r$.
The local luminosity $L$ is defined as the non-neutrino heat flux
passing through the surface of the spherical shell.
Since the heat flux is transported through thermal conduction,
one can write the equation of thermal transport as
\be
 \frac{L}{4\pi r^2} =
 -\kappa \sqrt{1-\frac{2Gm}{r}} e^{-\phi}
 \frac{\partial}{\partial r}(e^\phi T) \:,
\label{e:tc}
\ee
where $\kappa$ is the thermal conductivity.
Eqs.~(\ref{e:tb}) and (\ref{e:tc}) are partial differential equations
for luminosity $L(r,t)$ and temperature $T(r,t)$,
and all quantities $Q_h,Q_\nu,C_v,\kappa$
depend as well on the radial coordinate~$r$ and time~$t$.
On the stellar surface,
luminosity and temperature can be connected by
\be
 L_\gamma(t) = L(R,t) = \sigma_\text{SB} 4\pi R^2 T_s^4(t) \:,
\ee
where $L_\gamma$ and $T_s$ are photon luminosity
and effective surface temperature, respectively, and
$\sigma_\text{SB}$ is the Stefan-Boltzmann constant.

Once the interior of the star has become isothermal ($T_i$)
after about $10^2$ years,
isolated by a heat-blanketing atmosphere
with a effective surface temperature $T_s$,
the above equations simplify to the heat equation
\be
 C(T_i)\frac{d T_i}{d t} =
 - L_\nu^\infty(T_i) - \lgi(T_s) \:,
\label{e:heatINS}
\ee
where the relation $T_s(T_i)$ is determined by the model
of the heat-blanketing envelope,
see Sec.~\ref{s:envel}.
$C(T_i)$ is the total specific heat integrated over the whole star
and the quantities $L_\nu^\infty$, $\lgi$
are respectively the redshifted total neutrino luminosity
and the surface photon luminosity detected by a distant observer.

Observable quantities for astrophysical observation at large distance
are apparent radius $R_\infty$,
effective surface temperature $T_s^\infty$,
and apparent photon and neutrino luminosities
$\lgi$, $L_\nu^\infty$
\cite{Misner77},
\bal
 R_\infty &= R/\!\sqrt{1-2GM/R} \:,
\\
 T_s^\infty &= T_s \sqrt{1-2GM/R} \:, 
\\
 \lgi &=
 L_\gamma(1-2GM/R) =
 \sigma_\text{SB} 4\pi R^2 T_s^4 \big( 1-2GM/R \big) =
 \sigma_\text{SB} 4\pi R_\infty^2 (T_s^\infty)^4 \:,
\\
 L_\nu^\infty &=
 4\pi\int_0^R\!\!\!dr r^2
 {e^{2\phi(r)} \over \sqrt{1-2Gm(r)/r}}\, Q_\nu(r) \:,
\eal 
with $M=m(R)$ the gravitational mass of the star of radius $R$.
Hence, by solving Eqs.~(\ref{e:tb},\ref{e:tc}) or Eq.~(\ref{e:heatINS}),
one can obtain the dependence of the apparent photon luminosity
$\lgi$ (or of $T_s^\infty$) on the stellar age $t$.
This is the main goal of the cooling theory.

\subsection{Models of the crust}

The crust of an isolated NS born in a core-collapse supernova explosion
is formed during the cooling of the very hot and fully fluid proto-NS
\cite{Haensel90,Zdunik11}.
In the initial state,
the composition of the outer layer corresponds to nuclear equilibrium,
because $T>10^{10}\,$K and therefore nuclear reactions are sufficiently rapid.
The hot plasma crystallizes while cooling,
and a standard assumption is that during the process of cooling and
crystallization the plasma keeps the nuclear equilibrium.
Consequently, when matter becomes strongly degenerate,
the structure and EoS of the crust can be well approximated by
cold catalyzed matter,
the ground state of the matter at $T=0$.
The structure of this crust may be calculated
to various degrees of sophistication,
as reviewed in more detail in Sec.~\ref{s:crust}.
In practice, the contribution of the crust to the neutrino cooling
is usually negligible.

\subsection{Heat-blanketing envelope}
\label{s:envel}

Within a few hundred years the redshifted temperature inside a newly born NS
for densities larger than
$\rho_b \approx 10^{10}\gc3$
becomes uniform due to the high thermal conductivity.
However, in the NS atmosphere the heat transport is dominated by the photons
and in between there exists a thin layer,
which has a low thermal conductivity,
since the electrons are not highly degenerate
and the large density strongly prevents photon transport.
This results in high temperature gradients in the envelope,
that is a few hundred meters thick.

Therefore, a variety of models
are devoted solely to the precise modeling of the heat-blanketing envelope
\cite{Gudmundsson83,Potekhin97,Potekhin03,Shternin07},
in the plane-parallel and stationary approximation,
resulting in a relation between the surface temperature $T_s$
and the internal temperature $T_i$ at $\rho_b$.
Some models consider various compositions for the envelope and, in particular,
different abundances of light elements such as hydrogen, helium, carbon,
resulting from the accretion of matter \cite{Potekhin97}
or binary mixtures of these \cite{Beznogov16}.
The higher the abundances of light elements,
the higher the surface temperature for a given $T_i$.
As these abundances are unknown for a given NS,
in the simulations one usually considers two limiting cases corresponding
to the absence of light elements (non-accreted envelope)
and a maximum amount of them (fully-accreted envelope),
in order to estimate the possible range of predictions.

\begin{figure}[t]
\vspace{-42mm}
\centerline{\includegraphics[scale=0.66]{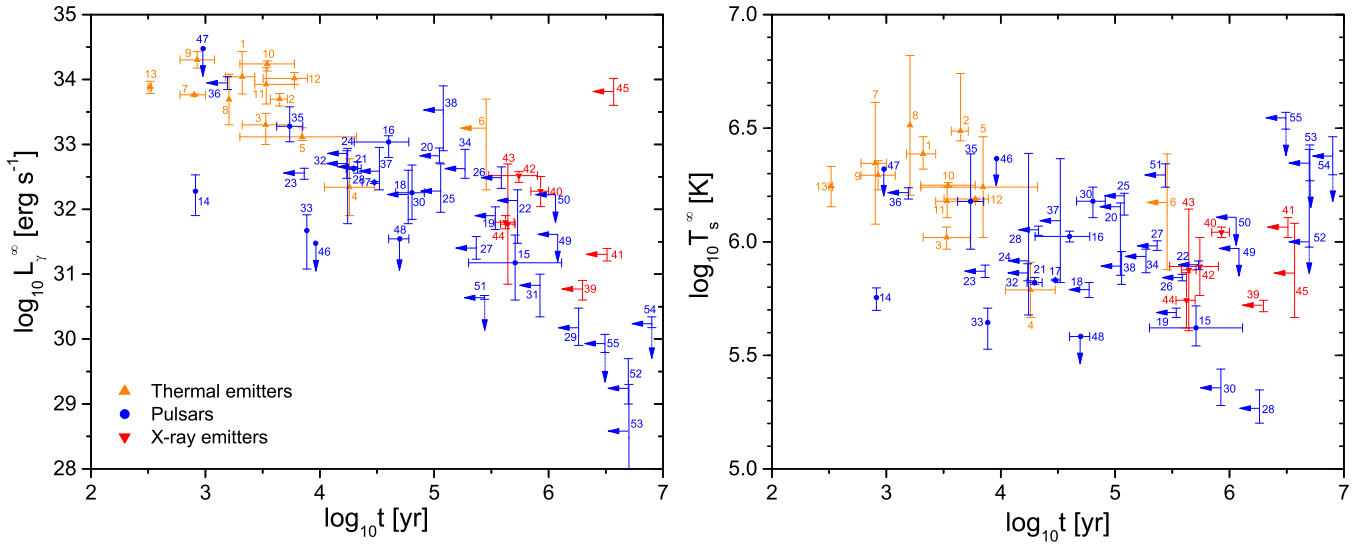}}
\vspace{-30mm}
\caption{
Luminosity (left panel) and temperature (right panel)
vs age of 55 isolated NSs.
Adapted from Ref.~\cite{Potekhin20}.
Numbers refer to Table~1 of that reference.
See also the online database
{\tt http://www.ioffe.ru/astro/NSG/thermal/cooldat.html}
(where object 1 has been withdrawn).
}
\label{f:lt}
\end{figure}

\subsection{Luminosity data}
\label{s:lum}

The observable luminosity of isolated NSs is basically the photon luminosity,
since the neutrino luminosity is far too weak to be detected directly.
However,
due to the unknown composition of the atmosphere just discussed,
the unknown distance, or the interstellar absorption,
it is difficult to obtain accurate results of photon luminosities.
Even so, these data are still essential to
impose constraints on the theoretical study of NS cooling.
Of course, also the ages of neutron stars need to be known.
They can be extracted from the spin period and its time derivative,
which are the characteristic ages
$t \equiv P/2\dot{P}$ \cite{Manchester09},
or from kinetic properties of the stars (proper motion for instance),
which are the kinetic ages.
The kinetic ages are favoured where possible,
and characteristic ages are treated as upper limit in most cases.
Compilations of these data have been given in
\cite{Page04,Yakovlev04,Page09,Tsuruta09,
Vigano13,Beznogov15a,Potekhin15,Potekhin20} 
and are continuously updated.
In Fig.~\ref{f:lt} we reproduce the most recent set
\cite{Potekhin20}
of the bolometric luminosities $\lgi$
and surface temperatures $T_s^\infty$
of 55 NSs.
It has too be stressed
and can (not) be seen in the figure
that weak sources are simply too faint to be
observed and therefore these data are subject to an important and difficult to
quantify selection effect,
which strongly hampers the interpretation of theoretical cooling models
that rely on unbiased data.

\subsection{Microphysical ingredients}
\label{s:micro}

\subsubsection{Nucleon effective masses}

The most prominent input information regarding single-particle propagation
are the in-medium nucleon ($n,p$) effective masses $m^*=k_F/[de(k)/dk]$,
related to the density of states $m^* k_F/\pi^2$.
As most other microphysics input depends on them,
they affect in several instances the computation of NS cooling.
Note that these are already in-medium modified quantities,
whereas other ingredients
(matrix elements, cross sections, ...)
are still often approximated by vacuum results,
as will be discussed in the following.

In principle, they should therefore be computed in a consistent manner
together with the EoS.
However, since other ingredients of the cooling calculations
(matter composition, matrix elements, pairing properties)
are far more uncertain and grossly affect the final results,
this is not always done and bare nucleon masses are simply used,
or different approximate but inconsistent values
are used in the various components of the cooling simulations.
In any case, in several theoretical approaches for the EoS
the effective masses can be calculated straightforwardly
\cite{Lesinski06,Ma04,Baldo14},
even at finite temperature \cite{Donati94,Tan16,Shang20}.
We refer to the specialized review \cite{Baoan18}.

\subsubsection{Specific heat and thermal conductivity}
\label{ss:heat}

The main contribution to the overall specific heat $C(T_i)$
in Eq.~(\ref{e:heatINS}) comes from the NS core,
i.e.~from the neutrons, protons, and electrons.
If the core is non-superfluid,
most of the specific heat comes from the neutrons,
as indicated by the simplest result for a Fermi gas
at density $n$ and `low' temperature $T\ll e_F$,
\be
 C = \frac{\pi^2}{2} \frac{nT}{e_F} \:.
\ee
For the less important crust contributions there are several refinements
\cite{Potekhin15}.
If the neutrons and protons are superfluid,
then below their superfluid critical temperature $T_c$
their values become exponentially suppressed
\cite{Levenfish94,Yakovlev99}
and $C$ is dominated by the electrons \cite{Shternin07}.

The local thermal conductivity $\kappa$ in Eq.~(\ref{e:tc})
is a sum of contributions of the different constituents of matter.
Here we only illustrate qualitatively the dominant neutron contribution
in the core
\cite{Flowers79,Baiko01,Shternin07,Shternin13,Potekhin15,Schmitt18},
neglecting the coupling to protons,
\be
 \kappa \approx \frac1T \frac{5\pi^2}{192} \frac{m^2 n}{{m^*}^4 \sigma} \:,
\ee
where $\sigma$ is the effective cross section at the root of the phenomenon.
In-medium effects
\cite{Wambach93,Sedrakian94,Schulze97,Schnell98,Shternin13,Baldo14}
provide an effective mass $m^*<m$,
but also modify the effective cross section
in the opposite way, $\sigma^*>\sigma$.
The overall effect is an enhancement of the conductivity
\cite{Sedrakian94,Zhang10,Benhar10,Shternin13,Shternin20},
which remains dominant compared to the lepton conductivity
\cite{Gnedin95,Shternin07,Shternin08}
for temperatures $T\lesssim10^8\,$K.
However, the theoretical results of different groups differ
by even an order of magnitude,
see the discussions in \cite{Baiko01,Shternin13}.
As a further complication,
superfluidity affects the thermal conductivity in a highly nontrivial way
and can either enhance or reduce it \cite{Baiko01,Schmitt18}.

The quantitative understanding of the thermal conductivity
is only important for young NSs, $t\lesssim{\cal O}(10^2\,$yr),
where thermal relaxation is not yet reached in the interior.
For example,
a {\em reduction} of the thermal conductivity was assumed in
the medium-modified one-pion-exchange (MOPE) model of
\cite{Blaschke04,Blaschke12,Blaschke13,Grigorian16}  
in order to explain the apparent accelerated cooling of the Cas~A NS
\cite{Ho09,Heinke10,Elshamouty13,Posselt13,Ho15,Posselt18}
with an age of about 330 years.

\subsubsection{Neutrino emissivity}
\label{s:rates}

In this section we briefly recall the main neutrino emission mechanisms
in the NS core,
which dominate the total neutrino luminosity $L_\nu^\infty$,
and the relevance of the nucleon effective masses and other in-medium effects,
following closely the detailed treatment given in Ref.~\cite{Yakovlev01}
and updates in \cite{Potekhin15,Schmitt18}.
Only the rates for the non-superfluid scenarios will be given in this section,
for which the dependence on the effective masses is via the general factor
\cite{Baldo14}
\be
 M_{ij} \equiv  \left( \frac{\rho_p}{\rho_0} \right)^{1/3} \!\!
 \left(\frac{m_n^*}{m_n}\right)^i \left(\frac{m_p^*}{m_p}\right)^j \:.
\label{e:m}
\ee
In the presence of superfluidity the dependence becomes highly nontrivial
and requires detailed calculations~\cite{Yakovlev01}. 
In the following all emissivities $Q_\nu$ are given in units of
erg$\,$cm$^{-3}$s$^{-1}$.

\begin{figure}[t]
\centerline{\includegraphics[scale=0.6]{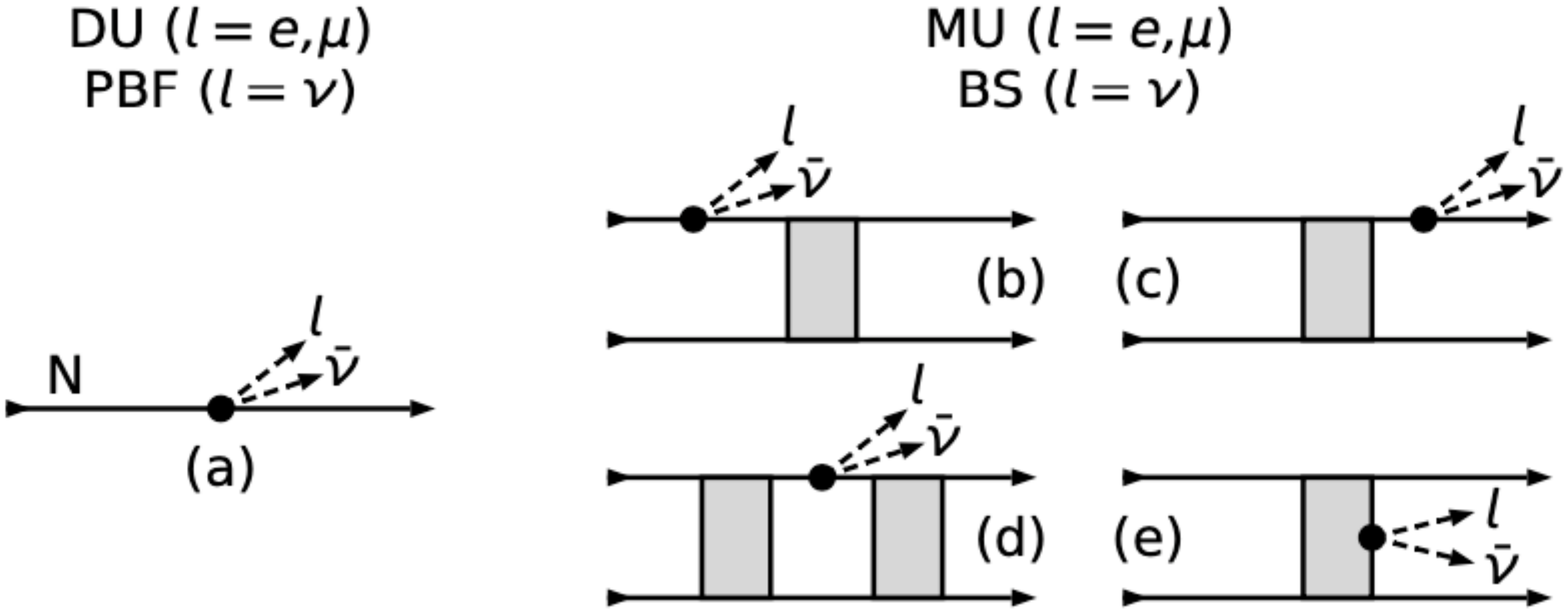}}
\vspace{-15mm}
\caption{
Elementary diagrams for neutrino emission.
}
\label{f:dia}
\end{figure}

{\par\medskip\noindent\em Direct Urca}

In the absence of pairing
three main mechanisms are usually taken into account,
which are the direct Urca (DU), the modified Urca (MU),
and the NN bremsstrahlung (BS) processes,
as sketched in Fig.~\ref{f:dia}.
By far the most efficient mechanism of NS cooling is the DU process,
Fig.~\ref{f:dia}(a),
which is in fact the in-medium neutron beta decay followed by
its inverse reaction
\cite{Boguta81,Lattimer91}:
\be
 n \rightarrow p + e^- + \bar{\nu}_e
 \qquad \text{and} \qquad
 p+e^- \rightarrow n+\nu_e \:.
\label{e:DU}
\ee
However, the energy and momentum conservation imposes a density threshold
on this process.
The result for $npe$ NS matter is given by
\be
 Q^{(DU)} \approx
 4.0\times10^{27} M_{11} T_9^6
 \mathop\Theta(k_F^{(p)} + k_F^{(e)} - k_F^{(n)}) \:,
\label{e:du}
\ee
where $T_9$ is the temperature in units of $10^9\,$K.
If muons are present,
then the corresponding DU process may also become possible,
in which case the neutrino emissivity is increased by a factor of 2.
Thus the DU process appears in a step-like manner,
once the proton fraction exceeds a value of 13--14~\%
(slightly dependent on the muon fraction at onset).
As the proton fraction $x$ is determined by the nuclear symmetry energy,
\be
 (3\pi^2\rho x)^{1/3} \approx 
 \mu_e = \mu_n - \mu_p
 \approx 4(1-2x) E_\text{sym}(\rho) \:,
\ee
EoSs with a large enough symmetry energy feature DU cooling.
This is the case for many microscopic EoSs,
see Fig.~\ref{fig:EOS} for illustration.

{\par\medskip\noindent\em Modified Urca}

If the DU process is kinematically forbidden or strongly reduced,
various less efficient neutrino processes may be operating in the NS core.
While the former is a pure weak reaction,
those processes are driven by strong interactions,
and are consequently affected by much greater theoretical uncertainties
that will be discussed in the following.
The two main ones with an emissivity $Q_\nu\propto T^8$ are the MU processes,
Fig.~\ref{f:dia}(b-e),
\be
 N + n \rightarrow N + p + e^- + \bar{\nu}_e
 \qquad \text{and} \qquad
 N + p + e^- \rightarrow N + n + \nu_e \:,
\label{e:MU}
\ee
where $N=n,p$ is a spectator nucleon that ensures momentum conservation.
Since five degenerate fermions are involved instead of three,
the efficiency is significantly reduced compared to the DU process.
The emissivities of the MU processes in the neutron and proton branches,
employing for the long-range part of the in-medium NN interaction
(gray block in Fig.~\ref{f:dia})
the dominant one-pion-exchange (OPE) contribution and
for the short-range part an effective contribution
in the framework of Landau theory,
are given by \cite{Friman79,Yakovlev95,Yakovlev01,Potekhin15}
\bal
 Q^{(Mn)} &\approx 8.1\times10^{21} M_{31} T_9^8 \alpha_n \beta_n \:,
 \label{e:mun}
\\
 Q^{(Mp)} &\approx 8.1\times10^{21} M_{13} T_9^8 \alpha_p \beta_p
 \big( 1 - k_F^{(e)}\!/4k_F^{(p)} \big)
 \mathop\Theta(3k_F^{(p)} + k_F^{(e)} - k_F^{(n)}) \:,
\label{e:mup}
\eal
where the factors $\alpha_n,\alpha_p$
take into account the momentum-transfer dependence
of the squared reaction matrix element under the Born approximation,
and $\beta_n,\beta_p$ include the non-Born corrections
due to NN interaction effects,
which are not described by the OPE \cite{Yakovlev01}.
The currently adopted values are
$\alpha_p=\alpha_n=1.13$ and
$\beta_p=\beta_n=0.68$.
The main difference between the proton branch and the neutron branch
is the threshold character,
although usually irrelevant.
If muons are present in the dense NS matter,
the equivalent MU processes become also possible,
and accordingly several modifications should be included in
Eqs.~(\ref{e:mun},\ref{e:mup}),
as discussed in Ref.~\cite{Yakovlev01}.

As stated above, these MU rates have been derived using a very simple
approximation (free OPE + Landau parameters)
for the in-medium NN interaction in the relevant diagrams
Fig.~\ref{f:dia}(b,c),
and are consequently density independent
apart from the effective-mass corrections and kinematic factors.
This has been attempted to improve in several works.

In particular,
in the medium-modified OPE (MOPE) model of
\cite{Senatorov87,Migdal90,Voskresensky01,Voskresensky18}
one asserts that
the emissivity receives a strong and dominant density-dependent correction
due to the softening of the pion mode
(reduction of the in-medium pion mass).
Then at $\rho>\rho_0$, diagrams of type Fig.~\ref{f:dia}(e) dominate
(not included in the OPE result),
which describe in-medium conversion of a virtual charged pion to a neutral pion
with the emission of a real lepton pair.
The modification factor for the medium-modified Urca reaction
with respect to the free OPE result is  
\cite{Schaab97,Blaschke04,Grigorian05,Blaschke12,Blaschke13,Kolomeitsev15}
\be
 \beta^{MMU}(\rho) \approx
 3 \Big( \frac{\rho}{\rho_0} \Big)^{10/3}
 \frac{[\Gamma(\rho)/\Gamma(\rho_0)]^6}{[\tilde\omega_\pi/m_\pi]^8} \:,
\label{e:mmu}
\ee
where
$\Gamma(\rho)\approx 1/[1+1.6(\rho/\rho_0)^{1/3}]$
is an estimate of the effect when
replacing the bare $\pi N\!N$ vertex by the dressed one,
computed from Landau-Migdal parameters,
and
\be
 \tilde\om^2 = \min{[-D_\pi^{-1}(\om=0,k)]} \:
\ee
is the effective pion gap in the medium,
computed from the dressed pion propagator $D_\pi(\om,k)$.
The density dependence of $\tilde\om$ is model dependent
and also related to a possible pion condensation at high density.
With typical parameters,
one obtains enhancement by a factor of $\sim3$ at $\rho_0$
and up to $5000$ at $3\rho_0$.
Note, however, that this enhancement strongly depends on the uncertain values
of the pion gap and the vertex correction at large density.

On the contrary,
when replacing the OPE approximation by the in-medium $T$-matrix
\cite{Blaschke95},
a {\em reduction} to about 1/4 of the OPE result was found.

Recently \cite{Shternin18},
a further `kinematic' in-medium effect was explored,
namely that for densities above the DU threshold
the propagator of the virtual nucleon after emission of the lepton pair
in Fig.~\ref{f:dia}(b)
might develop a pole that is unaccounted for in the standard treatment
leading to Eqs.~(\ref{e:mun},\ref{e:mup}).
This correction of the MU rates leads to a significant enhancement
by a factor of several,
in particular close to the DU threshold
$\rho \approx \rho_\text{DU}$,
and thus provides a more smooth switch-on of the DU process.

The problem of an accurate computation of the MU rates
can currently be considered unresolved,
as no formalism incorporating consistently
(also together with the computation of the EoS)
the various in-medium effects has yet been attempted.
It seems, however, that the simple expressions Eqs.~(\ref{e:mun},\ref{e:mup})
constitute underestimates of the real MU rates.

{\par\medskip\noindent\em Bremsstrahlung}

In the absence of the DU process,
the standard neutrino luminosity of $npe$ matter is determined
not only by the MU processes but also by the BS processes in NN collisions,
also described by Fig.~\ref{f:dia}(b-e),
\be
 N+N \rightarrow N+N+ \nu+\bar{\nu} \:,
\ee
with $N$ a nucleon and
$\nu$, $\bar{\nu}$ an (anti)neutrino of any flavor.
These reactions proceed via weak neutral currents and produce neutrino pairs
of any flavor \cite{Friman79,Yakovlev95}.
Contrary to the MU, an elementary act of the NN BS
does not change the composition of matter.
The BS has accordingly no thresholds associated with
momentum conservation and operates at any density in uniform matter.
In analogy with the MU process,
the emissivities depend on the employed model of NN interactions
and in OPE+Landau approximation are \cite{Yakovlev01}
\bal
 Q^{(Bnn)} &\approx 2.3 \times 10^{20} M_{40} T_9^8
 \alpha_{nn} \beta_{nn} {(\rho_n/\rho_p)^{1/3}} \:,
\label{e:bsnn}\\
 Q^{(Bnp)} &\approx 4.5 \times 10^{20} M_{22} T_9^8
 \alpha_{np} \beta_{np} \:,
\label{e:bsnp}\\
 Q^{(Bpp)} &\approx 2.3 \times 10^{20} M_{04} T_9^8
 \alpha_{pp} \beta_{pp} \:.
\label{e:bspp}
\eal
As for the MU rates,
the dimensionless factors $\alpha_{N\!N}$ come from the estimates of the
squared matrix elements at $\rho = \rho_0$:
$\alpha_{nn}=0.59$, $\alpha_{np}=1.06$, $\alpha_{pp}=0.11$.
The correction factors $\beta_{N\!N}$ are taken as
$\beta_{nn}=0.56$, $\beta_{np}=0.66$, $\beta_{pp}=0.70$.
All three processes are of comparable intensity,
with $Q^{(Bpp)} < Q^{(Bpn)} < Q^{(Bnn)}$.
The simple estimates Eqs.~(\ref{e:bsnn},\ref{e:bsnp},\ref{e:bspp})
are about 50 times weaker than the OPE MU rates
Eqs.~(\ref{e:mun},\ref{e:mup}).
However, also in this case the in-medium effects might completely
change the picture:

As for the MU rates,
calculations show that the use of the realistic $T$-matrix instead of the OPE
leads to the suppression of the neutrino emissivity approximately
by a factor of four
\cite{Hanhart01,Timmermans02,VanDalen03,Li09,Li15},
although \cite{Blaschke95} found an even larger decrease by a factor of 10--20
in the medium.
Several medium effects to second order in the
low-momentum universal potential $V_\text{low\,k}$
were evaluated in \cite{Schwenk04},
yielding an overall reduction of the emissivity by a density-dependent factor
of about 2--4,
compared to the standard factor $\alpha_{nn}\beta_{nn}\approx0.35$.
In the MOPE model the in-medium correction for BS is
\cite{Blaschke04,Grigorian05,Blaschke12,Blaschke13,Kolomeitsev15}
\be
 \beta_{nn}^{MBS}(\rho) \approx 3 \Big(\frac{\rho}{\rho_0}\Big)^{4/3}
 \frac{[\Gamma(\rho)/\Gamma(\rho_0)]^6}{[\tilde\om_\pi/m_\pi]^3} \:.
\label{e:mbs}
\ee
Contrary to the $T$-matrix corrections,
significant enhancement at $\rho>\rho_0$ up to a factor of
$\beta_{nn}^{MBS}\sim100$ is obtained,
depending on the model adopted for the pion propagator.

We summarize that the strong nuclear interaction influences directly
several ingredients of the MU and BS neutrino emission rates,
and very large in-medium effects have been claimed in some publications.
However, conflicting results of both enhancement and suppression are reported
depending on the specific medium effect considered,
and no generally accepted view has yet emerged.
However, of even greater importance for the cooling theory are
the superfluid properties of nuclear matter,
as discussed in the following.

\subsection{Effects of superfluidity}
\label{s:pbf}

The neutrino cooling processes in NSs are dramatically affected by
the neutron and proton superfluidity
\cite{Yakovlev99,Yakovlev01,Lombardo01,Sedrakian06,Page13,Page14,
Chamel17,Haskell18,Sedrakian19},
and the knowledge of the pairing gaps $\Delta$ in the 1S0 and 3PF2 channels
of the NN interaction
in beta-stable matter is essential for a correct description
of the thermal evolution of a NS.
These superfluids are produced by the
$pp$ and $nn$ Cooper pairs formation
due to the attractive part of the NN potential,
and are characterized by a critical temperature $T_c \approx 0.567\Delta$
for isotropic gaps.
The pairing gaps reduce the phase space available for the cooling reactions
and also modify the neutrino-emission matrix elements by the effect of anomalous
propagators.
The latter feature is nearly always disregarded,
leading to universal superfluid modification factors
which do not depend on details of the strong-interaction matrix elements.
The occurrence of pairing leads to two relevant effects in NS cooling,
namely

(i) A strong reduction by several orders of magnitude when $T < T_c$
of the emissivity of the neutrino processes the paired component is involved in,
with a corresponding reduction of the specific heat of that component
(Sec.~\ref{ss:heat}).
For example, proton superfluidity in the core of a NS suppresses both Urca
processes (DU and MU) but does not affect the neutron-neutron BS.
The general expressions cannot be obtained analytically,
but fit formulas are available \cite{Yakovlev01,Gusakov02}.
For illustration we list the suppression factors of the emissivities
$R\equiv Q_\Delta/Q_0$ in the case of p1S0 pairing
in the low-temperature limit
$\nu \equiv \Delta/T \gg 1$
for the various neutrino reactions
\cite{Levenfish94b,Yakovlev95,Yakovlev01,Gusakov02}:
\bal
 R_{DU} &\approx 7.19\times10^{-4} \nu^{5.5} e^{-\nu} \:,
\\
 R_{Mn} &\approx 0.166\times10^{-5} \nu^{7.5} e^{-\nu} \:,
\quad
 R_{Mp} \approx 1.44\times10^{-5} \nu^7 e^{-2\nu} \:,
\\
 R_{Bnp} &\approx 0.516\,\nu e^{-\nu} \:,
\quad
 R_{Bpp} \approx 0.216\,\nu^2 e^{-2\nu} \:.
\eal
At intermediate temperatures, the suppressions follow power laws, though.
Note that in the absence of neutron pairing,
the $Bnn$ reaction becomes the dominant unsuppressed cooling mechanism.

(ii) Onset of the ``Cooper pair breaking and formation'' (PBF) process
with associated neutrino-antineutrino pair emission,
\be
 N \ra N + \nu + \bar\nu \:,
\ee
which is in fact a neutral-weak-current variant of the DU process
Fig.~\ref{f:dia}(a),
and kinematically forbidden in normal matter.
However in superfluid matter it becomes unblocked
\cite{Flowers76,Voskresensky87,Yakovlev01} 
and the emissivity can be written as
\be
 Q_N^{(PBF)} \approx 3.5 \times 10^{21}
 \frac{m_N^*k_F}{m_N^2} T_9^7 a_N F_N(y) 
\ee
with $y\equiv\de(T)/T$.
For both 1S0 and 3P2 pairings,
the emission of neutrinos by the PBF process can occur
through axial and vector channels.
The computation of the factor $a_N=a_N^V+a_N^A$
from the elementary superfluid weak interaction matrix element
and its renormalization due to the strong interaction
is very intricate and has been tackled during the last decade
\cite{Yakovlev01,Kundu04,Leinson06a,Leinson06b,Sedrakian07,Kolomeitsev08,
Leinson08,Page09a,Steiner09,Leinson09,Leinson10,Kolomeitsev10,Kolomeitsev11,
Sedrakian12,Leinson13,Shternin15,Leinson16},
with the current state of results for p1S0 and n3P2 pairing:
\be
 a_p \approx  g_V^2 v_F^4  +  g_A^2 v_F^2      
\\ \ , \quad
 a_n \approx g_A^2/2 \approx 0.8  
\:,
\ee
where $g_V=1$ and $g_A=1.26$ are the nucleon weak vector
and axial-vector coupling constants.
Thus, the PBF process for singlet pairing is a relativistic effect
($v_F = k_F/m^* \ll 1$)
and much weaker than the other cooling processes.
However, the PBF contribution for triplet pairing
is important and even dominant for certain temperature ranges,
as determined by the universal control function
\be
 F(y) = y^2 \int_0^\infty\!\! dx\,
 \Bigg( \frac{x^2+y^2}{1+e^{\sqrt{x^2+y^2}}} \Bigg)^2 \:.
\ee
Qualitatively this process starts once the temperature decreases to $T_c$
of a given type of baryons,
becomes maximally efficient when $T \approx 0.8\,T_c \approx 0.45\de$,
and then is exponentially suppressed for $T \ll T_c$~\cite{Yakovlev01}.
Thus during the cooling evolution and in regions of the star
where locally $T(r,t) \approx 0.4\de(r)$,
the n3P2 PBF process might be the dominant cooling reaction.


\subsection{Nuclear pairing gaps}
\label{s:gap}

As evidenced in the previous section,
the most important nuclear physics input
for the superfluid properties of stellar matter
are the pairing gaps $\de_i(\rho)$ in the different partial waves.
Usually the most important ones are the p1S0 and n3P2 pairing channels,
while the n1S0 (crust only) gap is much less relevant
for the cooling \cite{Wei20},
and the p3P2 gap is normally disregarded due to its uncertain properties
at extreme densities.

Microscopically the gaps are determined by the irreducible NN interaction
in the different partial waves,
apart from the s.p.~properties (EoS).
In the simplest BCS approximation,
and detailing the more general case of pairing in the coupled 3PF2 channel,
the pairing gaps are computed by solving the (angle-averaged) gap equation
\cite{Amundsen85,Baldo92,Takatsuka93,Elgaroy96,Khodel98,Baldo98}
for the two-component $L=1,3$ gap function,
\be
  \left(\!\!\!\begin{array}{l} \de_1 \\ \de_3 \end{array}\!\!\!\right)\!(k) =
  - {1\over\pi} \int_0^{\infty}\!\! dk' {k'}^2 {1\over E(k')}
  \left(\!\!\!\begin{array}{ll}
   V_{11}\!\! & \!\!V_{13} \\ V_{31}\!\! & \!\!V_{33}
  \end{array}\!\!\!\right)\!(k,k')
  \left(\!\!\!\begin{array}{l} \de_1 \\ \de_3 \end{array}\!\!\!\right)\!(k')
\label{e:gap}
\ee
with
\be
  E(k)^2 = [e(k)-\mu]^2 + \de_1(k)^2 + \de_3(k)^2 \:,
\ee
while fixing the (neutron or proton) density,
\be
  \rho = {k_F^3\over 3\pi^2}
   = 2 \sum_k {1\over 2} \left[ 1 - { e(k)-\mu \over E(k)} \right] \:.
\label{e:rho}
\ee
Here $e(k)=k^2\!/2m+U(k)$ is the s.p.~energy,
$\mu \approx e(k_F)$ is the chemical potential
determined self-consistently from Eqs.~(\ref{e:gap}--\ref{e:rho}),
and
\be
   V^{}_{LL'}(k,k') =
   \int_0^\infty \! dr\, r^2\, j_{L'}(k'r)\, V^{IS}_{LL'}(r)\, j_L(kr)
\label{e:v}
\ee
are the relevant matrix elements
($I=1$ and
$S=1$; $L,L'=1,3$ for the 3PF2 channel,
$S=0$; $L,L'=0$ for the 1S0 channel)
of the bare NN potential $V$.
The relation between (angle-averaged) pairing gap at zero temperature
$\de \equiv \sqrt{\de_1^2(k_F)+\de_3^2(k_F)}$
obtained in this way and the critical temperature of superfluidity is then
$T_c \approx 0.567\de$.
(If no angle average is performed, the prefactor varies slightly,
see, e.g.~\cite{Yakovlev99,Yakovlev01}, 
but the angle-average procedure is usually an excellent approximation
\cite{Baldo92,Papakonstantinou17}).

\begin{figure}[t]
\vspace{-0mm}
\centerline{\includegraphics[angle=270,scale=0.85]{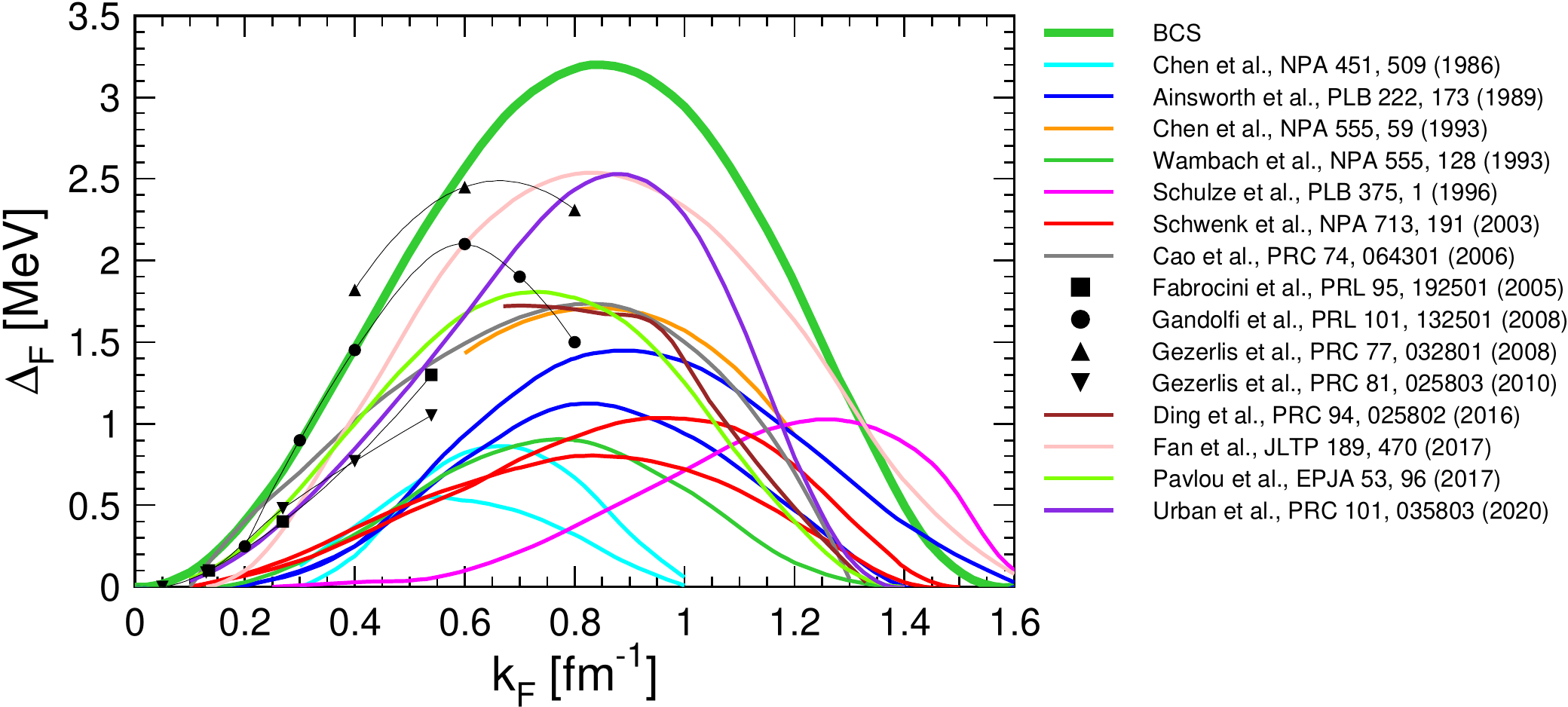}}
\vspace{-0mm}
\caption{
Theoretical predictions for the 1S0 gap in pure neutron matter
\cite{Chen86,Ainsworth89,Chen93,Wambach93,Schulze96,Schwenk03,Cao06,
Fabrocini05,Gandolfi08,Gezerlis08,Gezerlis10,Ding16,Pavlou17,Fan17,Urban20}.
Curves represent traditional approaches and
markers denote Monte Carlo calculations of different kind.
\off{
Chen \cite{Chen86},             
Ainsworth \cite{Ainsworth89},   
Chen \cite{Chen93},             
Wambach \cite{Wambach93},       
Schulze \cite{Schulze96},       
Schwenk \cite{Schwenk03},       
Cao \cite{Cao06},               
Fabrocini \cite{Fabrocini05},   
Gandolfi \cite{Gandolfi08},     
Gezerlis \cite{Gezerlis08},     
Gezerlis \cite{Gezerlis10},     
Ding \cite{Ding16},             
Fan \cite{Fan17},               
Pavlou \cite{Pavlou17},         
Urban \cite{Urban20}            
}}
\label{f:gap}
\end{figure}

However, in-medium effects might strongly modify these BCS results,
as both the s.p.~energy $e(k)$
(effective mass $m^*$ and quasiparticle strength $Z$)
and the interaction kernel $V$ itself
are affected by TBF (Sec.~\ref{s:tbf})
and polarization corrections.
It turns out that in the p1S0 channel all these corrections lead
to a suppression of both magnitude and density domain of the BCS gap
\cite{
Chen86,      
Ainsworth89, 
Wambach93,   
Chen93,      
Schulze96,   
Schulze01,   
Schwenk03,   
Zhou04,      
Fabrocini05, 
Cao06,       
Baldo07,     
Gandolfi08,  
Gezerlis08,  
Margueron08, 
Gezerlis10,  
Ding16,      
Pavlou17,    
Fan17,       
Drischler17, 
Ramanan18,   
Urban20},    
see Fig.~\ref{f:gap} for a compilation of previous results
and, e.g.~\cite{Lombardo01,Gezerlis14,Pavlou17,Sedrakian19}
for further reviews.
The results are widespread,
which certifies the extreme sensitivity of the gap to details of interaction,
approximations, and theoretical approach.

The situation is even worse for
the gap in the n3P2 channel,
which already on the BCS level depends on the NN potential
\cite{Takatsuka93,Baldo98,Khodel98,Maurizio14,Srinivas16,Drischler17},
as at high density there is no constraint by the NN phase shifts.
Furthermore TBF act generally attractive in this channel,
but effective mass and quasiparticle strength reduce the gap
and polarization effects on $V$ might be of both signs,
in particular in asymmetric beta-stable matter
\cite{
Khodel04,            
Zhou04,              
Schwenk04b,          
Dong13,              
Ding16,              
Drischler17,         
Papakonstantinou17,  
Ramanan20}.          
Note that most theoretical investigations consider only pure neutron matter.
Thus, due to the high-density nature of this pairing,
the various medium effects might be very strong and competing,
and there is still no reliable quantitative theoretical prediction for this gap.

\subsection{Cooling simulation scenarios}

As detailed in the preceding sections,
the sophistication of cooling simulations has evolved with time and we briefly
report here the main developments.

\begin{figure}[t]
\vspace{-0mm}
\centerline{\includegraphics[scale=0.43]{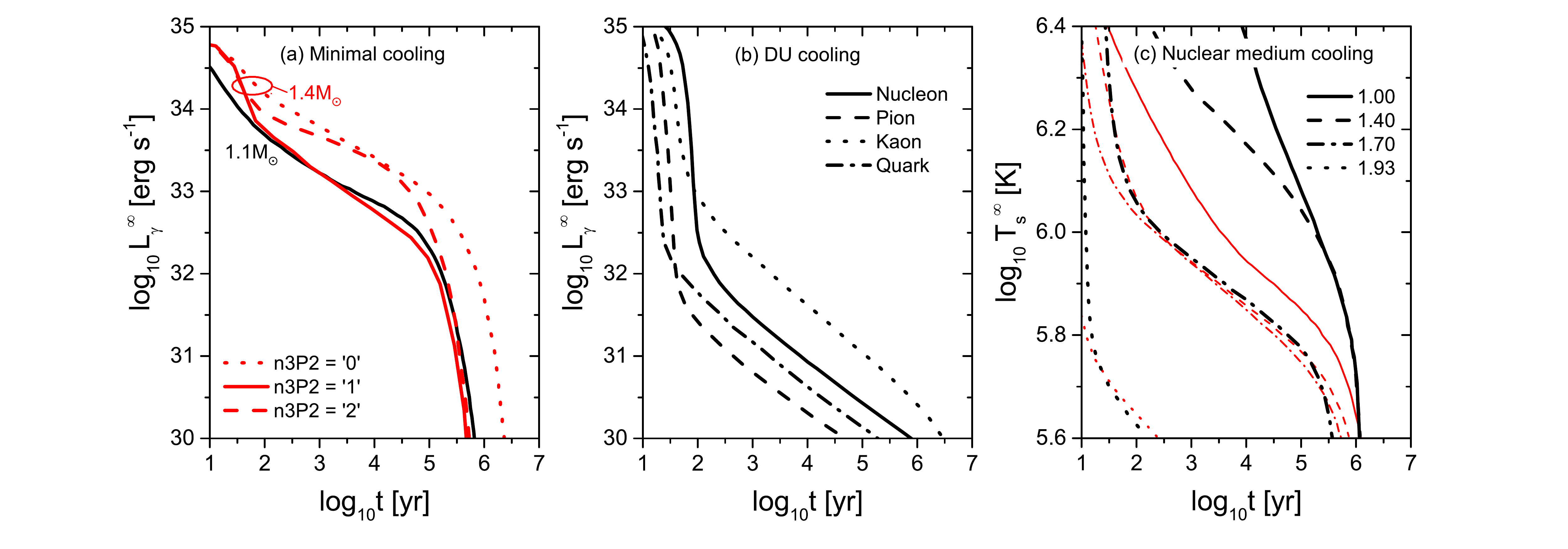}}
\vspace{-6mm}
\caption{
Example cooling diagrams for various cooling scenarios: 
{\bf (a)} Minimal cooling scenario
with the APR EoS \cite{Akmal98}
($M_\text{DU}=2.03\ms$)
and heavy-element envelopes.
For the cooling curves of $M=1.4\ms$,
three choices of the n3P2 gaps are considered
(`0', `1', `2'; the size of the n3P2 gap increases with the number).
Adapted from Ref.~\cite{Page04}. 
{\bf (b)} Including different DU processes
with the G$_{300}$ EoS \cite{Glendenning89} 
($M_\text{DU}=1.38\ms$)
inside a $1.4\ms$ NS,
except the kaon case whose mass is $1.8\ms$.
Adapted from Ref.~\cite{Schaab96}. 
{\bf (c)} Nuclear medium scenario
with the HHJ (parameterized softened APR) EoS \cite{Heiselberg99}
($M_\text{DU}=1.84\ms$)
for several choices of the NS mass.
The thin red curves indicate the cooling without pairing,
while the thick black curves employ both neutron and proton gaps,
but the n3P2 gap suppressed by a factor 10.
Adapted from Ref.~\cite{Blaschke04}.  
}
\label{f:cool0}
\end{figure}

\begin{figure}[t]
\vspace{-14mm}
\centerline{\includegraphics[scale=0.6]{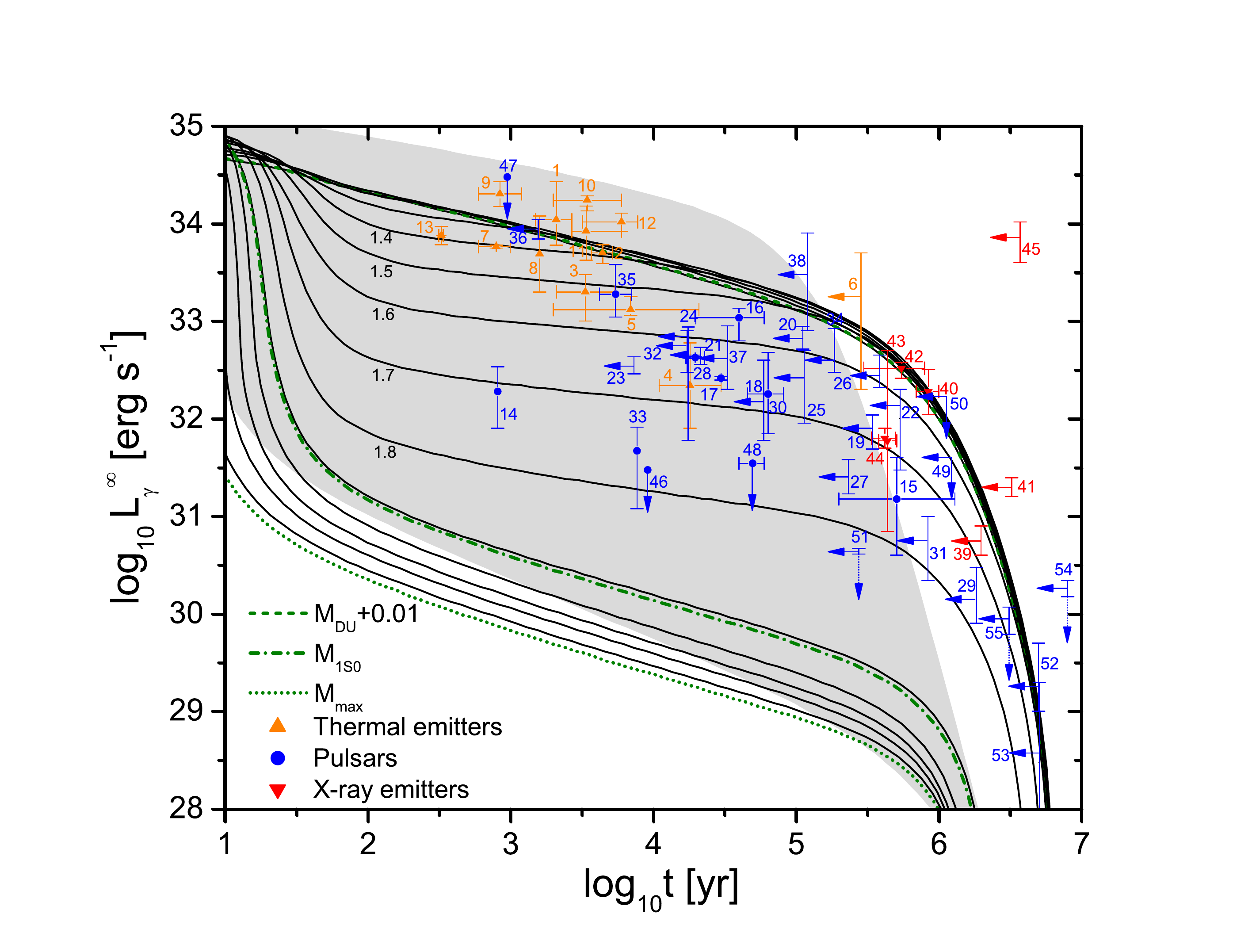}}
\vspace{-6mm}
\caption{
Cooling diagram for the BHF V18 EoS \cite{Lu19}
with p1S0 BCS gap and no n3P2 pairing,
for different NS masses $M/\!\ms=1.0,1.1,\ldots,2.3$
(decreasing black curves).
The dashed green curve
(hardly visible)
marks the NS mass $M_\text{DU}+0.01\ms=1.02\ms$
at which the DU process has just set in,
the dash-dotted green curve marks the NS mass $M_\text{1S0}=1.92\ms$
for which the p1S0 gap vanishes in the center of the star,
and the dotted green curve corresponds to $M_\text{max}=2.34\ms$.
The black curves are obtained with a Fe atmosphere and
the shaded areas cover the same results obtained with a light-elements
($\eta=10^7$) atmosphere.
The data points are from \protect\cite{Potekhin20}.
Adapted from \cite{Wei19}.
}
\label{f:cool}
\end{figure}

\subsubsection{Minimal cooling}

The occurrence of superfluidity in NS matter is generally accepted
and its effects have to be considered in the NS cooling simulations.
A proposed extension of the standard cooling scenario without superfluidity,
the so-called `minimal cooling scenario' \cite{Page04,Page09a},
includes the neutrino emission from the PBF processes
as the fastest cooling reaction (see Sec.~\ref{s:pbf})
and assumes that other enhanced neutrino emissions
(various DU processes) are not allowed.
This assumption is justified if stellar matter has a too small proton fraction
($\lesssim 13\%$),
and furthermore no exotic matter is present in the core.
Within minimal cooling scenarios,
the MU processes are suppressed by neutron and proton pairing gaps,
and the n3P2 PBF process dominates cooling in the entire neutrino cooling era.
This remains valid even with different models of pairing gaps \cite{Page04}.
The p1S0 PBF reaction is also present in the core of NSs, however,
its effect is practically negligible
\cite{Page04,Page09a,Wei19}.

A detailed discussion of the minimal cooling is given
in Refs.~\cite{Page04,Page09a}.
An example of cooling curves with PBF cooling \cite{Page09a}
is shown in panel (a) of Fig.~\ref{f:cool0}.
With PBF cooling NSs have a lower luminosity compared to the one
without superfluidity,
and this depends on the sizes of the gaps.
In addition, the cooling curves have a weak dependence on the stellar mass.
This results in a difficulty to describe some young cold NSs and old hot NSs.
Accordingly, for the minimal cooling model to be consistent with data,
it has to be fine-tuned to assure the most efficient cooling
\cite{Page06,Page09a,Grigorian16,Beloin18},
although fine-tuning the n3P2 gap \cite{Page11}
allows also to reproduce the
speculated accelerated cooling of the supernova remnant Cas~A
\cite{Ho09,Heinke10,Elshamouty13,Posselt13,Ho15,Posselt18}.

\subsubsection{Importance of DU cooling}

The efficient DU process seems to be required
in a consistent cooling model
in order to reproduce both relatively hot
(XMMU~J1731-347, no.~10 in Fig.~\ref{f:lt})
and relatively cold
(PSR~J0205+6449, PSR~B2334+61, PSR~J0007+7303, PSR~B1727-47; nos.~14,33,46,48)
cooling objects in the current database.
Apart from the nucleonic DU process,
possible exotic matter in the NS core,
e.g.~hyperons \cite{Grigorian18,Grigorian18b,Raduta19},
pion or kaon condensation \cite{Lattimer91,Blaschke04},
and deconfined quarks \cite{Page00,Grigorian05,Popov06b},
could also provide DU processes and result in a fast cooling,
as illustrated in panel (b) of Fig.~\ref{f:cool0}.

The possibility of a high enough proton fraction to allow the
nucleonic DU process was realized early in schematic models
\cite{Boguta81,Lattimer91,Page92}.
In fact many modern microscopic nuclear EoSs do reach easily the
required proton fractions for the DU process
\cite{Schaab96,Akmal98,Li08,Burgio10,Li12,Taranto13}
and are able to describe current cooling observations
of isolated NSs \cite{Taranto16,Wei19,Wei20},
as well as the reheated cooling of the accreting NSs in
X-ray transients in quiescence \cite{Beznogov15b,Fortin18,Potekhin19},
provided the DU process is dampened by the presence of nuclear pairing
throughout a sufficiently large (but not complete) set of the NS population.
In this framework a successful consistent modeling of all cooling data
(in particular the unusually hot XMMU~J1731-347 \cite{Klochkov15})
requires a p1S0 gap extended over a large enough density/mass range
and a very small or vanishing n3P2 gap
\cite{Blaschke04,Grigorian05b,Blaschke12,Blaschke13,
Beznogov15a,Beznogov15b,Taranto16,Beznogov18,Wei19,Wei20,Potekhin20}. 
This is opposite to the minimal cooling scenario,
where n3P2 PBF cooling is required as the only fast cooling reaction.

Fig.~\ref{f:cool} shows an example obtained with the BHF V18 EoS
(Sec.~\ref{s:bhfv18eos})
\cite{Wei19},
featuring a threshold mass of DU cooling $M_\text{DU} = 1.01\ms$,
so that practically all cooling curves involve nucleonic DU cooling.
The main effect of superfluidity on NSs with $M > M_\text{DU}$
is the strong quenching of the DU process,
and thus a substantial reduction of the total neutrino emissivity.
Hence those stars have a higher luminosity than in the non-superfluid case.
On the other hand, if $M > M_\text{1S0} = 1.92\ms$
(corresponding to a NS with the p1S0 gap vanishing in the center),
the complete blocking of the DU process disappears
and the star cools very rapidly again.
One observes results in line with these features in the figure,
namely between $M_\text{DU}$ and $M_\text{1S0}$
there is a smooth dependence of the luminosity on the NS mass
for a given age.
All objects in the database can be explained by choosing a proper atmosphere
model in each case.

\subsubsection{Medium-modified Urca processes}

As discussed in Sec.~\ref{s:rates},
the standard cooling rates disregard various in-medium effects.
The authors of the ``nuclear medium cooling'' approach
\cite{Schaab97,Voskresensky01,Blaschke04,Voskresensky18,Shternin18}
advocate very strong modifications of the different cooling processes,
in particular a strong density-dependent enhancement of the MU and BS reactions,
Eqs.~(\ref{e:mmu},\ref{e:mbs}).
This causes a smooth crossover from the standard to the enhanced cooling
scenario for increasing NS masses
and can describe reasonably well the young cold NSs
without the inclusion of DU processes and relevant gaps
\cite{Blaschke04,Blaschke13,Grigorian16}.
Panel (c) of Fig.~\ref{f:cool0} shows that
cooling curves of $1.0\ms$ (solid curve) and $1.4\ms$ (dashed curve) NSs
are widely separated compared to minimal cooling
where the curves overlap.
But, superfluidity is required for describing hot NSs.
The slow cooling and intermediate cooling objects
(for instance object 2, RX~J0822-43)
cannot be reproduced without gaps (thin curves).
However, the inclusion of the n3P2 gap results in a too efficient PBF process
and a too fast cooling,
which prevents to explain at least several of the cooling data.
One requires thus a strong suppression of the n3P2 gap
\cite{Blaschke04,Blaschke13,Grigorian05b,Grigorian16,Grigorian18},
just as in the models involving DU cooling.

\subsection{Epilogue} 

Accurate quantitative modeling of NS cooling with account of all
possible effects,
even on the purely nucleonic level,
is very complicated and has so far only been achieved in rather simplified ways.
We reiterate here only two of the most critical issues:

(i)
The elementary cooling rates, heat capacities, thermal conductivities, etc.,
have been computed based on very simplified in-medium NN interactions
(e.g.~OPE with severe kinematic averages).
Going beyond this approximation in one way or another
($T$ or $G$ matrix, in-medium pion propagation, kinematic corrections,
relativistic corrections, ...)
seems to indicate very important in-medium corrections
(potentially of several orders of magnitude),
but no consistent approach
(using an in-medium interaction computed consistently
based on the same NN interaction as for the nuclear EoS)
has yet been presented to our knowledge.
Therefore most current theoretical results may in general be considered only
order-of-magnitude estimates.

(ii) An even more important regulator of cooling rates
are the neutron and proton pairing gaps in NS matter.
While there is some general agreement on a substantial suppression
of the p1S0 gap,
the high-density n3P2 gap is currently theoretically not under control.

The main issue in cooling simulations is whether or not
the proton fraction in the stellar matter is high enough to allow DU cooling.
The former is the case for many modern microscopic EoSs.
Then cooling is determined practically only by properties of the DU process.
Evidence has been put forward for a theoretically-supported nucleonic scenario
in which the DU process is active for high-mass stars,
and damped by p1S0 pairing for intermediate-mass stars,
while n3P2 pairing is not allowed.
However, alternative scenarios are still possible
and can not generally be excluded by the current limited amount of cooling data.

Going beyond the nucleonic level,
the cooling can be affected by the presence of hyperons
(see Sec.~\ref{s:hyp} and,
e.g.~\cite{Tsuruta09,Grigorian18,Raduta19} and references therein)
or deconfined quarks
(\cite{Page00,Grigorian05,Popov06b,Wei20} and references therein),
by pion or kaon condensation
(see \cite{Lattimer91,Yakovlev01,Blaschke04} for review and references),
by emission of axions
(e.g.~\cite{Sedrakian19b} and references therein),
and other effects.
These scenarios are currently even more uncertain and unexplored
than the purely nucleonic one.

\section{Conclusions}
\label{s:conc}

In this review,
we have presented the current status of the EoS for compact objects,
and discussed the different theoretical approaches,
both microscopic and phenomenological,
for constructing the EoS.
Only nucleonic and hyperonic degrees of freedom were taken into account.
In spite of the tremendous progress in the last few years
regarding the actual constraints on the EoS,
coming from both nuclear physics experiments and astrophysical observations,
especially thanks to the recent GW detection,
it is not possible yet to seriously exclude one or more
scenarios depicted by the different theoretical models.
Recently a lot of attention has been devoted
to the results on NS mass and radius from the NICER mission,
as well as the tidal polarizability measured for the GW170817 event,
along with the observation of the large mass
of the millisecond pulsar MSP J0740+6620,
$M=2.14^{+0.10}_{-0.09}\ms$.
These constraints are at the moment the more stringent for selecting the EoS,
whereas those coming from laboratory nuclear experiments
are less strict since often model dependent.

We expect that in the near future a new generation of telescopes and projects
such as eXTP \cite{Watts19}
and advanced GW detectors such as LIGO and Virgo
can provide more and more precise data that can
significantly contribute to probe the internal structure of compact objects,
thus disentangling the different theoretical models.
Besides that, we have shown that the cooling theory of NSs
is a rich field of physics
with a great potential to constrain the high-density stellar EoS.
Here the main issue is whether the EoS is stiff enough
to allow a high proton fraction and
the rapid DU cooling process that dominates all other mechanisms.
Currently theoretical models are strongly hampered by uncertainties mainly
related to a correct treatment of the in-medium interactions,
so that theoretical predictions must be considered much less reliable
than those related to the `ordinary' EoS at subnuclear densities.

Despite all efforts,
we remark that there are still several open issues in the EoS modelling,
which need to be addressed in the near future.
These include:
\begin{itemize}
\item The role of TBF in the nuclear-matter EoS if only nucleonic degrees
of freedom are considered;
\item The solution of the ``hyperon puzzle'' in microscopic approaches
and the more general question about the treatment of non-nucleonic
degrees of freedom;
\item The conditions for a phase transition to quark matter to take place;
\item The importance of thermal effects for the construction of the EoS;
\item The cooling mechanisms, in particular pairing, consistent with the EoS.
\end{itemize}

Answers to those questions will surely improve our understanding of
the various fundamental phenomena related to the astrophysics of NSs,
their internal constitution,
the explosion mechanisms of core-collapse supernovae,
the mass threshold for BH formation,
the dynamics of NS binary mergers,
and the nucleosynthesis of heavy elements.
Therefore the current theoretical uncertainties will require significant
efforts to be undertaken in these directions in the near future.




\bibliographystyle{elsarticle-num} 


\newcommand\mathplus{+}
\newcommand\mdash{--}
\newcommand{\araa}{Annu. Rev. Astron. Astrophys.\ }
\newcommand{\aap}{Astron. Astrophys.\ }
\newcommand{\apj}{Astrophys. J.\ }
\newcommand{\apjl}{Astrophys. J. Lett.\ }
\newcommand{\apjs}{Astrophys. J. Suppl.\ }
\newcommand{\aaps}{Astron. Astrophys. Suppl. Ser.\ }
\newcommand{\epja}{EPJA\ }
\newcommand{\mnras}{Mon. Not. R. Astron. Soc.\ }
\newcommand{\nat}{Nature\ }
\newcommand{\npa}{Nucl. Phys. A\ }
\newcommand{\nphysa}{Nucl. Phys. A\ }
\newcommand{\physrep}{Phys. Rep.\ }
\newcommand{\plb}{Phys. Lett. B\ }
\newcommand{\prc}{Phys. Rev. C\ }
\newcommand{\prd}{Phys. Rev. D\ }
\newcommand{\prl}{Phys. Rev. Lett.\ }
\newcommand{\ppnp}{Progr. Part. Nucl. Phys.\ }
\newcommand{\ssr}{Space Science Reviews\ }

\bibliography{part1,part2,part3}

\end{document}